\documentclass[preprint,aps]{revtex4}
\usepackage{color}
\usepackage{mathtools}
\usepackage{array}
\usepackage{tabularx}
\usepackage{diagbox}
\usepackage{appendix}
\usepackage{amsmath,bm,amssymb,amsfonts,dcolumn,color,graphicx,graphics,latexsym,epsfig}
\usepackage{rsfso}
\usepackage[compatibility=false]{caption}
\usepackage{subcaption}
\usepackage{hyperref}
\usepackage{natbib}
\usepackage{float}
\usepackage{booktabs}
\usepackage{tikz}
\usetikzlibrary{positioning}
\usepackage{pifont}
\newcommand{\cmark}{\ding{51}}%
\newcommand{\xmark}{\ding{55}}%

\begin{document}

\title{ Generalised Hayward spacetimes: Geometry, matter and
scalar quasinormal modes}

\author{Poulami Dutta Roy, Sayan Kar}
\email{poulamiphysics@iitkgp.ac.in,sayan@phy.iitkgp.ac.in}
\affiliation{Department of Physics, Indian Institute of Technology 
Kharagpur, 721 302, India}

\begin{abstract}
\noindent  
Bardeen's 1968 idea of a regular black hole spacetime
was revived by Hayward in 2006 through the construction of a new example of 
such a geometry. Later it was realised by Neves and Saa, that a wider, two-parameter class exists, with Bardeen and Hayward spacetimes as special cases.
In this article, we revisit and 
generalise the Hayward spacetime by applying the Damour-Solodukhin (DS)
prescription. Recalling the DS suggestion of a deformed
Schwarzschild spacetime where $g_{tt} = -\left (1-\frac{2M_1}{r}\right )$, $g_{rr} = \left (1-\frac{2M_2}{r}\right )^{-1}$ and $M_1\neq M_2$, we 
propose a similar deformation of 
the Hayward geometry. The
$g_{tt}$ and $g_{rr}$ in the original Hayward line element
remain functionally the same, {\em albeit} mutations introduced via
differently valued metric parameters, following the DS idea. This results in a plethora of spacetime geometries, known as well as new,
and including singular black holes, wormholes or regular black holes.
We first study the geometric features and matter content of each of such spacetimes
in some detail.
Subsequently, we find the scalar quasinormal modes corresponding to
scalar wave propagation in these geometries. We investigate how the 
real and imaginary parts of the quasinormal modes depend on the
values and ranges of the metric parameters used to classify the geometries. 
Finally, we 
argue how our 
results on this family of
spacetimes suggest their utility as black hole mimickers.
\end{abstract}


\maketitle

\newpage

\section {\bf Introduction}
\noindent In General Relativity, black holes are largely 
considered today as {\em acceptable} spacetimes primarily because 
one has been able to {\em `prove'} their existence theoretically
\cite{penrose_1965},\cite{hawking_1970}
and also {\em `see'} them in numerous observational contexts \cite{bhobs_1,bhobs_2,bhobs_3,bhobs_4,bhobs_5} including the recent imaging of a
supermassive black hole at the center of the Milky Way \cite{EHT_1,EHT_2}. As we know, black holes are characterised by their event horizon(s) and
the singularity. On the other hand, we have {\em naked singularities}
which are, by definition horizon-less. The Penrose cosmic censorship hypothesis \cite{penrose_1965} says that naked singularities cannot be there but neither a
proof of this conjecture in a general setting exists nor are we able to
ignore the numerous  counterexamples representing nakedly singular spacetimes with matter satisfying all the energy conditions. 
Still, studies on naked singularities and cosmic censorship continue
and seem to lie in a sort of {\em comfort zone} among researchers!

\noindent Contrast the above-mentioned types with
spacetimes wherein the singularity is generically {\em absent}, i.e. we are in 
the so-called non-singular domain. Such spacetimes can still have horizon(s)
or they may not--the former leading us to the regular black hole while the latter suggests wormholes.  Thus, within each category 
fixed by a spacetime being either singular or non-singular, one finds further
sub-classes defined via the presence or the absence of horizon(s). 
Eventually, this leads us to black holes and naked singularities on one side and regular black holes and wormholes on the other. In this article,
we intend to focus mainly on this non-singular class--with or without
horizons, i.e. regular black holes and wormholes. It remains a 
fact there appears to be a good deal of hesitancy in accepting 
such spacetimes as possibilities in General Relativity or in any
other metric theory of gravity. The reason behind is the
issue of matter--in particular, what is the stress energy that could
make such spacetimes possible? It is known that there are genuine
issues about energy condition violations for wormholes \cite{morris_1988,thorne_1988} and 
even though we have viable models for the matter required for
regular black holes, they are plagued by the
presence of the mass-inflation instability. The mass parameter exponentially increases, for such regular spacetimes, with a timescale determined by the surface gravity of the inner horizon. This instability is well studied for both regular and singular black holes in the literature \cite{ori_1991,poisson_1989,carballo_2018,carballo_2021}. Recently, some
authors have suggested possible ways of taming this instability in regular black holes \cite{carballo_2022,bonanno_2021} but it still remains an open issue which needs to be further addressed. These problems with non-singular spacetimes makes one unsure whether such geometries are indeed worth consideration, until any
observation emerges in support. 

\noindent Despite the above pros and cons which are associated with
any study on regular black holes or wormholes, we prefer to remain
optimistic and try to find paths and avenues which may perhaps 
provide new knowledge about their characteristics and signatures.
Let us now take a look at the approach we follow in constructing
model spacetimes which represent regular black holes or wormholes.
Much of this is inspired by the Damour--Solodukhin idea where 
a line element of the following form was suggested and studied,
\begin{eqnarray}
ds^2 = -\left (1-\frac{2M_1}{r}\right ) dt^2 +\frac{dr^2}{1-\frac{2 M_2}{r}}
+ r^2 d\Omega_2^2
\end{eqnarray}
The novelty here is in the $M_1$, $M_2$ which are different though,
functionally, the $g_{tt}$
and $g_{rr}$ are, as in Schwarzschild spacetime. However, such a 
variation changes the geometry quite drastically. Firstly, there is
a nonzero energy-momentum (matter) and secondly, one can avoid a horizon
and live with just a throat (depending on the allowed range of values of
$M_1$, $M_2$).

\noindent The takeaway from the Damour--Solodukhin case is this freedom of
varying the constants in the metric functions and looking into
its consequences. We choose to do this for the regular black hole
known as Hayward spacetime which is within a 
broader class of spacetimes of which the Bardeen geometry appears to be
another special case. As we will see below, our generalisation (following 
the Damour--Solodukhin prescription) results in a wide variety of
possible spacetimes.  We have not bothered much about `matter content'
in the sense of a Lagrangian description, while proposing
these generalisations. One may consider this as a deficiency
but, at the same time,
we proceed and try to find different features/characteristics
pertaining to the geometry, matter and perturbations associated with
them. It goes without saying that it would certainly be of interest to
precisely know about matter sources for these spacetimes and we hope to address such issues in future. 

\noindent Our work is organised as follows. In the next section (Section II)
and also in Section III
we discuss the variety of spacetimes in full detail.  Sections IV and V
deal with scalar waves, quasinormal modes and echoes. In Section VI we
higlight how we can distinguish between different types of spacetimes
through the quasinormal modes.  Finally, Section VII is a conclusion.

\noindent Unless otherwise stated we have used geometrised units $G=c=1$.

\section {\bf The spacetime geometries: early work and a proposal} 

\noindent In a pioneering conference paper of 1968, the concept of a regular black hole was first
introduced by Bardeen \cite{bardeen_1968} following the ideas of Sakharov \cite{sakharov_1966} and Gilner \cite{gilner_1966} who discussed a possible end state of gravitational collapse without a final singularity (for a review see \cite{ansoldi_2008} and references therein). These spherically-symmetric singularity free solutions, known 
today as the Bardeen regular black holes, opened a new window in the class of regular solutions in GR. The Bardeen regular black hole metric has the form,
\begin{equation}
    ds^2= -\Big(1-\frac{2 M r^2}{(r^2+g^2)^{3/2}}\Big) dt^2 + \frac{dr^2}{\Big(1-\frac{2 M r^2}{(r^2+g^2)^{3/2}}\Big)} + r^2 (d\theta^2 + sin^2 \theta d\phi^2)
\end{equation}
with $g$ being a constant and $M$ as the mass parameter. Setting $g \rightarrow 0$ reduces the solution to the standard Schwarzschild black hole. The above metric 
was later found as a magnetic solution to Einstein's equations coupled to nonlinear electrodynamics \cite{ayon-Beato_2000}. The parameter $g$ there, is physically interpreted as the monopole charge of the self-gravitating magnetic field.\\
Later, in \cite{neves_2014} a generalization of the Bardeen metric was introduced 
with a line element of the form
 \begin{equation}
     ds^2 = -\Big(1- \frac{2 m(r)}{r} \Big) dt^2 +\frac{dr^2}{1-\frac{2 m(r)}{r}} + r^2 (d\theta^2 + sin^2 \theta d\phi^2)
     \label{eq:gen_metric}
 \end{equation}
with
\begin{equation}
    m(r) = \frac{M_0}{\Big( 1+ (\frac{r_0}{r})^q\Big)^{p/q}}
\end{equation}
The $M_0$ and $r_0$ here are mass and length parameters respectively, with p and q being constants. This spacetime reduces to the Bardeen metric with a choice of $p=3,q=2$ and proper identification of $M_0$ and $r_0$. 

\noindent Another kind of regular black hole known today as the Hayward metric, 
first appeared as an independent class of regular solutions in \cite{hayward_2006}.
One can check that the Hayward spacetime may be 
 obtained from the broader class mentioned above \cite{neves_2014}, by 
choosing $p=q=3$. Hence, both the Bardeen and Hayward geometries 
arise as special cases of the generalised $(p,q)$ spacetimes suggested in 
\cite{neves_2014}. We will concentrate mostly on the Hayward variety in this
article.

\noindent The line element for the Hayward spacetime is explicitly written as
\begin{equation}
    ds^2 = - F(r) dt^2 + \frac{dr^2}{F(r)} + r^2 (d\theta^2 + \sin^2 \theta
    d\phi^2)
    \label{eq:hayward_BH}
\end{equation}
with 
\begin{equation}
    F(r) = 1 - \frac{2 M r^2}{r^3+ 2 M \ell^2}
\end{equation}
where $(\ell,M)$ are positive constants. The function $F(r)$ has the property that it reduces to the Schwarzschild form asymptotically
\begin{equation*}
    F(r) \sim 1 - \frac{2 M}{r} \hspace{0.2in} as \hspace{0.2in} r \rightarrow \infty
\end{equation*}
and being a regular spacetime, it behaves like a de-Sitter space for small values of $r$
, i.e. 
\begin{equation*}
    F(r) \sim 1 - \frac{r^2}{\ell^2} \hspace{0.2in} as \hspace{0.2in} r \rightarrow 0
\end{equation*}
The cosmological constant is identified as $\Lambda =3/\ell^2$ in the small $r$ limit. This regular black hole can also be sourced by nonlinear electrodynamics though in the weak field limit, standard Maxwellian electrodynamics is not recovered \cite{lemy_2018}. \\

\noindent Several articles in the literature have proposed generalized mass functions that result in different kinds of regular black hole solutions \cite{fan_2016,neves_2014,balart_2014,ayon-Beato_2004}, one of which is mentioned in eq.(\ref{eq:gen_metric}). Through our work, we propose another generalisation of the Hayward regular black hole. In our construction 
we retain the functional forms of the metric functions, 
but tweak the parameters therein to generate a wide range of diverse spacetimes that 
includes regular and singular black holes as well as wormholes.\\
In more concrete terms, we consider a spherically symmetric, static metric 
\begin{equation}
    ds^2 = - f_1(r) dt^2 + \frac{dr^2}{f(r)} + r^2 (d\theta^2 + \sin^2 \theta d\phi^2)
    \label{eq:metric}
\end{equation}
where $f_1(r)$ and $f(r)$ are chosen as 
\begin{equation}
    f(r) = 1- \frac{2 M r^2}{ r^3 + 2 M \ell^2}; \hspace{0.2in} f_1(r) = 1- \frac{2 M_1 r^2}{ r^3 + 2 M_1 \ell^2}
\end{equation}
The novelty in the metric above is that it yields a generalization of the Hayward regular black hole mentioned in eq.(\ref{eq:hayward_BH}) through the presence of a 
new parameter $M_1$ in $f_{1}(r)$ which is different from the $M$ in $f(r)$. 
The introduction of the parameter $M_1$ is in the same vein as was done in the Damour-Solodukhin (DS) construction\cite{damour_2007}. In the DS wormhole, alternatively known as the black hole foil, the metric deviates from the Schwarzschild black hole 
through the 
appearance of a parameter $\lambda$ in the $g_{tt}$ component of the metric. Effectively, for DS wormhole we have $g_{tt} =-\left (1-\frac{2M}{r}\right )$ and $g_{rr} = \left (1-\frac{2M_1}{r}\right )^{-1}$ with $M\neq M_1$. 

\noindent The mass parameter $M_1$, in the generalised Hayward case, 
extends the metric to a broader family that generates diverse types of spacetimes 
which includes the regular Hayward black hole.
Further simplification can be achieved by writing the parameters $M,M_1,\ell$ in dimensionless form using $\sigma=\frac{M_1}{M}$, $\kappa=\frac{\ell}{M}$ with the radial coordinate $x=\frac{r}{M}$. The line element then becomes
\begin{equation}
    ds^2 = - \Big( 1- \frac{2 \sigma x^2}{x^3+ 2 \sigma \kappa^2}\Big) dt^2 + \frac{dr^2}{\Big( 1- \frac{2 x^2}{x^3+ 2 \kappa^2}\Big)} + r^2 (d\theta^2 + \sin^2 \theta d\phi^2)
    \label{eq:dimensionless_metric}
\end{equation}
The benefit of writing the metric in this fashion (i.e. via dimensionless parameters)
will be evident in the following sections where we discuss numerous special cases. 
Thus, the final form which we use in this article is the one given in eq.(\ref{eq:dimensionless_metric}). Henceforth, we will use this line element
in all future discussions and special characteristics will be
linked to different values and ranges of $(\sigma,\kappa)$. Below, we indicate
the different aspects we elaborate upon in subsequent sections.

\subsection{Matter energy-momentum}

\noindent  Obtaining the Einstein tensor components (in the frame basis) 
for the above-stated line element and
using them in the Einstein field equations  $G_{ij} = 8\pi T_{ij}$, 
we write down $\rho$, $\tau$ and $p$ which comprise the diagonal elements in 
the energy-momentum tensor $T_{ij}$, in the frame basis. 
In order to get simplified expressions for the energy-momentum tensor components, we define the following quantities,
\begin{gather}
    \begin{split}
   & h_{1 \sigma} = x^3+ 2 \sigma \kappa^2 - 2 \sigma x^2 ; \hspace{0.5in} h_\sigma = x^3+ 2 \sigma \kappa^2\\
   & h_1 = x^3+ 2 \kappa^2 -2 x^2; \hspace{0.63in} h = x^3+ 2 \kappa^2\\
   & A = \frac{d}{dx} log \Big( \frac{h_{1 \sigma}}{h_\sigma}\Big) \hspace{0.8in} B = \frac{d}{dx} log \Big( \frac{h_{1}}{h}\Big)
    \end{split}
\end{gather}
Now we are in a position to write the $T_{ij}$ components (in the frame basis) making use of the newly defined quantities $h_{1 \sigma}, \, h_\sigma, \, h_1, \, h$ along with $A,B$. We obtain,
\begin{gather}
 \begin{split}
    &\rho = \frac{1}{8 \pi x^2 M^2} \Big( 1- \frac{h_1}{h} \Big) - \frac{1}{8 \pi M^2} \frac{h_1}{h} \frac{B}{x}\\
   & \tau = \frac{-1}{8 \pi x^2 M^2} \Big(1- \frac{h_1}{h}\Big) + \frac{1}{8 \pi M^2} \frac{h_1}{h} \frac{A}{x}\\
    &p = \frac{1}{8 \pi x^2 M^2} \frac{h_1}{h} \Big[ \frac{1}{2} A x + \frac{1}{4} A (A+B) x^2 + \frac{1}{2} x^2
 \frac{dA}{dx} + \frac{1}{2} B x\Big]   
 \label{eq:EC}
 \end{split}
\end{gather}
The expressions above will hold true for any $h_{1\sigma}$,$h_\sigma$, $h_1$ and $h$,
irrespective of their functional forms. Note also that $f_1 = \frac{h_{1\sigma}}{h_\sigma}$ and $f=\frac{h_1}{h}$.

\noindent It is easy to see that $h_1$, $h$ and hence $B$ and $\rho$ are independent of the parameter $\sigma$. On the other hand, the $\sigma$ dependence arises in $\tau$ and $p$ through the second term in $\tau$ and first three terms in $p$.\\
Some observations about the energy-momentum tensor for different spacetimes are in order. When we set $\sigma=1 \rightarrow A=B$, we obtain the condition for the Hayward regular black hole for which $\rho =-\tau$. The regular black hole is characterized with a de-Sitter core which corresponds to $\rho = -\tau = -p = \frac{3}{8 \pi \kappa^2 M^2}$ at $x=0$. As expected, at the center of the regular black hole, only the parameter $\kappa$ remains which is nothing but the redefined cosmological constant. The above condition satisfied by $\rho,\tau$ and $p$ at $x=0$ holds true even for spacetimes with $A \neq B \neq 0$. 
To further explore the status of the energy conditions, specific parameter values have to be chosen. The above complicated expressions (especially that for $p$) are much simplified in most of the cases we deal with. This will be done for each spacetime, separately, in the following section on classification.

\subsection{Embedding diagrams}

\noindent  The embedding of two dimensional sections of the
given spacetime, in a Euclidean background, helps us visualise the shape of 
each geometry, as associated with different parameter ranges. We begin by  considering the $t=constant$, $\theta =\pi/2$ spatial 2- D slice of the spacetime and  embed it in 3-D Euclidean space written in cylindrical coordinates. The 2-D spatial slice of the generalized metric from eq.(\ref{eq:metric}) gives
\begin{equation}
    ds^2 = \frac{dr^2}{f(r)} + r^2 d\phi^2 
\end{equation}
On the other hand, in cylindrical coordinates the 3-D flat space metric becomes
\begin{align}
    ds^{2} = d\zeta^{2} + \zeta^{2} d\psi^{2} + dz^{2}
    \label{eq:embed_1}
\end{align}
Comparing the two metrics (with $z=z(r)$ in eq.(\ref{eq:embed_1}), $\psi=\phi$,
$\zeta=r$), we arrive at the embedding functions as
\begin{gather}
    \zeta = r\\
    \frac{dz}{dr} = \sqrt{\frac{1}{f(r)}-1}
    \label{eq:embed}
\end{gather}

\noindent $z=z(r)$ needs to be integrated numerically for different values of $(\sigma,\kappa)$ which will be discussed in the forthcoming sections. It is to be noted that the embedding function depends only on the metric parameter $\kappa$ and all geometries having same value of $\kappa$ will have similar embedding diagram even though $\sigma$ values may be different.

\subsection{Defining the tortoise coordinate}

\noindent The radial coordinate ideally ranges from $[0,\infty)$ depending on the nature of the spacetime. But in most scenarios, it is convenient to work with tortoise coordinate $r_*$ defined as
\begin{equation}
    dr_*^2 = \frac{dr^2}{f_1(r) f(r)}
    \label{eq:tortoise}
\end{equation}
which ranges from $(-\infty,\infty)$. For wormholes, this mapping is especially helpful as it covers the two universes on both sides of the throat \cite{visser_1995}. Hence, in case of wormholes, $r = r_0$ (throat radius) is mapped to $r_* =0$. On the other hand, for black holes, the point $r_* = -\infty$ denotes the position of the horizon at some $r=$ constant value. For most values of parameter $(\sigma,\kappa)$, eq.(\ref{eq:tortoise}) is not integrable analytically . The tortoise coordinate is thus calculated numerically as an interpolating function using \textit{Mathematica}. The conditions imposed for the numerical computation of the tortoise coordinate are elaborated in the Appendix. 

\subsection{Classifying spacetimes: black holes and wormholes}

\noindent In our study we have been able to use  
a single line element which, for different parameter ranges,
may represent either black holes or wormholes. The essential criteria for classification into either of the two broad categories mentioned above will therefore have to be spelt out in full detail. In simple terms, how do we check, for some parameter
values, whether the geometry is a black hole or a wormhole? \\
To this end, we begin by finding the roots of the metric components $g_{tt}$ and $g^{rr}$. A black hole will have atleast one $r=constant$ null hypersurface which 
represents its event horizon. It may or may not harbour a singularity inside the horizon--the former being a 
singular black hole, while the latter represents a regular one. On the other hand, for a wormhole, we need a timelike hypersurface at some $r=r_0$ which will act as the throat and, around which, violation of the energy conditions will be unavoidable. Simultaneously, if the spacetime is regular for the entire range of $r$ namely $[r_0,\infty)$ then it can be classified as a wormhole. The violation of energy condition ensures the divergence of the congruence of geodesics which is the typical behavior associated with wormhole spacetimes. \\
Certain parameter choices might result in the spacetime having both a null and a timelike hypersurface. Their relative positions and the behavior of the other conditions will then be used 
to set up the classifying criteria.

\section{Classification: cases for various parameter ranges and features}

\begin{figure}[h]
 \centering
	\includegraphics[scale=0.58]{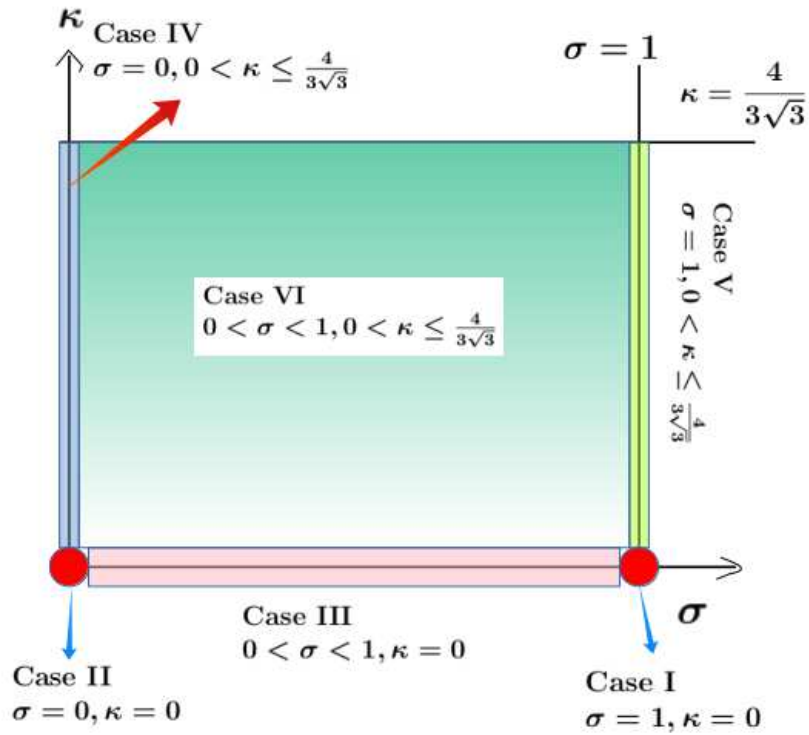}
 	\caption{Parameter space  ($\sigma$, $\kappa$) showing different spacetimes for specific, different ranges of the parameters.}
    \label{fig:parameter_space}
 \end{figure}
\noindent We will now explore the distinct geometries arising from specific ranges of the parameters $\sigma$ and $\kappa$. The Fig.(\ref{fig:parameter_space}) shows the six different spacetime classes obtained from our generalized metric. Along with checking the energy conditions satisfied by each spacetime, we will also study their geometric properties and embedding diagrams.

\subsection{\bf{Case I}: \texorpdfstring{$\sigma=1$, $\kappa=0$}{$σ=1$,$κ=0$} (Schwarzschild black hole)} 

\begin{equation}
    ds^2 = - \Big( 1- \frac{2 M}{r}\Big) dt^2 + \frac{dr^2}{\Big( 1- \frac{2 M}{r}\Big)} + r^2 (d\theta^2 + \sin^2 \theta d\phi^2)
\end{equation}

\noindent $\bullet$ \textbf{Nature of spacetime:} This choice simplifies the parameters as $M=M_1$ and $\ell=0$. The case then corresponds to the most trivial scenario, that of the Schwarzschild black hole with $f_1(r) = f(r) = 1- \frac{2 M}{r}$. The event horizon is at $r=2 M$ which is a null hypersurface as well as the infinite redshift surface. \\
\noindent $\bullet$ \textbf{Energy conditions:} The components of the energy-momentum tensor simply become zero since the Schwarzschild  black hole is a vacuum solution in GR. Hence, one need not worry about the energy conditions as they are trivially satisfied.  

\subsection{\bf{Case II}: \texorpdfstring{$\sigma=\kappa=0$}{$σ=κ=0$} (Schwarzschild wormhole)}

\begin{equation}
    ds^2 = - dt^2 + \frac{dr^2}{\Big( 1- \frac{2 M}{r}\Big)} + r^2 (d\theta^2 + \sin^2 \theta d\phi^2)
\end{equation}
\noindent $\bullet$ \textbf{Nature of spacetime:} The component $g_{tt} =-1$ makes the spacetime ultra-static and devoid of any horizon. The root of the metric function $g^{rr}$ gives $r=2 M$ as a timelike hypersurface which acts as a throat to the corresponding traversable wormhole. This results in the first wormhole geometry we encounter in the parameter space. The radial coordinate $r$ ranges from $[2 M , \infty)$ which denotes one copy of the universe while the tortoise coordinate $r_*$ covers both sides of the throat ranging from $(-\infty, \infty)$. The Ricci scalar for this wormhole is trivially 0 while the Kretschmann scalar remains finite for the entire range of $r$ indicating the absence of any singularity in the spacetime. \\
\noindent $\bullet$ \textbf{Energy conditions:} The energy-momentum tensor
components mentioned in eq.(\ref{eq:EC}) become, at the throat $r=2M$, $\rho(r=2 M) = 0$, $\tau(r=2 M) = -1/\left (8\pi (4M^2)\right )$ and $p(r=2 M)=1/\left ( 8\pi (8M^2)\right )$. Hence, the weak energy condition is violated at the throat, indicating the divergence of a congruence of geodesics.\\
\noindent $\bullet$ \textbf{Embedding diagram:} The embedding function $(z(r),\zeta(r))$ is obtained by integrating eq.(\ref{eq:embed}). This is the well-known Flamm's paraboloid.
$z(r)$ ($z(\zeta)$) is plotted in Fig. 2.
\begin{figure}[h]
 \centering
	\includegraphics[scale=0.7]{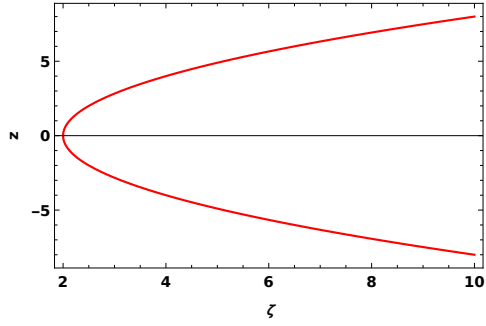}
 	\caption{Plot of $z(\zeta)$ showing embedding of the $t=constant$, $\theta=\pi/2$ slice of the Schwarzschild wormhole. }
 	\label{fig:embed}
 \end{figure}

\subsection{\bf{Case III}: \texorpdfstring{$0<\sigma<1,\kappa=0$}{$0<σ<1,κ=0$} (Damour--Solodukhin wormhole)}

\begin{equation}
    ds^2 = - \Big( 1- \frac{2 \sigma}{x}\Big) dt^2 + \frac{dr^2}{\Big( 1- \frac{2}{x}\Big)} + r^2 (d\theta^2 + \sin^2 \theta d\phi^2)
\end{equation}

\noindent $\bullet$ \textbf{Nature of spacetime:} Proceeding as before, we first determine the roots of the metric components $g_{tt}$ and $g^{rr}$ which gives $r=2 \sigma M$ and $r=2M$ respectively. While the $r=2M$ surface is timelike, the hypersurface corresponding to $r=2 \sigma M$ is null. In order to ascertain if the spacetime harbours any singularity, we calculate the curvature scalars and check their behaviour. It is observed that the Ricci and Kretschmann scalars remain finite at $r=2M$ but diverge only at $r=2 \sigma M <2M$.\\
\noindent $\bullet$ \textbf{Energy conditions:} It is found that the $\rho + \tau < 0$ at $r=2M$ which indicates violation of energy condition. The spacetime is thus a wormhole. The radial coordinate ranges from $[2M,\infty)$. The spacetime is regular everywhere in this domain of $r$. With proper choice of coordinates, this wormhole metric can be identified as the Damour-Solodukhin wormhole which is well-studied in the 
literature \cite{damour_2007}.\\
\noindent $\bullet$ \textbf{Embedding diagram: } Since the embedding will be independent of value of $\sigma$, the embedding diagram will be identical to the one shown in Fig.(\ref{fig:embed}) for the previous case  (the value of $\kappa$ here, is still 0).

\subsection{\bf{Case IV}: \texorpdfstring{$\sigma=0,\kappa \neq 0$}{$σ=0, κ ≠ 0$} 
(Hayward wormhole)}

\begin{equation}
    ds^2 = -  dt^2 + \frac{dr^2}{\Big( 1- \frac{2 x^2}{x^3+ 2 \kappa^2}\Big)} + r^2 (d\theta^2 + \sin^2 \theta d\phi^2)
\end{equation}

\noindent $\bullet$ \textbf{Nature of spacetime:} The metric tensor component $g_{tt} = -1$ indicates the ultra-static nature of this spacetime. Thus, the spacetime does not harbour any null surface and hence no horizon. The possibility of the spacetime being a black hole is 
thus ruled out completely. We then find the roots associated to $g^{rr}=0$ which exist provided $\kappa \leq \frac{4}{3 \sqrt{3}}$. The demand for the $g^{rr}$ component to have atleast one real, positive root sets the range for the parameter $\kappa$. Although calculating the discriminant of the cubic equation and setting it greater than zero gives the required restriction on $\kappa$, this condition simply refers to the equation having three distinct real roots. If the discriminant is less than 0, then the equation being cubic, will have two complex conjugate and one real root. The single real root might be positive or negative. So a better way to formulate the limit on $\kappa$ is by studying the behaviour of the cubic function in the range $x = [0,\infty)$. To perform this for any general $\kappa$, we first find its behaviour near $x=0$ and $x \rightarrow \infty$. Both $f(0)$ and $f(\infty)$ come out to be positive quantities. Now, taking derivative we get the critical points as $x=0, 4/3$. At $x=0$, the function is a maximum and at $x=4/3$, it is a minimum. If the value of the function $f(4/3)$ is negative, then the function is bound to intersect the x-axis at two points. Also, from Descartes rule of sign change we know the function can have two or zero real positive roots. Therefore, if $f(4/3)<0$, the function must have two positive real roots as desired. Simple algebra shows that the above condition reduces to $\kappa\leq \frac{4}{3 \sqrt{3}}$ in order for the function to have 
atleast one real positive root.\\
The larger root of $g^{rr}$ say $r=r_0$, which is also a timelike hypersurface, gives the position of the throat making the spacetime a wormhole. The scalars, Ricci and Kretschmann, remain finite for the entire range of $[r_0,\infty)$ making the spacetime regular everywhere. This wormhole geometry can be called a `Hayward wormhole' because of its similarity 
with the Hayward regular black hole metric with the only difference being its ultra-static nature.\\
\noindent $\bullet$ \textbf{Energy conditions:} In order to ensure the wormhole nature of the spacetime, divergence of a geodesic congruence is essential, at least at the throat, for all values of $\kappa$. The energy-momentum tensor components of eq.(\ref{eq:EC}) for this case becomes
 \begin{equation}
    \begin{split}
        \rho (x) = \frac{12 \kappa^2 }{8 \pi M^2 (x^3+ 2 \kappa^2)^2}\\
        \rho(x)+ \tau(x) = \frac{-2 (x^3 -4 \kappa^2)}{8 \pi M^2(x^3+ 2 \kappa^2)^2}\\
        \rho(x)+p(x) = \frac{(x^3 +8 \kappa^2)}{8 \pi M^2(x^3+ 2 \kappa^2)^2}     \end{split}
 \end{equation}
where we use the dimensionless radial coordinate $x =r/M$. The WEC will be satisfied provided the quantity $\rho(x)+\tau(x)$ is always positive. 
At the throat of the wormhole $x=x_0$, $\rho + \tau = -\frac{(3 x_0 -4)}{8 \pi M^2 (x_0^3+ 2 \kappa^2)}$. Now from the previous discussion about finding the roots of $f(r)$, we know that the larger root of $f(r)$ (i.e. the throat) to be greater than $x=4/3$ which indicates $\rho+\tau$ to be negative for all $\kappa$. A specific case arises when $\kappa = \frac{4}{3 \sqrt{3}}$ so that at throat $x_0 =4/3$, $\rho + \tau$ is 0. The violation then occurs a bit away from throat where $\rho + \tau$ is negative. Hence, WEC is always violated atleast at the location of the throat or near it, as is expected in a wormhole geometry. 

\begin{figure}[h]
 \centering
	\includegraphics[scale=0.7]{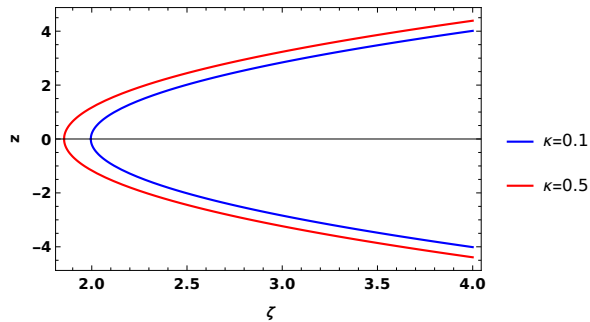}
 	\caption{Plot of $z(\zeta)$ showing embedding of the $t=constant$, $\theta=\pi/2$ slice of the Hayward wormhole for different $\kappa$ values. }
	\label{fig:Hayward_WH_embed}
 \end{figure}

\noindent $\bullet$ \textbf{Embedding diagram:} Once again, to construct the embedding diagram of the $t=$ constant hypersurface for this wormhole geometry, we integrate eq.(\ref{eq:embed}) for different values of $\kappa$ and plot them as shown in Fig.(\ref{fig:Hayward_WH_embed}). The throat radius of the wormholes also change with $\kappa$.

\begin{figure}[h]
 \centering
	\includegraphics[scale=0.7]{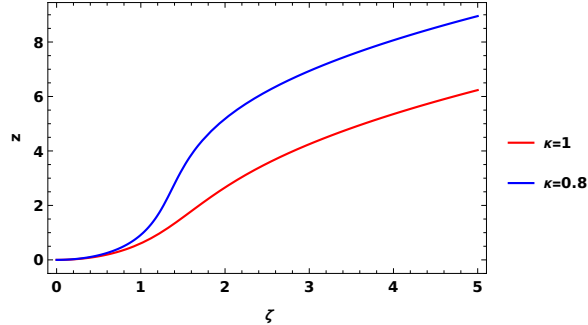}
 	\caption{The embedding diagram for Case IV spacetime with $\kappa> \frac{4}{3 \sqrt{3}}$ where the function can be smoothly extended to $r=0$. }
 	\label{fig:embed_kappa_1}
 \end{figure}

\noindent One might wonder about the case when $\kappa > \frac{4}{3 \sqrt{3}}$. In the absence of any real positive root to the metric component $g^{rr}$, the spacetime will be regular and can be smoothly extended to $r=0$. The corresponding embedding diagram obtained by integrating eq.(\ref{eq:embed}) for $\kappa > \frac{4}{3 \sqrt{3}}$ indicates a non-trivial behaviour as shown in Fig.(\ref{fig:embed_kappa_1}). The embedding function $z(\zeta)$, for such parameter ranges,  undergoes a change in gradient which is unique and needs to be explored through detailed future studies (see Fig.(\ref{fig:embed_kappa_1})). At present, we will only be interested in cases with $\kappa \leq \frac{4}{3 \sqrt{3}}$ where there is a well-defined horizon or a throat.

\subsection{\bf{Case V}: \texorpdfstring{$\sigma=1,\kappa \neq 0$}{$σ=1, κ ≠ 0$} (regular Hayward black hole)}

\begin{equation}
    ds^2 = - \Big( 1- \frac{2 x^2}{x^3 + 2 \kappa^2}\Big) dt^2 + \frac{dr^2}{\Big( 1- \frac{2 x^2}{x^3+ 2 \kappa^2}\Big)} + r^2 (d\theta^2 + \sin^2 \theta d\phi^2)
\end{equation}

\begin{figure}[h]
 \centering
	\includegraphics[scale=0.7]{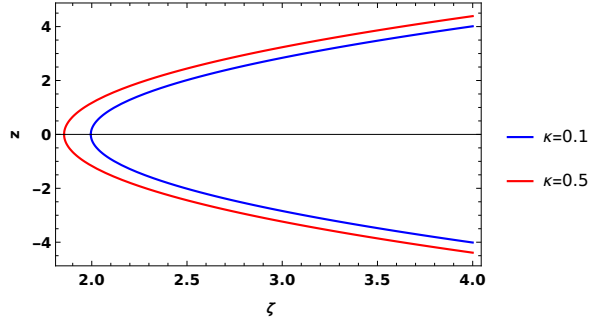}
 	\caption{Plot of $z(\zeta)$ showing embedding of the $t=constant$, $\theta=\pi/2$ slice of the Hayward regular black hole. }
 	\label{fig:Hayward_BH_embed}
 \end{figure}

\noindent $\bullet$ \textbf{Nature of spacetime:}  We notice that here $f_1(r)= f(r)$ i.e. the two mass functions are equal. For $f(r)$ to have roots, $\kappa \leq \frac{4}{3 \sqrt{3}}$. Equating $g_{tt}=0$ we get the larger root $r=r_{h} (x=x_{h})$ as a null hypersurface as well as the infinite redshift surface. The point $r=r_{h}$ gives the location of the horizon in this spacetime --thus it is a black hole. Extending the radial coordinate $r$ beyond the event horizon to $r=0$ shows that the curvature scalars remain finite even at $r=0$ indicating the spacetime to be a regular black hole. One notes that this spacetime is just
the  well-known Hayward regular black hole proposed in eq.(\ref{eq:hayward_BH}). \\
\noindent $\bullet$ \textbf{Energy conditions:} Calculation of the energy-momentum tensor components show the energy conditions to be satisfied for this regular black hole geometry.\\
\noindent $\bullet$ \textbf{Embedding diagram:} We once again integrate eq.(\ref{eq:embed}) for these parameter values and plot the embedding function for different values of $\kappa$ as shown in Fig.(\ref{fig:Hayward_BH_embed}). Similar to the previous case, the position of horizon changes with $\kappa$.

\subsection{\bf{Case VI}: \texorpdfstring{$0<\sigma<1$, $0< \kappa \leq \frac{4}{3 \sqrt{3}}$}{$0<σ<1,0< κ ≤ 4/3√3$ } (Hayward--Damour--Solodukhin wormhole)} 

\begin{equation}
    ds^2 = - \Big( 1- \frac{2 \sigma x^2}{x^3+ 2 \sigma \kappa^2}\Big) dt^2 + \frac{dr^2}{\Big( 1- \frac{2 x^2}{x^3+ 2 \kappa^2}\Big)} + r^2 (d\theta^2 + \sin^2 \theta d\phi^2)
\end{equation}

\noindent $\bullet$ \textbf{Nature of spacetime:} This is the most general version of our spacetime with $\sigma$ and $\kappa$ taking all possible combination of values in the specified range of parameter space. We begin by finding the roots of $f(r)$ (which we denote as $r = r_0$) and $f_1(r)$ (denoted as $r=r_h$). As discussed earlier, we will classify the spacetime depending on the position of $r_h$, $r_0$ along with the behaviour of energy conditions and curvature scalars. If $r_0> r_h$, then at $r=r_0$, $g_{tt}$ remains positive indicating the hypersurface to be timelike. \\
\noindent $\bullet$ \textbf{Energy conditions:} In eq.(\ref{eq:EC}), we find the quantities $ \tau$ and $p$ to be inversely dependent on $ h_{1\sigma} = x^3 - 2 x^2 \sigma + 2 \kappa^2 \sigma$, which is related to the $g_{tt}$ component of the metric. 
Hence, they will not be well-behaved at the roots of $f_1(r)$ i.e. at $r=r_h$, which implies that $r=r_h$ cannot be considered as a part of the spacetime. Thus, the possibility of the spacetime being a black hole is eliminated.\\
Then we proceed to investigate the behaviour of the energy conditions 
for different parameter ranges. The energy density $\rho(x)$ is always positive and hence will satisfy the condition $\rho \geq 0$. Next, we probe the behaviour of $\rho + \tau$ which has a form
\begin{equation}
    \rho + \tau = \frac{1}{8\pi M^2} \frac{h_1}{h} \left ( \frac{A-B}{x}\right )
\end{equation}
The location of the throat is $h_1=0$. One can easily show (using the
expressions for A and B given earlier) that, at the
throat, $\rho +\tau = \frac{-1}{8\pi M^2} \frac{{h_1}'}{x_0 h} = \frac{-3 ( x_0 -4/3)}{8 \pi M^2 (x_0^3+2 \kappa^2)}$. Now, following the discussion in Case IV on finding the condition $\kappa \leq \frac{4}{3 \sqrt{3}}$, we see the two positive roots of $f(r)$ to lie on both sides of $x =4/3$. The throat $x=x_0$ corresponds always to the larger root of $f(r)$ making $x_0 > \frac{4}{3 } $. Although for $\kappa = \frac{4}{3 \sqrt{3}}$, at $x_0 =4/3$, $\rho+\tau$ becomes 0, it stays negative away from the throat. Thus $\rho +\tau$ is negative for all parameters leading to violation of energy conditions for the wormhole, as expected. 
 	\label{fig:kappa_0.1_EC}
 
\noindent $\bullet$ \textbf{Embedding diagram:} The embedding diagram for this class of wormholes will be similar to the ones shown in Fig.(\ref{fig:Hayward_WH_embed}) as they only depend on $\kappa$. The curvature scalars, Ricci and Kretschmann, have also been found to be well-behaved in the entire range of the radial coordinate.\\

\noindent For this general case, another possibility opens up, that with $\kappa> \frac{4}{3 \sqrt{3}}$. As discussed earlier, $g^{rr}$ will not  have any roots and hence the possibility of the existence of a throat is absent. But what about the roots of $g_{tt} = 0$? We have explicitly checked that $f_1(r)$ too does not have any real, positive root for any value of $\sigma$ corresponding to $\kappa > \frac{4}{3 \sqrt{3}}$. Analytically, we can calculate the condition for $f_1(r)$ to have real, positive roots. The condition one will arrive at is- $\sigma\leq 0$ or $\kappa \leq \frac{4 \sigma}{3 \sqrt{3}}$. Since both the conditions are not satisfied for this parameter range, $f_1(r)$ will not have any real, positive root for $\kappa> \frac{4}{3 \sqrt{3}}$. Hence, in that parameter range, the spacetime is simply regular everywhere and beyond our area of interest. The embedding diagrams will be similar to Fig.(\ref{fig:embed_kappa_1}) irrespective of the value of $\sigma$.\\

\noindent We can further extend the parameter space and consider spacetimes with $\sigma >1, \kappa \geq 0$ which will result in a naked singularity at $r=r_h$ (root of $f_1(r)$). Interestingly, the energy conditions also get violated near $r=r_h$ providing an example of a scenario with singularity and energy condition violating matter. This corroborates the fact that spacetime singularities are indeed focal points of geodesic congruence but focusing does not necessarily imply a spacetime singularity. One can also consider $\sigma<0$ parameter values which corresponds to negative values of parameter $M_1$. Such $\sigma<0$ spacetimes are wormhole geometries irrespective of the value of $\kappa$. The $\sigma>1$ and $\sigma<0$ parameter cases will not be considered further in this paper. We hope to discuss the
related issues to such spacetimes in our future explorations.

\section{Scalar wave propagation and effective potentials}

\noindent We have obtained six physically relevant spacetimes that range from singular and regular black holes to various wormhole geometries, as discussed in the previous section. We will focus on this broad class of six metrics 
with positive parameter ($\sigma$, $\kappa$) values. The stability of these spacetimes will be probed by calculating the quasi-normal modes 
associated with the propagation of a massless test scalar field. Quasi-normal modes are complex characteristic frequencies excited in a perturbed object, through which the object looses energy to attain equilibrium. These modes were first studied for black holes by Vishveshwara in the 1970s \cite{vishveshwara_1970_1,vishveshwara_1970_2}. Since then they have been investigated extensively in the literature \cite{kokkotas_1999, maggiore_2018,cardoso_thesis,konoplya_2011,berti_2009} for various spacetimes and are especially relevant in the context of gravitational wave observations \cite{ferrari_2007,berti_2005,ligo_2016,barausse_2014,ghosh_2021}. Apart from  the stability analysis, the QNMs are of use in parameter estimation 
of diverse astrophysical objects which arise as remnants in various scenarios\cite{berti_2007,gossan_2012,carullo_2018,isi_2019,christensen_2022}. They also help us in testing the validity of the underlying theory of gravity
in which the geometry appears as a solution\cite{ligo_2016,konoplya_2016_1,bhattacharyya_2017,dreyer_2003} .
QNMs are relevant in proving the existence of event horizons in black hole spacetimes \cite{cardoso_2016,cardoso_2019} and in determining the shape of various wormholes \cite{konoplya_2018,pdr_2020,pdr_2021}. In the same way, for our general metric, we aim to utilize the scalar QNMs to be found in each of the above background 
spacetimes, as a tool for characterising the associated geometry.  \\
\noindent We begin by recalling the Klein-Gordon equation in a given curved geometry, i.e.
\begin{equation}
    \Box \Phi =0
\end{equation}
\noindent As our background spacetime is spherically symmetric and static
(Eqn. (7)), we use the following ansatz to decompose $\Phi$ in terms of spherical harmonics, where the indices of $Y(\theta, \phi)$ have been suppressed for simplicity. For stability analysis we use the radial coordinate $r$ and assume $M=1$. Thus, with
\begin{align}
  \Phi(t,r,\theta,\phi)=Y(\theta,\phi)\frac{u(r)e^{-i \omega t}}{r}.
\end{align} 
and the general metric ansatz (eq.(\ref{eq:metric})) the Klein-Gordon equation
separates. The radial equation
turns out to be a Schr\"{o}dinger-like equation, written using the tortoise coordinate $r_*$, i.e.,
\begin{align}
  \frac{d^2 u}{dr_*^2}+[\omega^2 - V_{eff}]u=0
  \label{eq:radial_eqn}
 \end{align}
where
\begin{align}
    V_{eff}(r) = \frac{m(m+1) f_1(r)}{r^2} + \frac{f(r) f_1'(r)+ f'(r) f_1(r)}{2 r}
    \label{eq:potential}
\end{align}
is the effective potential and $m$ is the azimuthal number arising from the separation of variables. The characteristics of each member spacetime
in our family of six,
will be reflected in the corresponding effective potentials. In this section, we will try to analyse $V_{eff}$ and plot it as a function of $r$ for the black hole cases and as a function of $r_*$, for the wormhole geometries. Use of the tortoise coordinate for wormholes helps in visualizing the wormhole 
as a two-sheeted geometry with a connecting throat. 
\begin{figure}[!h]
\begin{subfigure}{0.4\textwidth}
	\includegraphics[width=0.95\linewidth]{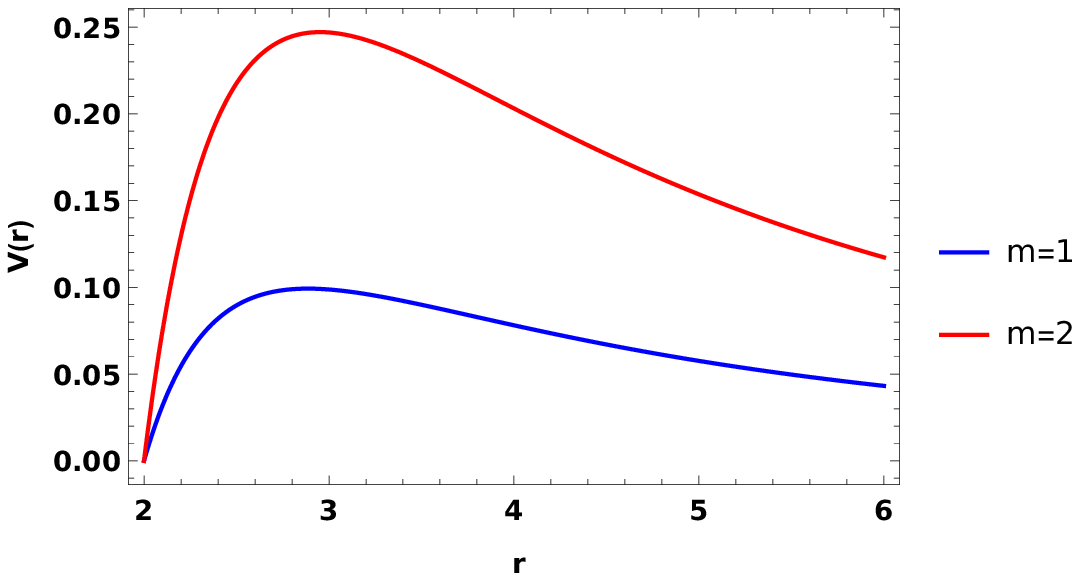}
 	\caption{$\kappa=0,\sigma =1$ (Case I)}
 \end{subfigure}
 \begin{subfigure}{0.4\textwidth}
 	\includegraphics[width=0.95\linewidth]{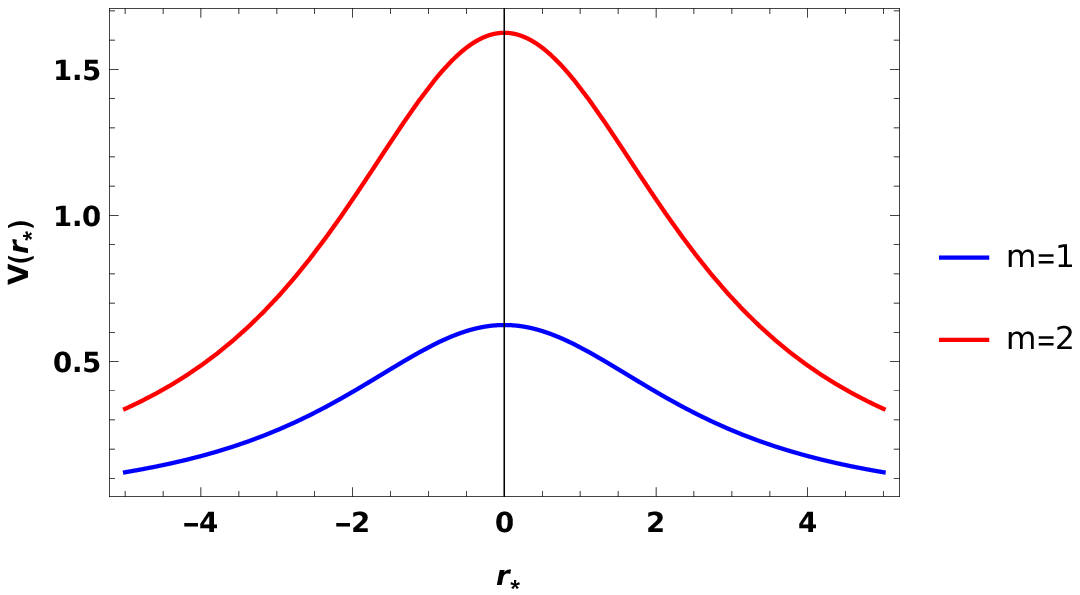}
 	\caption{ $\kappa=0,\sigma=0$ (Case II)}
 	\end{subfigure}
 \begin{subfigure}{0.4\textwidth}
 	\includegraphics[width=0.95\linewidth]{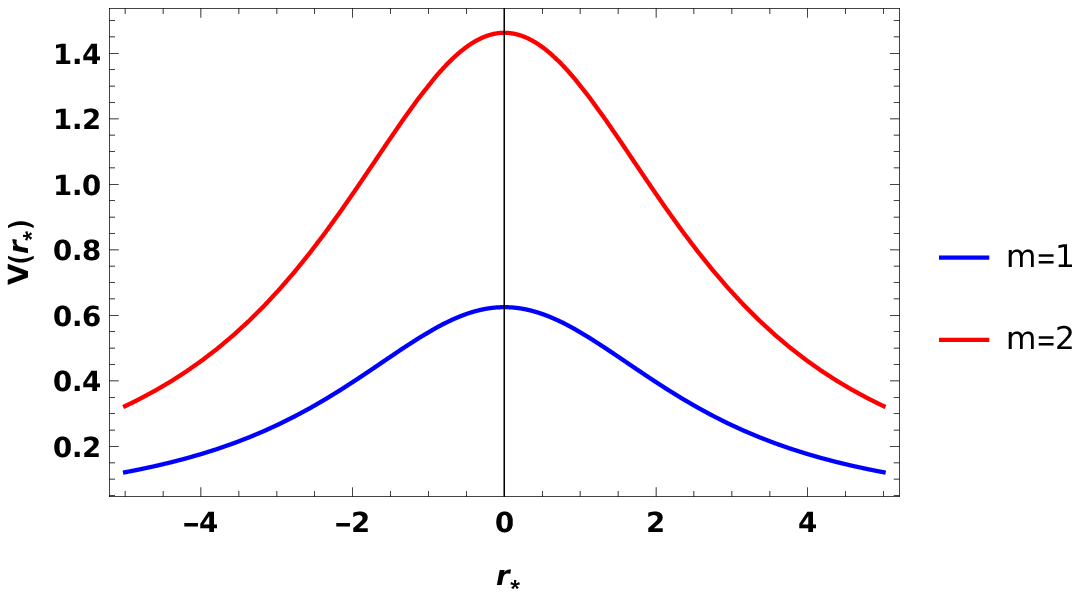}
 	\caption{ $\kappa=0,\sigma=0.1$ (Case III)}
 	\end{subfigure}	
  \begin{subfigure}{0.4\textwidth}
 	\includegraphics[width=0.95\linewidth]{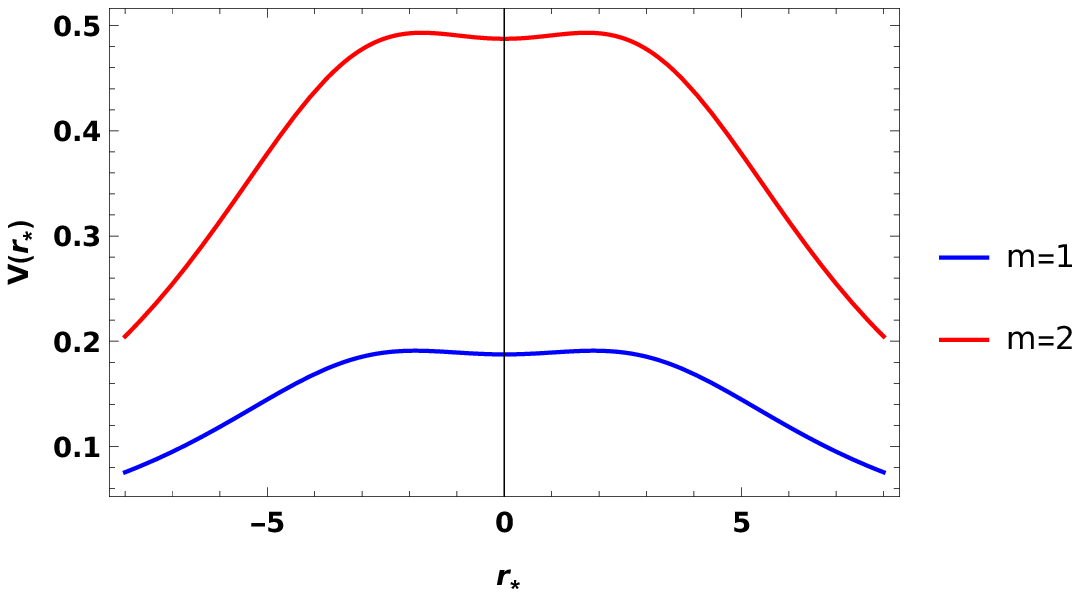}
 	\caption{ $\kappa=0,\sigma=0.7$ (Case III)}
 	\end{subfigure}	
 	\begin{subfigure}{0.4\textwidth}
 	\includegraphics[width=0.95\linewidth]{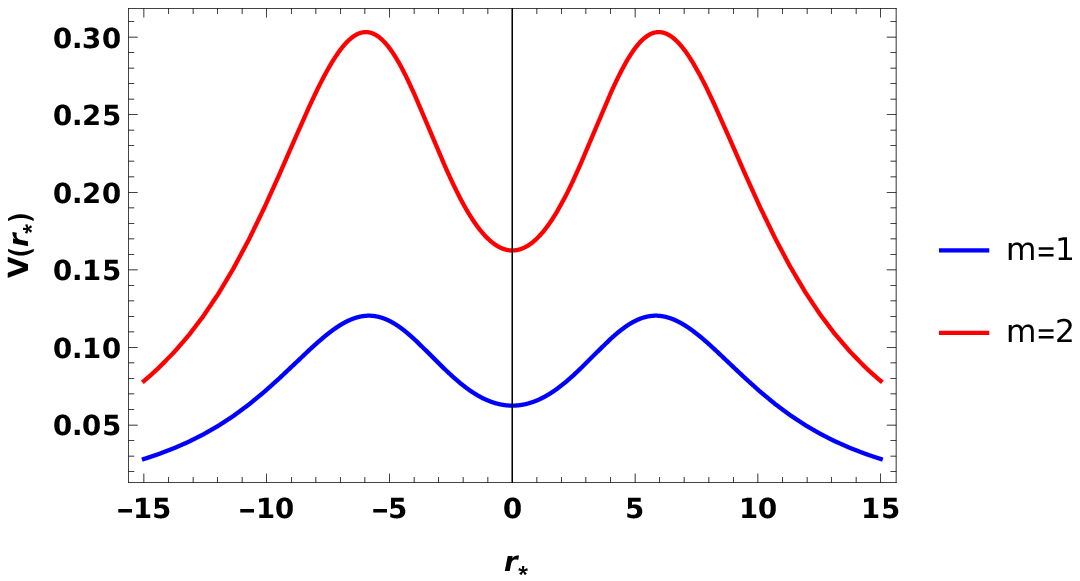}
 	\caption{ $\kappa=0,\sigma=0.9$ (Case III)}
 	\end{subfigure}	
 	\begin{subfigure}{0.4\textwidth}
 	\includegraphics[width=0.95\linewidth]{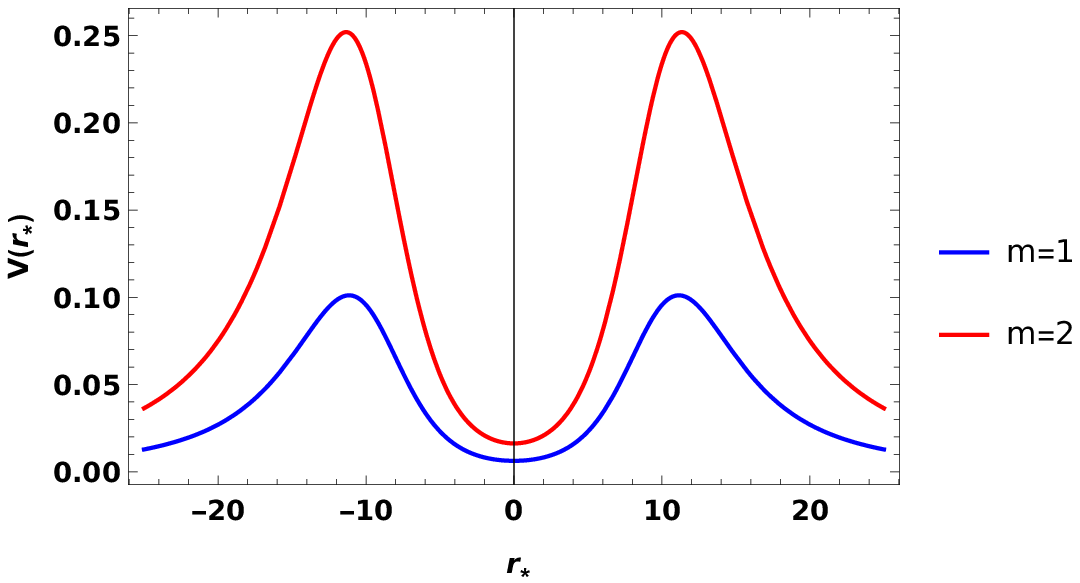}
 	\caption{ $\kappa=0,\sigma=0.99$ (Case III)}
 	\end{subfigure}	
 	\begin{subfigure}{0.4\textwidth}
 	\includegraphics[width=0.95\linewidth]{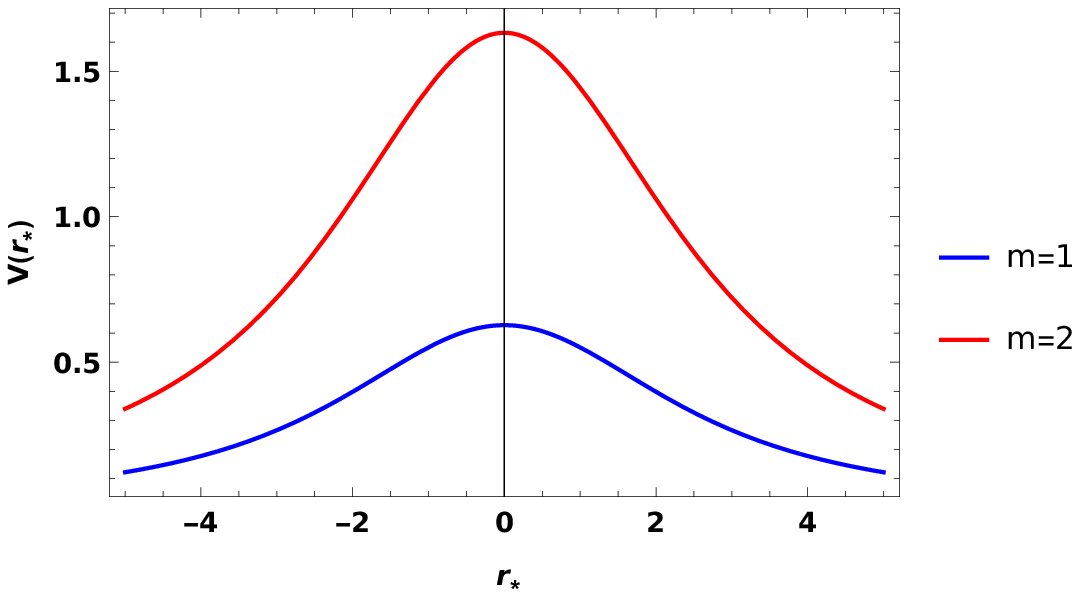}
 	\caption{ $\kappa=0.1,\sigma=0$ (Case IV)}
 	\end{subfigure}	
 	\begin{subfigure}{0.4\textwidth}
 	\includegraphics[width=0.95\linewidth]{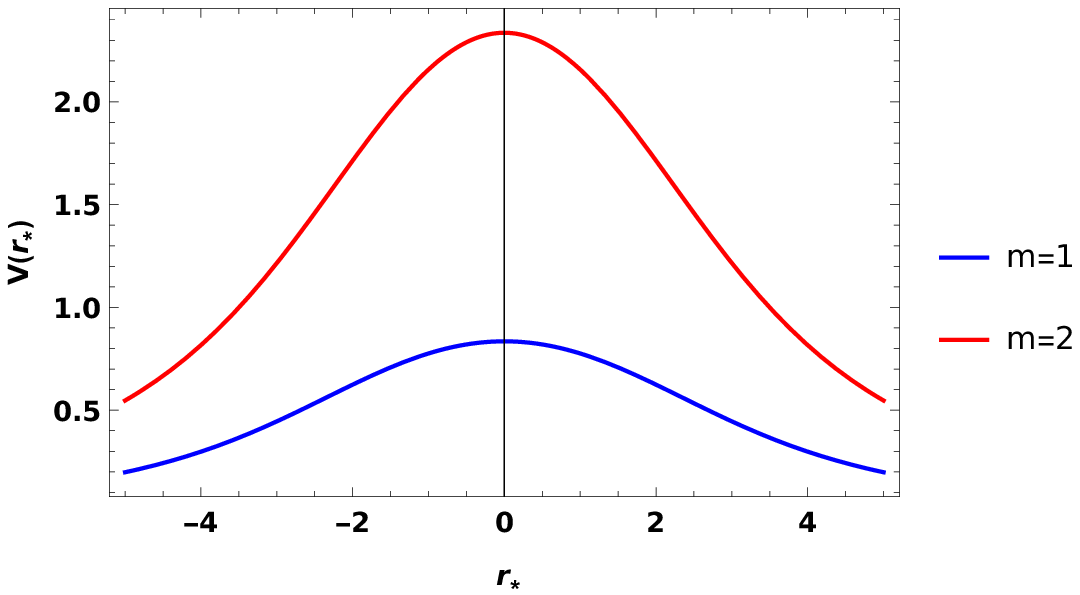}
 	\caption{ $\kappa=0.7,\sigma=0$ (Case IV)}
 	\end{subfigure}	
 	\caption{Variation of $V_{eff}$ for different parameter values of $(\sigma,\kappa)$ and angular momentum number $m$ (Case I-IV). We observe that the double barrier nature of the potential depends on the value of $\sigma$. Same $\kappa$ wormholes will have same throat radius but the tortoise coordinate will be different because of its dependence on $\sigma$.}
 	\label{fig:potential_plots}
 \end{figure} 

\begin{figure}[!h]
 	\begin{subfigure}{0.4\textwidth}
 	\includegraphics[width=0.95\linewidth]{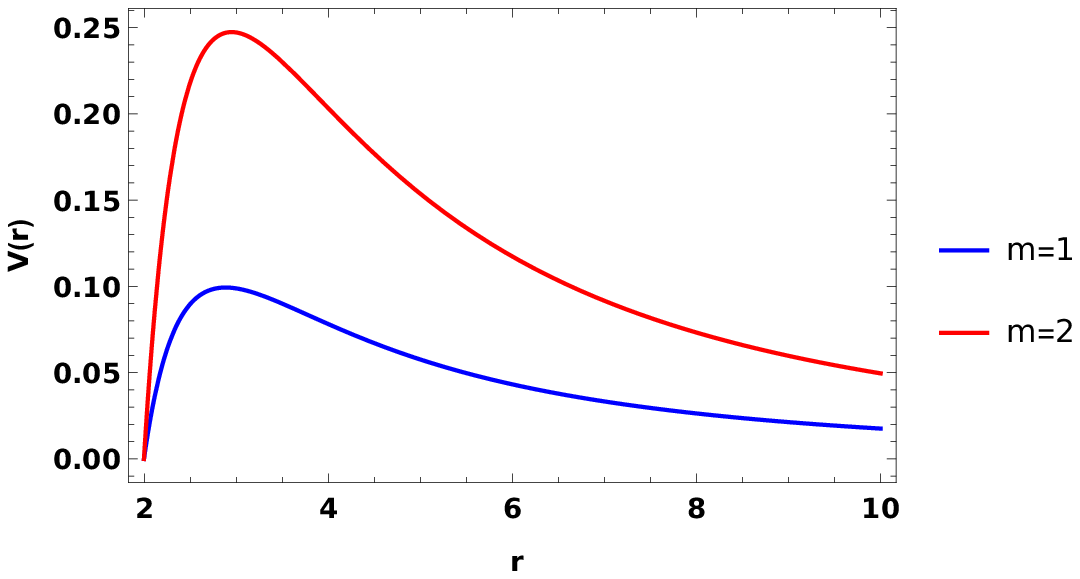}
 	\caption{ $\kappa=0.1,\sigma=1$ (Case V)}
 	\end{subfigure}	
 	\begin{subfigure}{0.4\textwidth}
 	\includegraphics[width=0.95\linewidth]{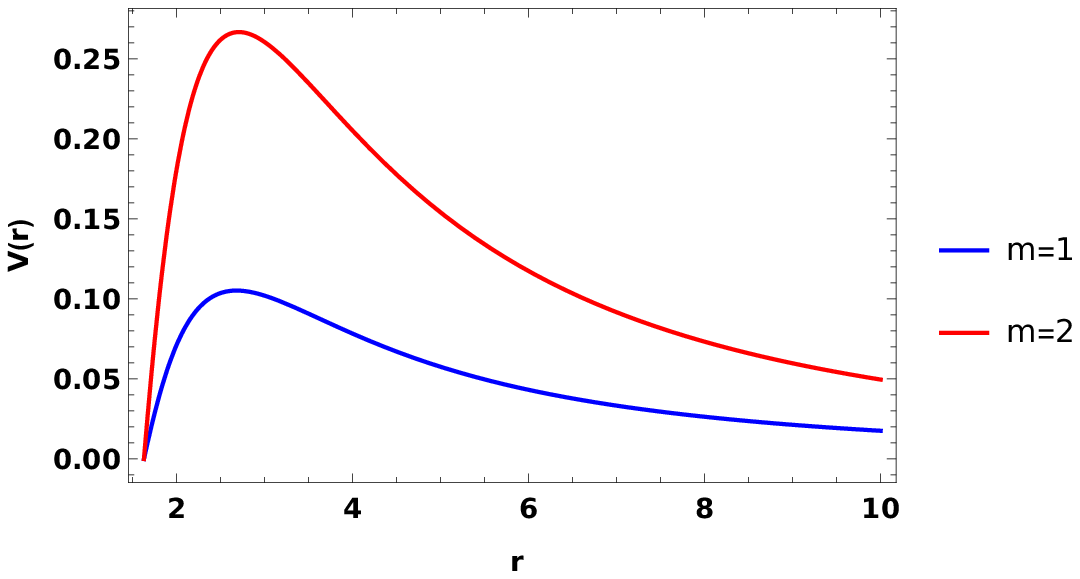}
 	\caption{ $\kappa=0.7,\sigma=1$ (Case V)}
 	\end{subfigure}	
 	\begin{subfigure}{0.4\textwidth}
 	\includegraphics[width=0.95\linewidth]{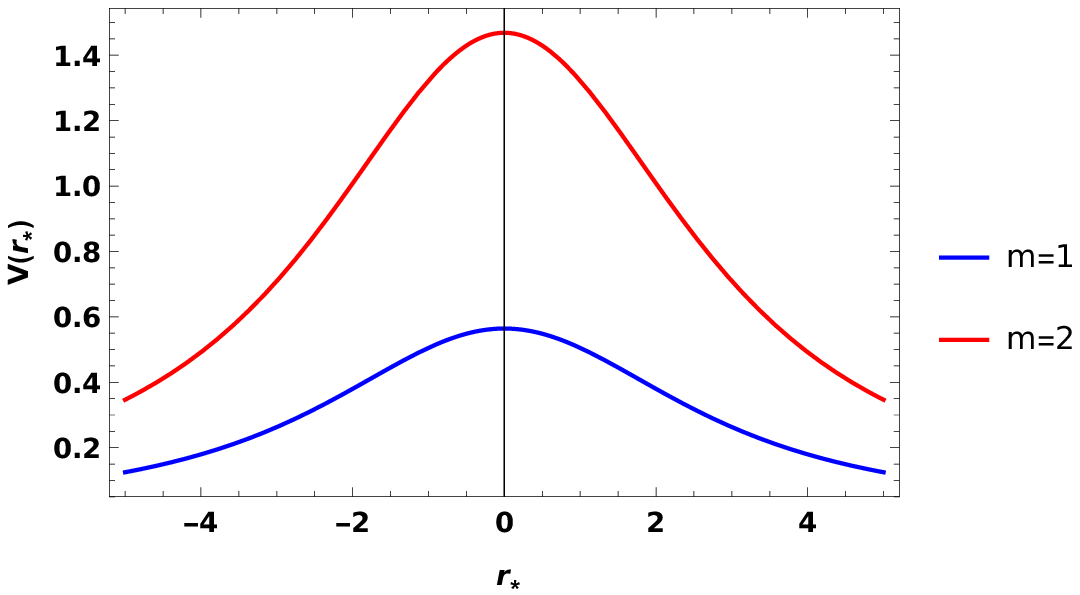}
 	\caption{ $\kappa=0.1,\sigma=0.1$ (Case VI)}
 	\end{subfigure}	
 	\begin{subfigure}{0.4\textwidth}
 	\includegraphics[width=0.95\linewidth]{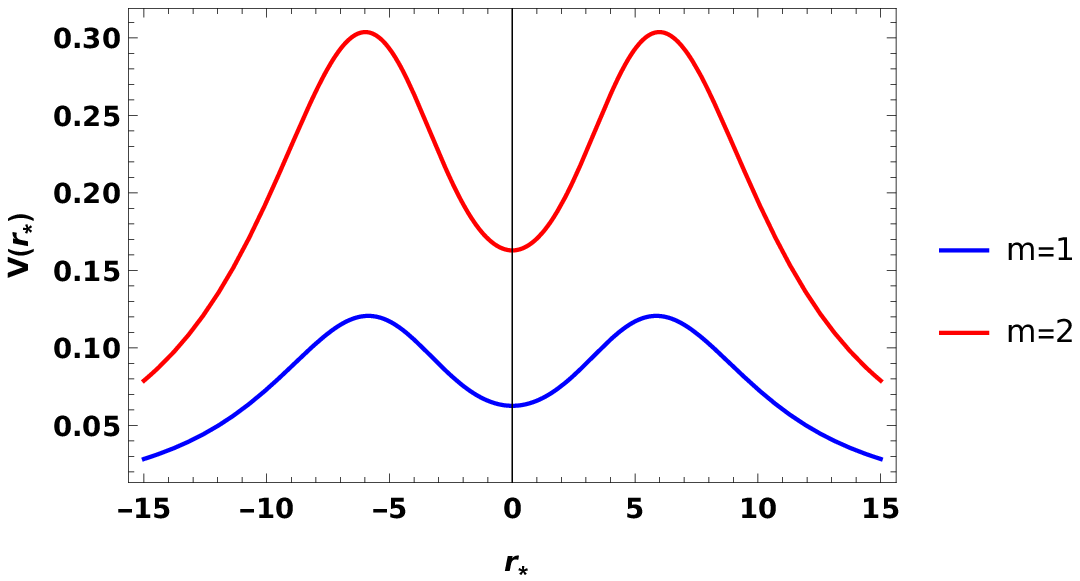}
 	\caption{ $\kappa=0.1,\sigma=0.9$ (Case VI)}
 	\end{subfigure}	
 	\begin{subfigure}{0.4\textwidth}
 	\includegraphics[width=0.95\linewidth]{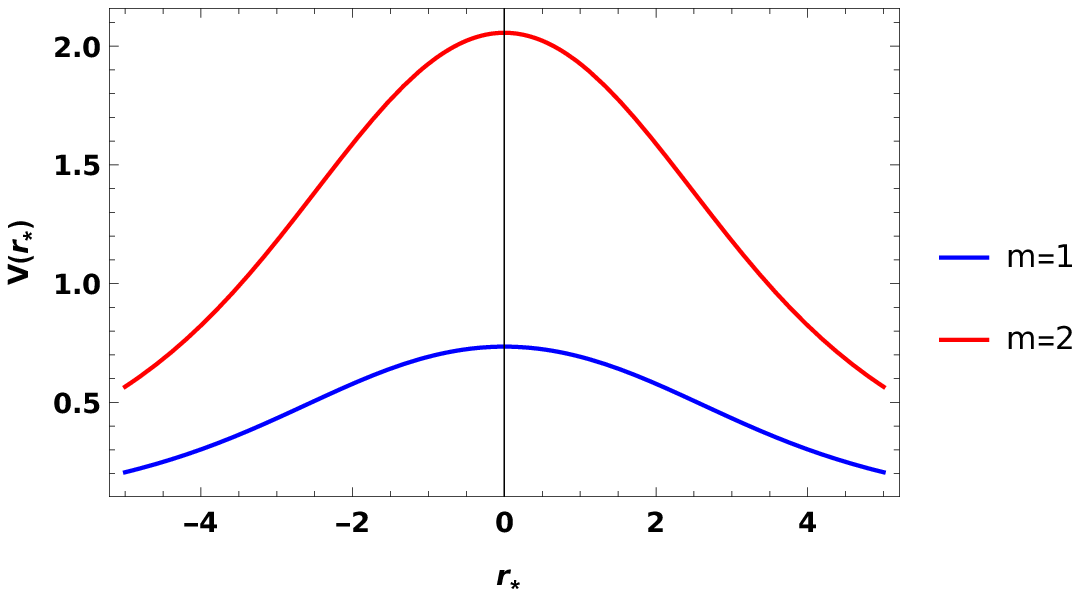}
 	\caption{ $\kappa=0.7,\sigma=0.1$ (Case VI)}
 	\end{subfigure}	
 	\begin{subfigure}{0.4\textwidth}
 	\includegraphics[width=0.95\linewidth]{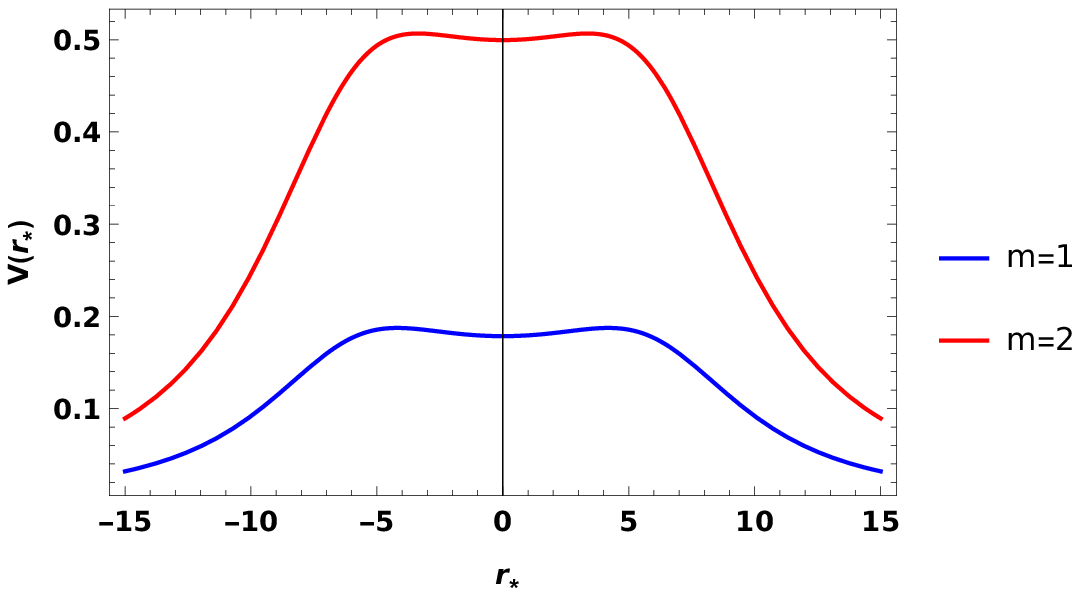}
 	\caption{ $\kappa=0.7,\sigma=0.75$ (Case VI)}
 	\end{subfigure}	
 	\caption{Plots of effective potential $V_{eff}$ for Case V and VI. Once again the double barrier potential is observed for values of $\sigma$ close to 1.}
 	\label{fig:potential_plots1}
\end{figure} 	

\noindent From the multiple plots shown in Fig.(\ref{fig:potential_plots}) and (\ref{fig:potential_plots1}), it is evident that the generalized metric harbours both single and double barrier potentials depending on the choice of the parameters $(\sigma,\kappa)$. The double peak potential is observed in wormholes belonging to Case III and VI as the value of parameter $\sigma$ approaches (but is never equal to) one. For the Damour-Solodukhin wormhole, $\sigma \rightarrow 1$, and it can be thought as two copies of the Schwarzschild black hole potential. This feature has been explored in other works as well \cite{bueno_2018}. \\
To get a better understanding of the nature of potential especially for the wormhole geometries, we test whether the potential has a maximum or a minimum at the throat i.e. $r_* =0$. If $r_*=0$ is a minimum, then the potential is atleast a double-barrier. We have, at $r_*=0 (f(r)=0)$, and with $x =r/M$,
\begin{equation}
     \frac{d^2 V_{eff}}{d r_*^2} \Big|_ {f=0} = \frac{f'(x) f_1(x)}{2 M^2} \Big[ f_1(x) \Big( \frac{-f'(x)}{2 x^2} + \frac{f''(x)}{2 x} -\frac{2 m (m+1)}{x^3} \Big) + f_1'(x) \Big( \frac{f'(x)}{x} + \frac{m(m+1)}{x^2} \Big) \Big]
\end{equation}
For the Case III wormhole with $\kappa =0$, the throat lies at $x=2$ and is a minimum if,  $V''_{eff}>0$. Using the expressions for the metric functions
with arbitrary $\sigma$ ($0<\sigma<1$) and $\kappa=0$ we obtain, for a minimum, the requirement
$\sigma > \frac{3+4m(m+1)}{5+6m(m+1)}$. Note that for $m=0$ one can have a minimum as long as $\sigma>0.6$. For $m=1$, $\sigma< 0.647$, the potential remains a single barrier with a maximum at the throat, whereas for $\sigma>0.647$ we have a
minimum at the throat and a double barrier. This generic behaviour, i.e. the
existence of a critical $\sigma$, holds true for all $m \geq 0$. Different values of $m$ will have different $\sigma$ at which the transition occurs from single to double barrier potential. Similar analysis holds for general $(\sigma,\kappa)$ values too, but since the analytic expressions for the throat radius are
cumbersome, we have verified the behaviour numerically (not stated here) for each parameter set. For black hole cases, the maximum will occur at some point away from the horizon and can be obtained by finding the critical points of $V_{eff}(r)$. Note that the condition in (29) is valid for arbitrary $f(r)$, $f_1(r)$ and may be used to know (without writing the scalar wave equation) 
if a geometry will exhibit a single or a double
barrier effective potential for scalar wave propagation.  

\noindent Table \ref{tab:summary} summarizes the properties of the spacetimes belonging to each parameter set. 

 \begin{table}[H]
  \centering
  \begin{tabular}{|c|c|c|c|c|c|c|}
      \hline
     &  Case I & Case II & Case III & Case IV & Case V & Case VI \\
    \hline 
  Type & BH & WH & WH & WH & Regular BH & WH\\
  \hline
  Violation of EC & \xmark & \cmark & \cmark & \cmark & \xmark & \cmark \\
  \hline
  Singularity & \cmark & \xmark & \xmark & \xmark & \xmark & \xmark \\
  \hline
  Single barrier potential & \cmark & \cmark & \cmark & \cmark & \cmark & \cmark\\
  \hline 
  Double barrier potential & \xmark & \xmark & \cmark & \xmark & \xmark & \cmark\\
 
  \hline
\end{tabular}
\caption{ Summary of various properties of spacetimes for different parameter ranges of $(\sigma,\kappa)$.}
\label{tab:summary}
\end{table} 

\section{Stability analysis: quasi-normal modes and echoes}

\noindent After obtaining the form of the effective potential, we calculate the corresponding scalar quasi-normal modes associated with each case. The calculation of the QNM frequencies can be executed using multiple techniques available in the literature. One of the most convenient analytical techniques for calculating QNMs is the WKB method initially discussed in \cite{schutz_1985,iyer_1987}. Later, a number of articles extended the technique to higher orders, using Pade corrections, to achieve better convergence \cite{konoplya_wkb_2003,konoplya_wkb_2019,matyjasek_2017}. In our spacetimes here, 
the WKB method is effective in obtaining analytical results for potentials with a single barrier. Hence, QNMs cannot quite be obtained using the WKB method for some specific wormhole geometries where the effective perturbation
potential is a double barrier.

\noindent To handle the double barrier cases and to verify the results obtained from the WKB method for single barrier potentials, we implement numerical techniques for calculating QNMs \cite{franzin_2022,konoplya_2011}. We will make use of the direct integration method and Prony extraction scheme to obtain the QNM frequencies numerically. Let us go through the essence of the two methods one-by-one. Direct integration or the shooting method was proposed by Chandrasekhar and Detweiler \cite{chandrasekhar_1985} where the differential equation eq.(\ref{eq:radial_eqn}), written in terms of the radial coordinate $r$, is integrated numerically. For black holes, purely outgoing boundary conditions are implemented at spatial infinity and at the horizon. The integration is then done from the horizon to a finite intermediate point where the solution and its derivative is matched with the solution obtained at infinity. The matching condition gives discrete values of $\omega$ which are the QNMs. Numerical errors may occur in the method, as the integration has to be done a small distance away from horizon to ensure smooth behaviour of all functions. Also, spatial infinity is taken at a finite value. These errors due to numerical limitations can be minimised by choosing suitable precision and accuracy for the integration scheme. The robustness of the method should be tested by slightly changing the matching point but keeping the result unchanged. On the other hand, for wormhole geometries, ideally, the integration needs to be done from $r_* \rightarrow -\infty$ to $\infty$. All wormholes, arising from our general metric, are symmetric about their throat at $r_*=0$ and will have even and odd classes of solutions for the perturbation equation. The integration is thus performed on positive $r_*$, i.e., from $r=r_0$ ($r_* =0$) to $\infty$ and the condition on the solution at the throat decides whether it is symmetric or antisymmetric, on reflection (i.e. $r_*\rightarrow -r_*$). For our case, this method is especially suitable, as implementing this technique does not require the evaluation of the tortoise coordinate. Additional numerical errors might be introduced due to $r_*$, as it can only be obtained numerically for our metric in the form of an interpolation function. Avoiding the use of $r_*$ in direct integration removes one possible source of numerical error. \\
The Prony extraction scheme \cite{konoplya_2011,berti_2007_1} begins by first obtaining the time domain (TD) evolution of an initial Gaussian pulse in the background of our spacetime. Instead of using eq.(\ref{eq:radial_eqn}), we keep the time dependence so that the equation becomes
\begin{align}
    \frac{\partial^{2}\psi}{\partial t^{2}} - \frac{\partial^{2}\psi}{\partial r_*^{2}} + V_{eff}(r_*) \psi =0 .
    \label{eq:QNM}
\end{align} 
The wave equation for $\psi(r_*,t)$ is recast using light cone coordinates $(u,v)$ such that $(du=dt-dr_*, dv= dt+dr_*)$. The appropriate discretization scheme is discussed in \cite{konoplya_2011,gundlach_1994}. With proper initial conditions defined on the null grid, the wave equation is integrated to obtain the time domain profile. A detailed study of the ringdown profile indicates the presence of QNM frequencies while the damped signal will establish the stability of the corresponding spacetime under the given perturbation. The signal is fitted with a superposition of damped sinusoidal frequencies, in the Prony extraction method and the most dominant quasi-normal mode is thereby extracted. \\
Apart from proving stability, the time domain signal plays crucial role in studying the behaviour of a wormhole geometry characterized especially with a double barrier effective perturbation potential. The double peaks  cause multiple reflection of the incident wave producing a phenomenon known as `echoes'. After the initial prompt ringdown, the signal gets dominated by the repetitive echoes of decreasing amplitude \cite{cardoso_2016,maggio_2020,bueno_2018}. The quasinormal modes of the wormhole starts to influence the ringdown only at very late times when the echoes have died down. These wormholes thus possess long-lived dominant modes that come into play at late times. Such a distinctive feature of the ringdown is only shown by a double potential barrier. Through our work we will observe how the ringdown pattern changes for different spacetimes. We also note the differences in the 
approach to stability for black holes and
various wormhole geometries. 
\noindent The phenomenon of `echoes' have been widely studied for various classes of spacetimes \cite{mark_2017,conklin_2017,pdr_2020,pdr_2021,yang_2021,fang_2021,dey_2020} and are prime indicators of any new physics close to the event horizon of black holes\cite{abedi_2017,cardoso_2016_echoes,uchikata_2019,nakano_2017,oshita_2020,wang_2018,cardoso_2019_1,price_2017,oshita_2020_1}. They also play a significant role as a distinguishing feature of exotic compact objects (ECO) in the context of {\em black hole mimicker} analysis \cite{cardoso_2019,bronnikov_2020,churilova_2019}.\\
We now move on to studying the ringdown profile and obtain the scalar QNMs for each case separately.

\subsection {\bf {Case I:} \texorpdfstring{$\sigma=1$, $\kappa=0$}{$σ=1$,$κ=0$ }(Schwarzschild black hole)}

\begin{figure}[h!]
    \centering
    \includegraphics[width=0.42\linewidth]{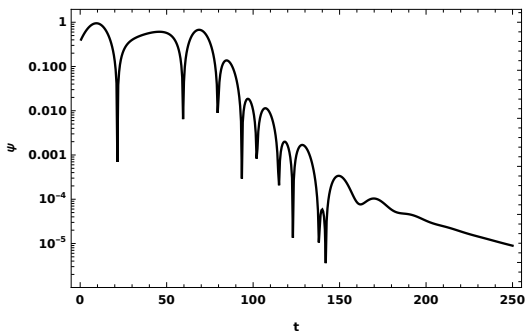}
    \caption{Typical Schwarzschild black hole ringdown profile as observed at $r_* = 30$ with initial Gaussian profile $\psi =e^{- \frac{(u-10)^2}{100}}$ and grid spacing $h=0.5$.}
    \label{fig:Schwarzschild_TD}
\end{figure}

\noindent For the Schwarzschild black hole, the time domain signal (see Fig.(\ref{fig:Schwarzschild_TD})) is dominated by the quasi-normal modes of the black hole \cite{chandrasekhar_1985,nollert_1992,nollert_1993} which are also the modes excited from the perturbation of the photon ring \cite{cardoso_2009}.

\subsection{\bf {Case II}: \texorpdfstring{$\sigma=0$, $\kappa=0$}{$σ=κ=0$} (Schwarzschild wormhole)} 

\begin{figure}[h]
\begin{subfigure}{0.45\textwidth}
	\includegraphics[width=0.83\linewidth]{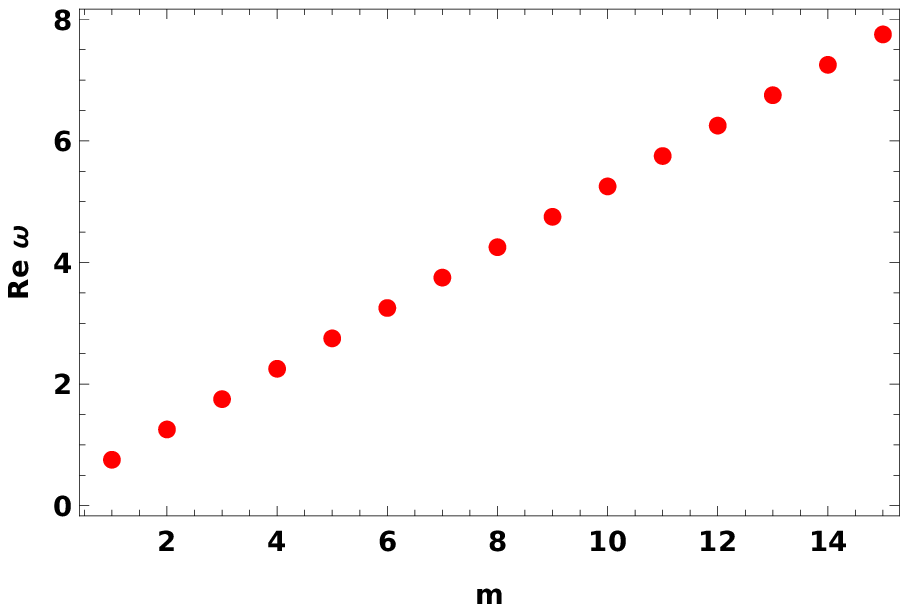}
 	\caption{Variation of Re($\omega$) (in $M^{-1}$) with $m$.}
 \end{subfigure}
 \begin{subfigure}{0.45\textwidth}
 	\includegraphics[width=0.88\linewidth]{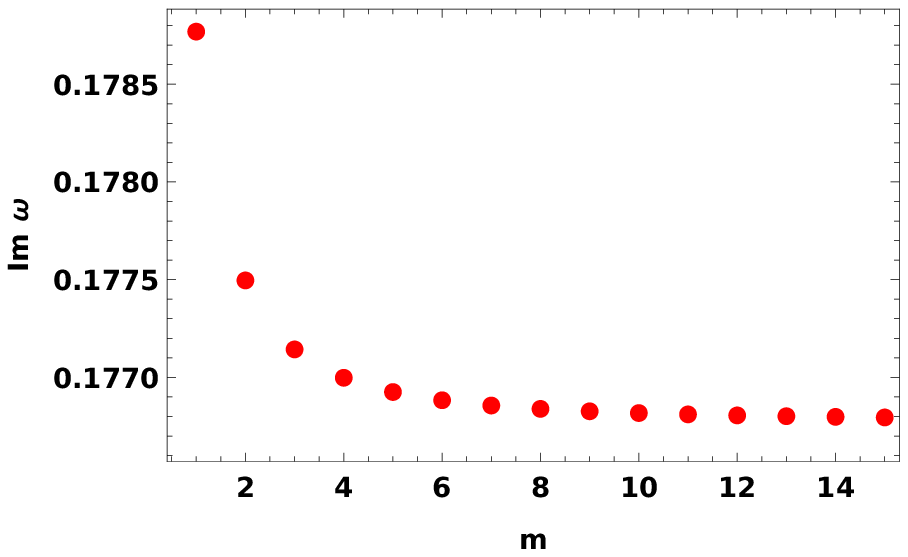}
 	\caption{ Variation of Im($\omega$) (in $M^{-1}$) with $m$.}
 	\end{subfigure}
 	\caption{QNM variation for Case II wormhole with angular momentum $m$. The Im($\omega$) decreases with increasing $m$ as observed in most wormhole geometries.}
 	\label{fig:CaseII_QNM}
 \end{figure}

\noindent The wormhole with these parameters possesses quasinormal modes that will be function of angular momentum number $m$ and the mass parameter $M$. Taking the eikonal limit (large $m$),  $\omega \approx \Big(\frac{m}{2} \sqrt{1-\frac{i}{\sqrt{2}m}} \Big) M^{-1}$ and hence the imaginary component tends to $Im(\omega) \rightarrow 0.17677 \, M^{-1}$, as visible in the plot of Fig.(\ref{fig:CaseII_QNM}). 

\subsection{ \bf{Case III}: \texorpdfstring{$0<\sigma<1,\kappa=0$}{$0<σ<1,κ=0$ } (Damour-Solodukhin wormhole)} 

\begin{figure}[h]
\begin{subfigure}{0.45\textwidth}
	\includegraphics[width=0.85\linewidth]{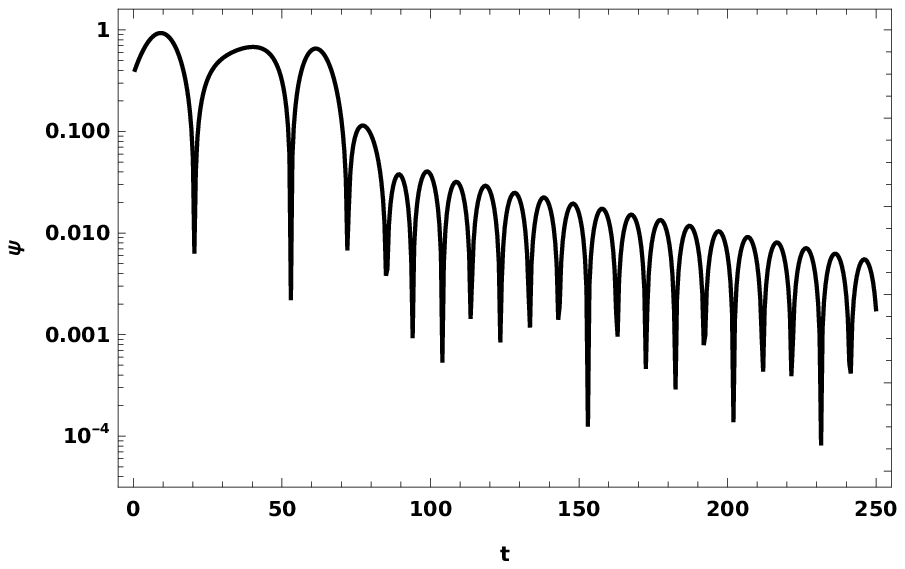}
 	\caption{$\sigma =0.9$}
 \end{subfigure}
 \begin{subfigure}{0.45\textwidth}
 	\includegraphics[width=0.85\linewidth]{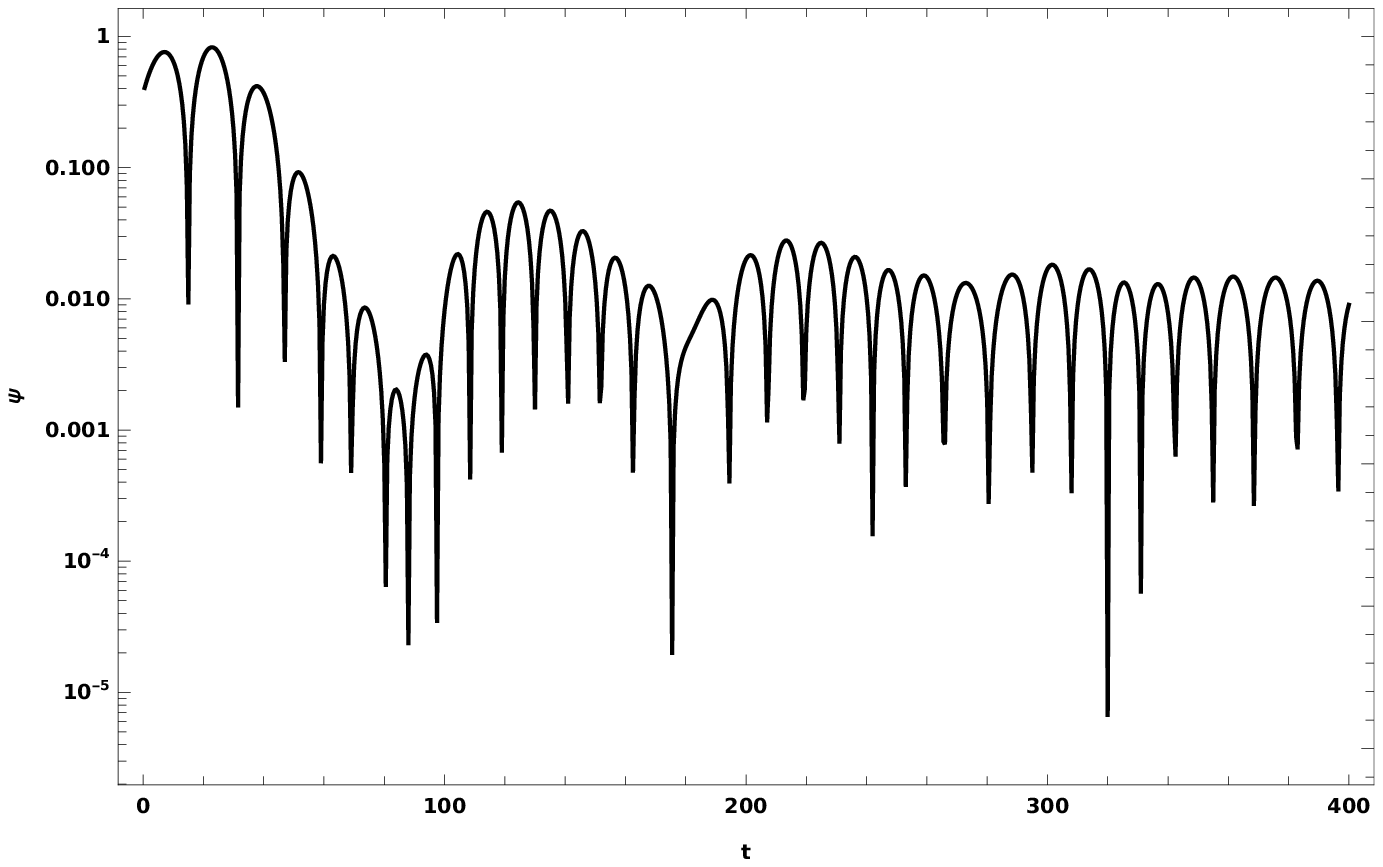}
 	\caption{ $\sigma =0.9999$}
 	\end{subfigure}
 	\caption{TD profile for Damour-Solodukhin wormhole with $\sigma$ values close to 1 observed at $r_* =30$ with initial Gaussian profile $\psi =e^{- \frac{(u-10)^2}{100}}$, grid spacing $h=0.5$ and $m=1$.}
 	\label{fig:DS_sigms_large}
 \end{figure}

\noindent The Damour-Solodukhin wormhole deserves attention since both single and double potential barriers occur, depending on the value of $\sigma$. When $\sigma =0.9$, the two potential peaks are well separated and the profile is dominated by the wormhole QNM just after the initial ringdown has ended (see Fig.(\ref{fig:DS_sigms_large})). For even higher $\sigma$ values, say $\sigma =0.9999$, the potential peak separation widens further and leads to significant change in the ringdown behaviour. The initial ringdown for $\sigma =0.9999$ wormhole occurs at $t \approx 50-90$ and resembles that of the Schwarzschild black hole but is soon followed by well-defined echoes. Extracting the frequency, using Prony method, from $t \approx 50 -90$ gives a QNM of the order $0.292267 -i 0.122944$ $M^{-1}$ which is very close to value fof the Schwarzschild black hole dominant mode. The echoes start appearing after the prompt ringdown from $t \approx 120$. The echo signal is dominated by modes of the order $0.285289 -i 0.03223$ $M^{-1}$ for $t \approx 130-180$ and $0.2297 -i 0.00403$ $M^{-1}$ for $t \approx 350-400$. Expectedly, modes with long damping period dominate the signal at late times. These modes are very close to higher overtone QNMs of the wormhole as also noted in \cite{bueno_2018}. The fundamental QNM of the wormhole is $0.09205 -i 10^{-6}$ $M^{-1}$ obtained using DI. 
The small magnitude of Im($\omega$) of the fundamental mode indicates that at late times the signal is dominated entirely by the fundamental wormhole QNM. Such behavior has been observed in other black hole/ wormhole transition spacetimes as well\cite{cardoso_2016,churilova_2019,yang_2021}.

\begin{table}[h]
  \centering
  \begin{tabular}{|c|c|}
      \hline
     m &  $\omega$ M (DI) \\ 
    \hline 
  1 & 0.32013 -i 0.012934  \\
  2 & 0.484687 -i 0.003451 \\
  3 & 0.6478119 -i 0.000712\\
 4 & 0.8088245 -i 0.0001199  \\
  5& 0.968591 -i 0.000018068\\
  \hline
\end{tabular}
\caption{\label{tab:CaseIII} Fundamental $\omega_{QNM}$ values for Case III wormhole with $\sigma=0.9$. DI is used as WKB cannot be applied due to the double barrier nature of the potential. The Im($\omega$) decreases rapidly with increasing $m$, indicating the large $m$ modes to be long-lived and dominating at late times.}
\end{table}

\noindent Note the time domain profile for wormholes with well-separated double barrier potentials needs to be computed for long time if one intends to extract QNM frequencies of the wormhole from the signal. This process is computationally exhaustive as high precision is required for computing the QNM frequencies with such low damping rate. The DI method thus becomes more suitable for these wormholes. To verify the authenticity of the DI result, we have checked its convergence over wide range of matching points. The QNM values for $\sigma=0.9$ is shown in Table \ref{tab:CaseIII} that indicate the low damping rate of the modes.

 \begin{table}[h]
  \centering
  \begin{tabular}{|c|c|c|}
      \hline
     $\sigma$ &  $\omega$ M (WKB) &  $\omega$ M (DI) \\ 
    \hline 
  0.01 & 0.751719 -i 0.177449 & 0.7516804 -i 0.1774696 \\
  0.1 & 0.718926 -i 0.165179 & 0.718891 -i 0.165198\\
  0.2 & 0.680852 -i 0.150609 & 0.6808197 -i 0.1506254\\
  0.3 & 0.640844 -i 0.1348974 & 0.64081047 -i 0.134899 \\
  0.4 & 0.598844 -i 0.11791406 & 0.598592 -i 0.11782059\\
  0.5 & 0.554175 -i 0.0986858 & 0.5538142 -i 0.0992029\\
  0.6 & 0.509473 -i 0.0676018 & 0.505933 -i 0.078929 \\
  0.7& - & 0.453844 -i 0.0571522\\
  0.8 & - & 0.394867 -i 0.0344919 \\
  0.9 & - & 0.3201304 -i 0.0129345 \\
  0.95 & - & 0.2655024 -i 0.0043492\\
  0.99 & - & 0.182275 -i 0.0003563\\
  \hline
\end{tabular}
\caption{\label{tab:n2} Fundamental $\omega_{QNM}$ for different Damour-Solodukhin wormholes (Case III) with $m=1$. The double barrier nature of potential becomes prominent as $\sigma \rightarrow 1$ due to which WKB method cannot be applied.}
\end{table}  

\begin{figure}[h]
\begin{subfigure}{0.48\textwidth}
	\includegraphics[width=0.95\linewidth]{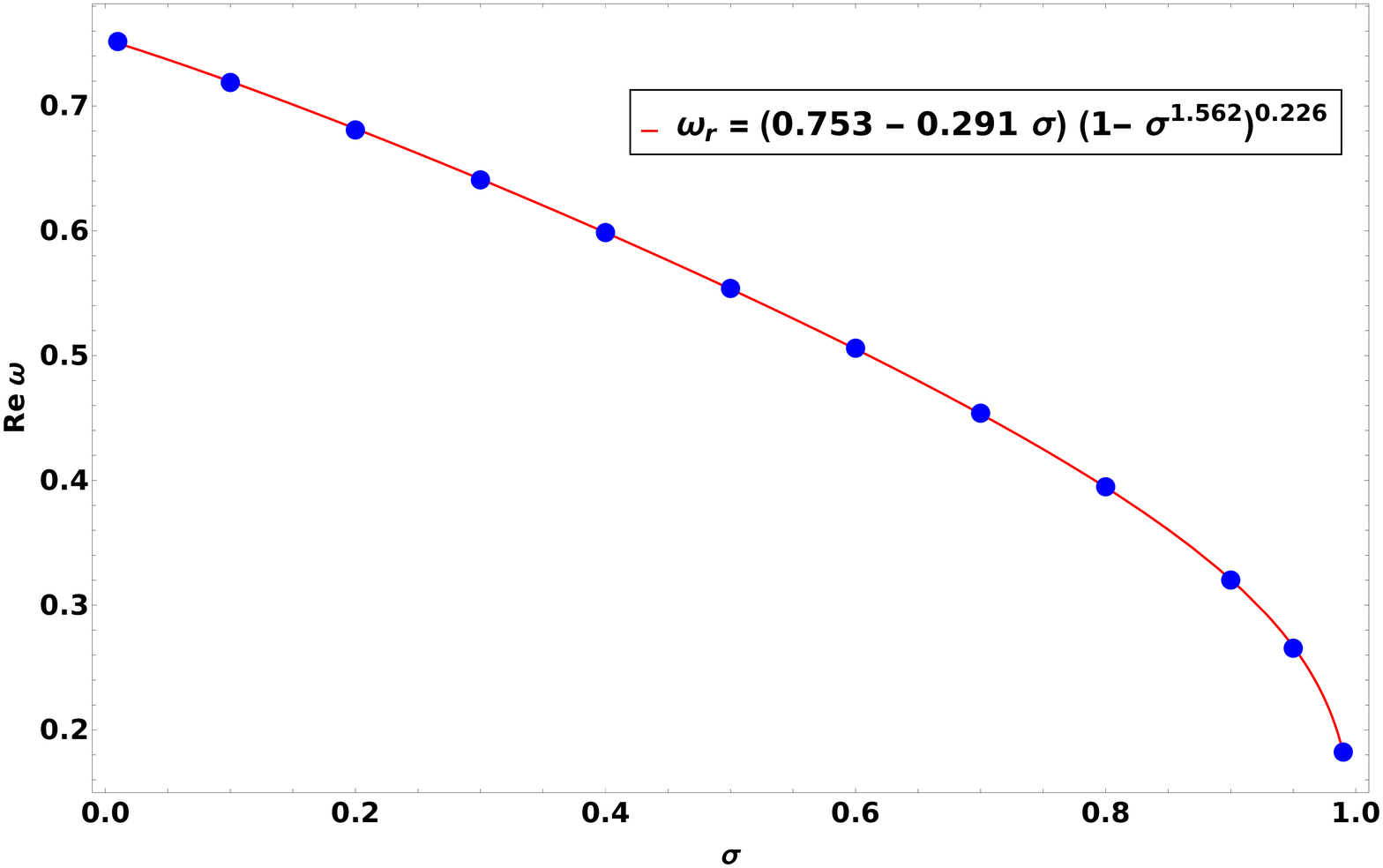}
 	\caption{Fitting $Re (\omega)$}
 \end{subfigure}
 \begin{subfigure}{0.48\textwidth}
 	\includegraphics[width=0.95\linewidth]{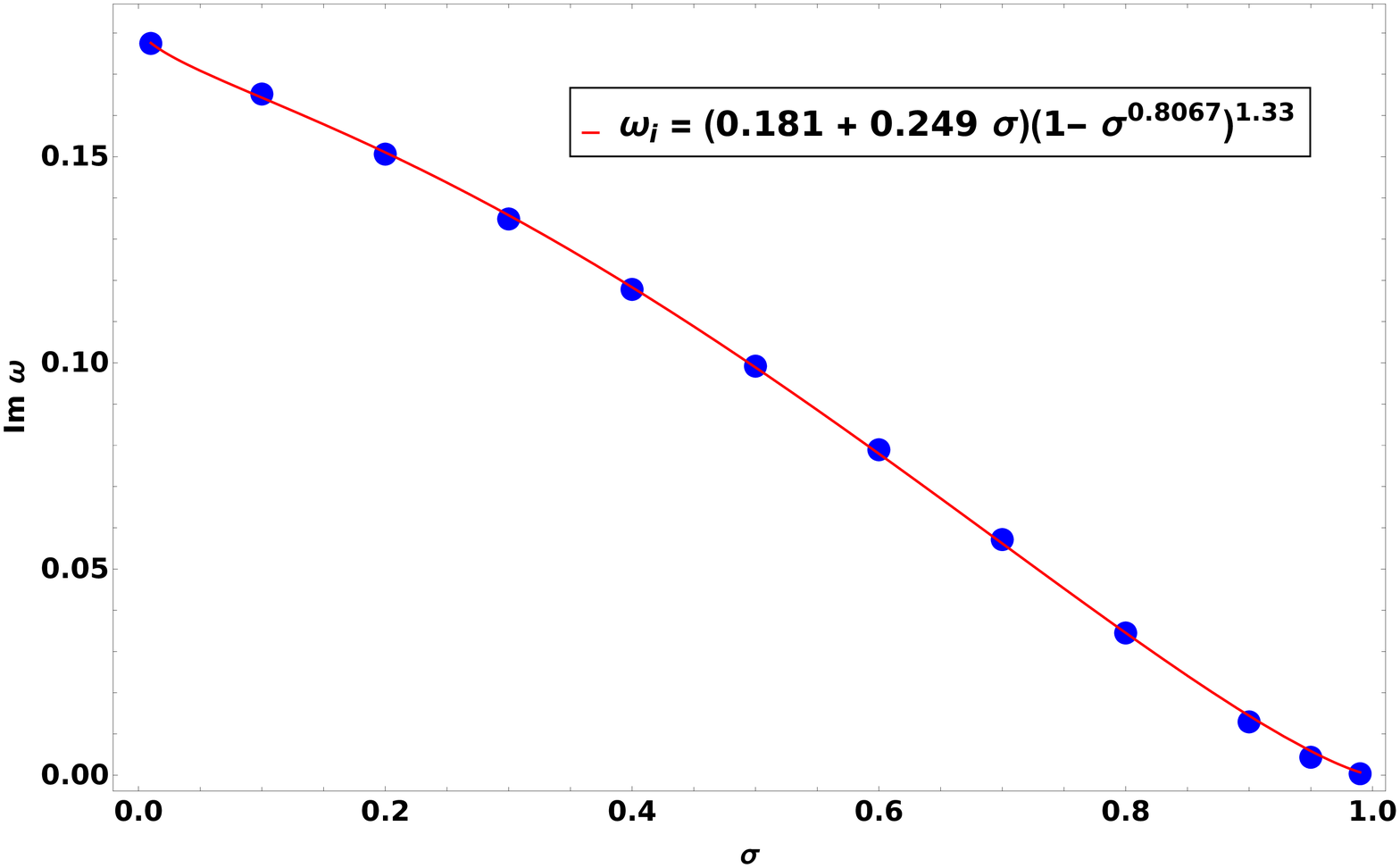}
 	\caption{ Fitting $Im (\omega)$}
 	\end{subfigure}
 	\caption{Analytical fitting function mentioned in the inset boxes in the figures indicate the dependence of the Re($\omega$) and Im($\omega$) (in $M^{-1}$) on $\sigma$ for Case III wormholes and is shown by the red curve for $m=1$.}
 	\label{fig:CaseIII_fit}
 \end{figure}

\noindent The QNMs will also depend on parameter $\sigma$, for any particular $m$ value, which are shown in Table \ref{tab:n2}. To visualize the dependence analytically, we have obtained the fitting functions for $\omega_r$ and $\omega_i$ as a function of $\sigma$ using the {\em NonlinearModelFit} feature of {\em Mathematica} and are denoted within the plots of Fig.(\ref{fig:CaseIII_fit}). 

\noindent Machine precision (precision upto 16 decimal places) has been used for obtaining the fitting functions in {\em Mathematica}. Fig.(\ref{fig:CaseIII_fit}) shows the best-fit model where the analytical red curve matches well with the data points shown in blue. The goodness of the fit is measured by calculating the $\chi^2$ value associated with the fit. The above fitting function for $\omega_r$ (shown in inset of Fig.(\ref{fig:CaseIII_fit})) gives a `p value' as 0.87 calculated using {\em Mathematica} which gives the probability of obtaining the observed results from the fit as $93\%$. The probability value is taken from the chi-square distribution table corresponding to 4 degrees of freedom \cite{chi_distribution}. This indicates that the model considered is suitable and can be accepted to represent the real parts of QNM frequencies. Similarly, for $\omega_i$ the probability is about $94\%$ ensuring the goodness of the fitting function. We can now use the fitting function to estimate the frequency and damping time corresponding to different wormhole parameters. In physical units, throat radius is $r_0 =2M$ and the oscillation frequency becomes

\begin{equation}
 \nu = \frac{c^3}{G M} \frac{f(\sigma)}{2 \pi} Hz = \frac{32390.9}{M} f(\sigma) \, Hz
\end{equation}
with the mass parameter $M$ expressed in units of solar mass $M_\odot$ and $G = 6.67 \times 10^{-11}$ $m^3 kg^{-1} s^{-2}$ , $c= 3 \times 10^8 m s^{-1}$. The mass parameter $M$ denotes the corresponding ADM mass of the wormhole \cite{shaikh_2018_1}. Let us consider wormhole with $\sigma=0.01$ and we take the detector sensitivity range from 10Hz- 10 kHz \cite{ligo_2020}. To obtain the throat radius ($r_0=2M$) in length units we multiply a factor $G/c^2$. The throat radius then becomes $7.16472 \times 10^3$ km and mass of wormhole $2430.25 M_\odot$ for frequency of 10Hz. Similarly, if we take $\nu=10 kHz$, then throat radius and mass is $7.164$ km and $2.43025 M_\odot$ respectively. Comparing with the masses of the astrophysical objects mentioned in the LIGO/VIRGO data catalogue, we deduce the mass of wormhole corresponding to 10 Hz frequency to be very large and although possible, is not of interest at the moment. So a realistic wormhole, if they exist in nature, may emit scalar/gravitational waves of a very high frequency of the order of $\sim$kHz. The analysis can be carried out for other
$\sigma$ values as well as for gravitational perturbations, but the results possibly remain qualitatively the same.\\
On the other hand, the damping time is given by the imaginary part of $\omega$   
\begin{equation}
    \tau = \frac{2 \pi G M}{c^3 g(\sigma)} sec
\end{equation}
If we consider $\sigma=0.01$ and take the mass of the wormhole as $2.43025 M_\odot$ obtained earlier, then damping time is of the order 0.000422 sec. Similar fitting functions can also be obtained for other mode values where the values of the constants $a,b,c,d$ will change accordingly. 

\subsection{\bf{Case IV}: \texorpdfstring{$\sigma=0,\kappa \neq 0$}{$σ=0, κ ≠ 0$ } (Hayward wormhole) }

\noindent The quasi-normal modes of the Hayward wormhole are calculated using the WKB method improved with Pade approximation along with the standard numerical routines as the potential is a single barrier for all parameter values. We begin by exploring the dependence of the QNM values on the angular momentum number $m$ for different $\kappa$ wormholes, where some interesting non-trivial behavior is observed. 

\begin{figure}[h]
    \centering
    \includegraphics[width=0.6\textwidth]{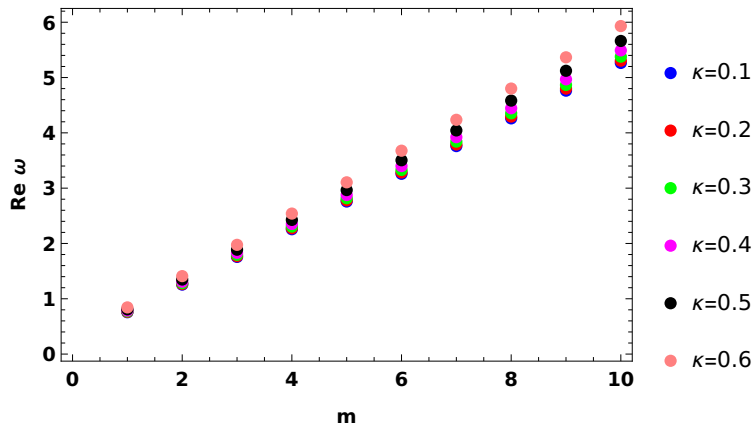}
    \caption{Variation of Re($\omega$)(in $M^{-1}$) with $m$ for different $\kappa$ Hayward wormholes.}
    \label{fig:CaseIV_real}
\end{figure}

\begin{figure}[h]
\begin{subfigure}{0.48\textwidth}
	\includegraphics[width=0.88\linewidth]{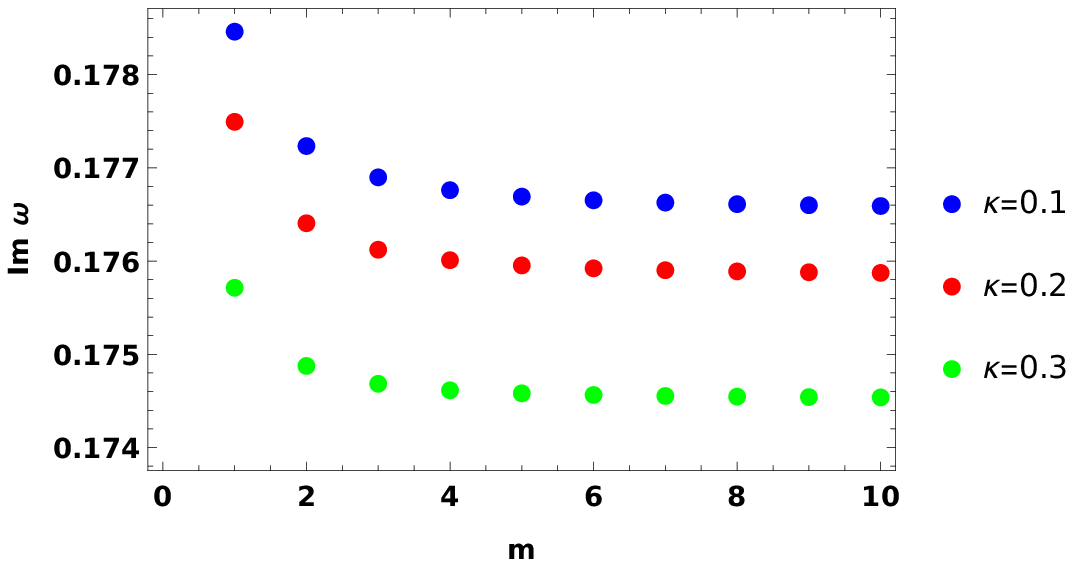}
 \end{subfigure}
  \begin{subfigure}{0.48\textwidth}
 	\includegraphics[width=0.88\linewidth]{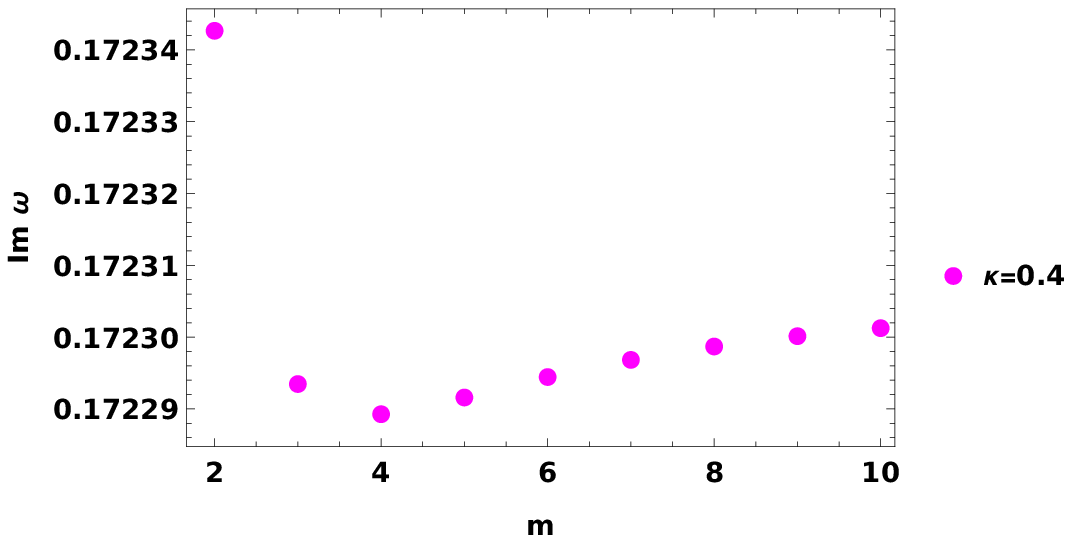}
 	\end{subfigure}
\vspace{0.2in}
 \begin{subfigure}{0.48\textwidth}
 	\includegraphics[width=0.88\linewidth]{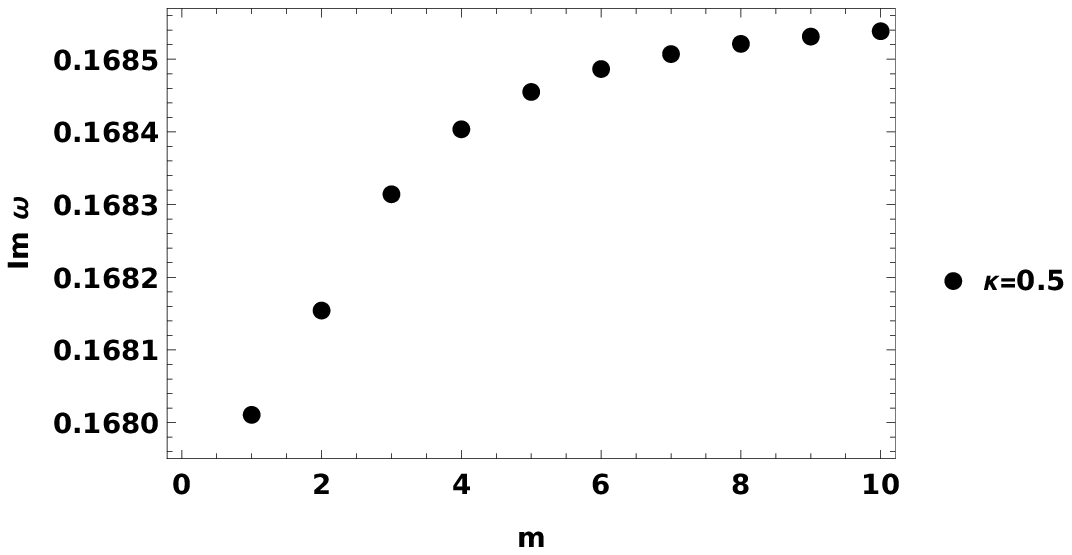}
 	\end{subfigure}
 \begin{subfigure}{0.48\textwidth}
 	\includegraphics[width=0.88\linewidth]{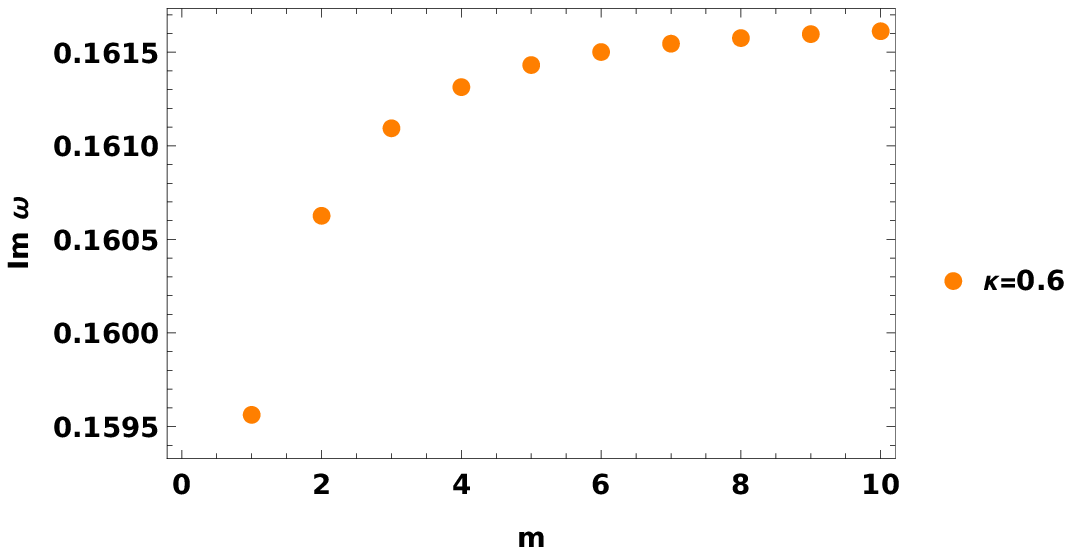}
 \end{subfigure}
 	\caption{Variation of imaginary component of QNM frequency (in units of $M^{-1}$) for different $\kappa$ Hayward wormholes with $m$.}
 	\label{fig:CaseIV_Im_QNM}
 \end{figure}

\begin{figure}
    \centering
    \includegraphics[width=0.4\linewidth]{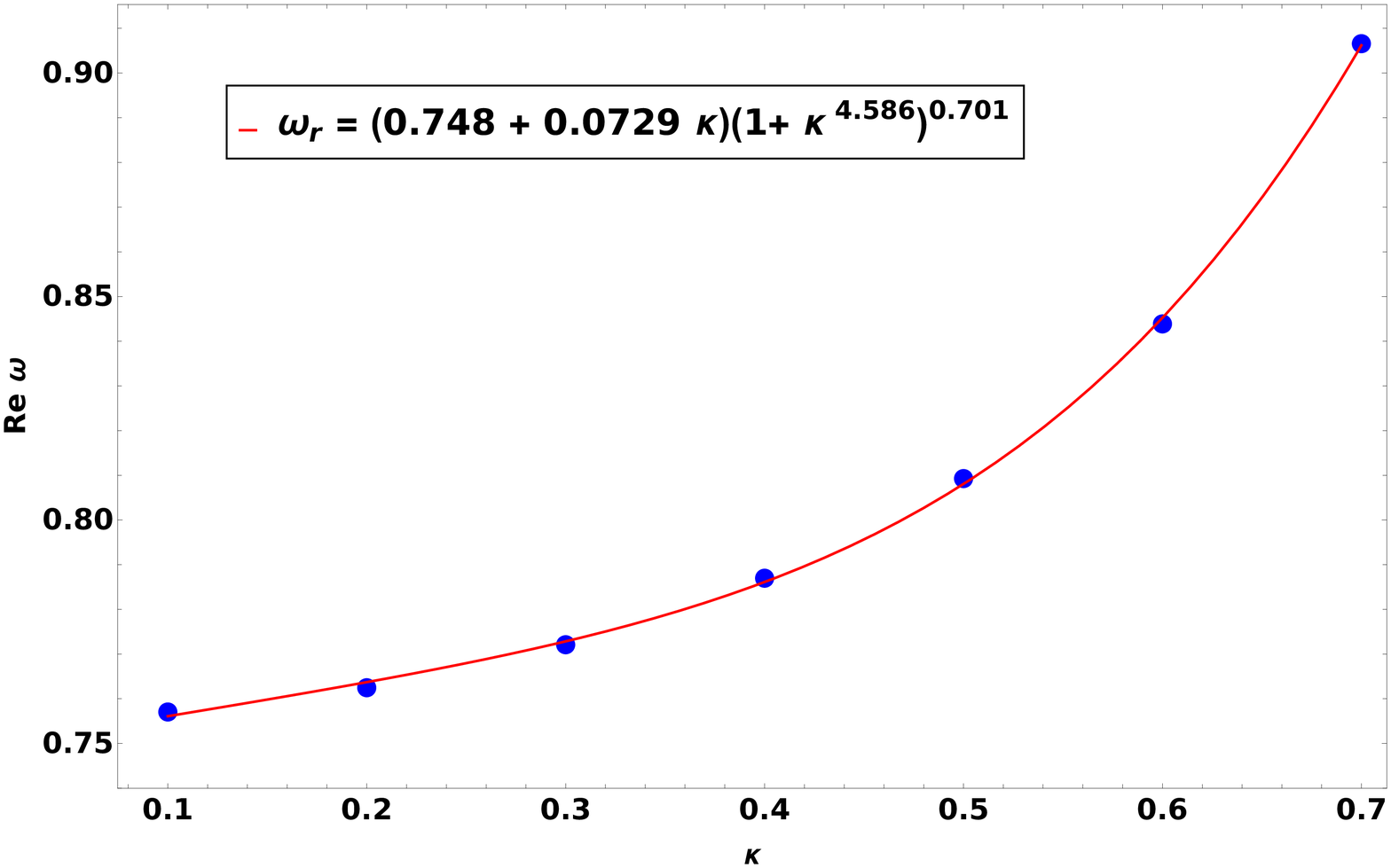}
    \caption{Analytical fitting function mentioned in the inset boxes in the plot indicates the dependence of the Re($\omega$) (in $M^{-1}$ units) on $\kappa$ for Hayward wormholes and is shown as the red curve with $m=1$. It matches well with the QNM values shown in blue.}
    \label{fig:CaseIV_real_fit}
\end{figure}

\begin{figure}[h]
\begin{subfigure}{0.43\textwidth}
	\includegraphics[width=0.9\linewidth]{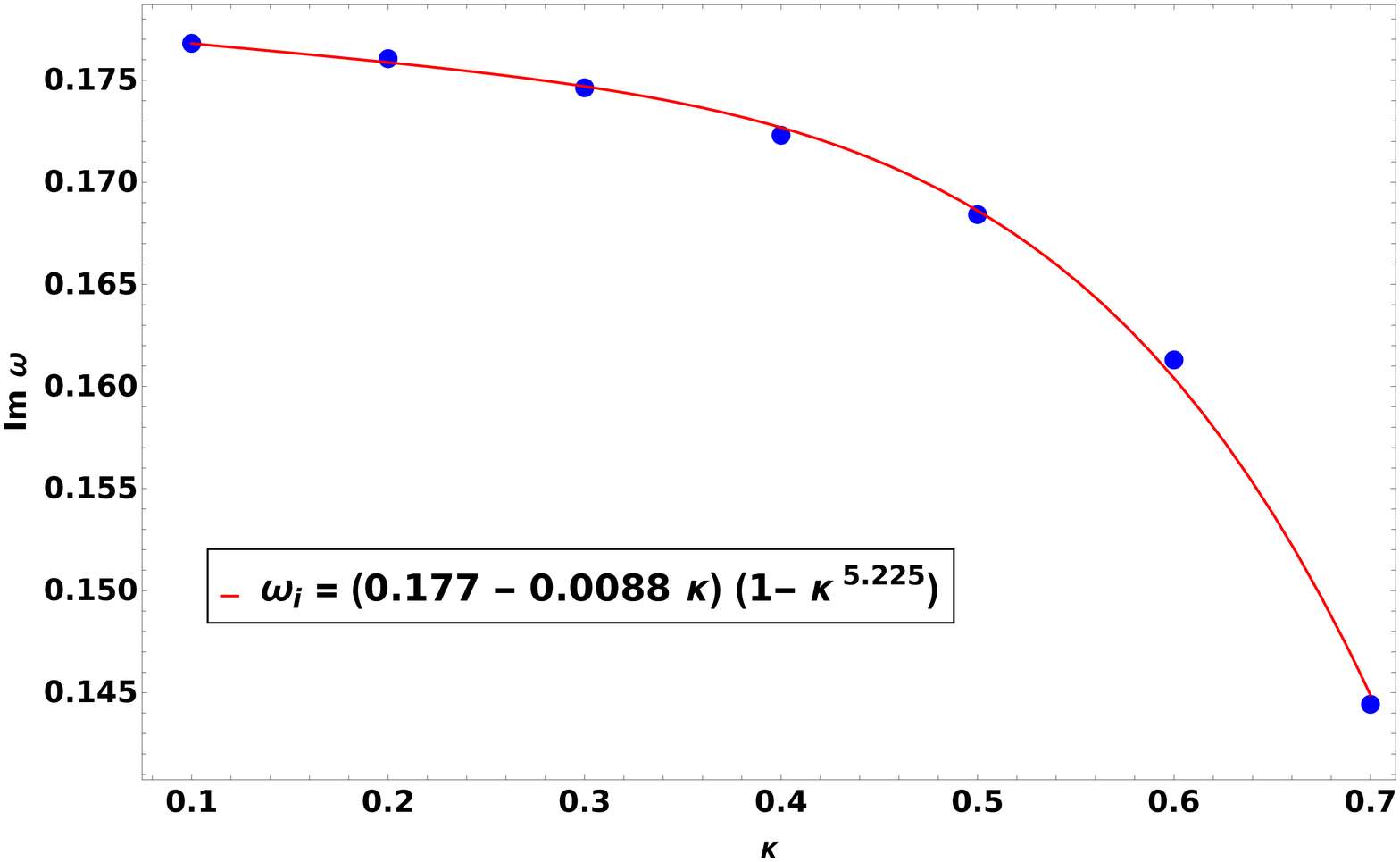}
 	\caption{Fitting $Im (\omega)$ for $m=4$}
 \end{subfigure}
 \begin{subfigure}{0.43\textwidth}
 	\includegraphics[width=0.9\linewidth]{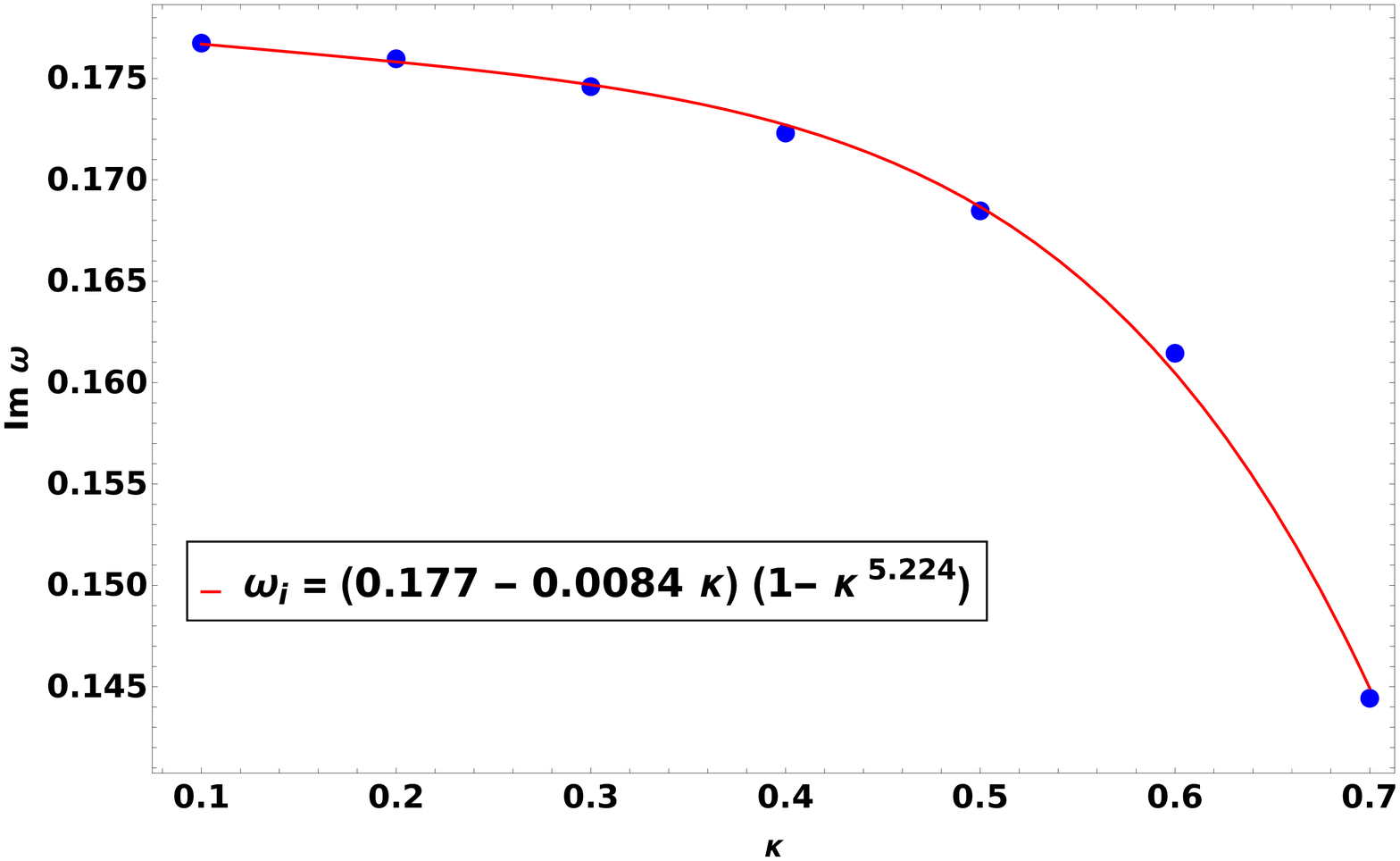}
 	\caption{ Fitting $Im (\omega)$ for $m=6$}
 	\end{subfigure}
 	\caption{Analytical fitting functions mentioned in the inset boxes in the plots indicate the dependence of the Im($\omega$) (in $M^{-1}$ units) on $\kappa$ for Hayward wormholes and is shown as the red curve with $m=4$ and $m=6$. }
 	\label{fig:CaseIV_im_fit}
 \end{figure}

\noindent Fig.(\ref{fig:CaseIV_real}) shows the real $\omega$ to be increasing with increasing $m$ which holds true for all values of $\kappa$. On the other hand, imaginary $\omega$ decreases with increasing $m$  until $\kappa$ is small (see Fig.(\ref{fig:CaseIV_Im_QNM})). As we consider higher values of $\kappa$, the $\omega_i$ starts to increase with $m$ where the intermediate values of $\kappa$ show the transitional behaviour. Such behaviour of the imaginary part of $\omega$ with respect to $m$ has been observed in other wormhole geometries as well \cite{kim_2018,pdr_2021}.\\
Next, we study the dependence of the QNMs on parameter $\kappa$ and calculate the corresponding fitting functions as discussed in the previous section. 
For $\omega_r$, $m=1$ we obtain the fitting function using the {\em NonlinearModelFit} in {\em Mathematica}. Such fitting model has a probability of $94\%$ from chi-square estimation and hence is taken as a good fit. The behaviour will hold true for all $m$ values but with different values of constants in the fitting function (see Fig.(\ref{fig:CaseIV_real_fit})). To get an estimate on the mass and throat radius of the wormholes for different $\kappa$, we consider 
\begin{equation}
    \nu = \frac{c^3 f(\kappa)}{2 \pi G M}
\end{equation}
where $f(\kappa)$ is the fitting function and $M$ denotes the ADM mass in solar mass units. The throat radius $x_0$ needs to be multiplied by a factor of $GM/c^2$ to get the radius in length unit. For $\kappa=0.01$ and $\nu=10$ Hz, $M= 2427.76 M_\odot$ and throat radius is $7.15 \times 10^3$ km. Similarly, for 10 kHz frequency we get the mass of the order of $2.4 M_\odot$ and throat radius 7.15 km. The analysis can be repeated for other $\kappa$ values as well.\\
For the imaginary component of QNM, we derive the fitting function 
where the values of the constant change for different $m$. Although the dependence on $m$ changes for different $\kappa$, we find the fitting function mentioned within the plots of Fig.(\ref{fig:CaseIV_im_fit}) to work well for all $m$ values. For both $m=4$ and $6$, chi-square fitting gives a probability of about $96\%$. This happens because the value of $\omega_i$ decreases with increasing $\kappa$ irrespective of $m$. The fitting function matches well with the QNM values evaluated for two different $m$, $m=4$ and $6$, as shown in Fig.(\ref{fig:CaseIV_im_fit}).

\subsection{\bf{Case V}: \texorpdfstring{$\sigma=1,\kappa \neq 0$}{$σ=1, κ ≠ 0$ } (Regular Hayward black hole)} 

\begin{figure}[h]
\begin{subfigure}{0.46\textwidth}
	\includegraphics[width=0.9\linewidth]{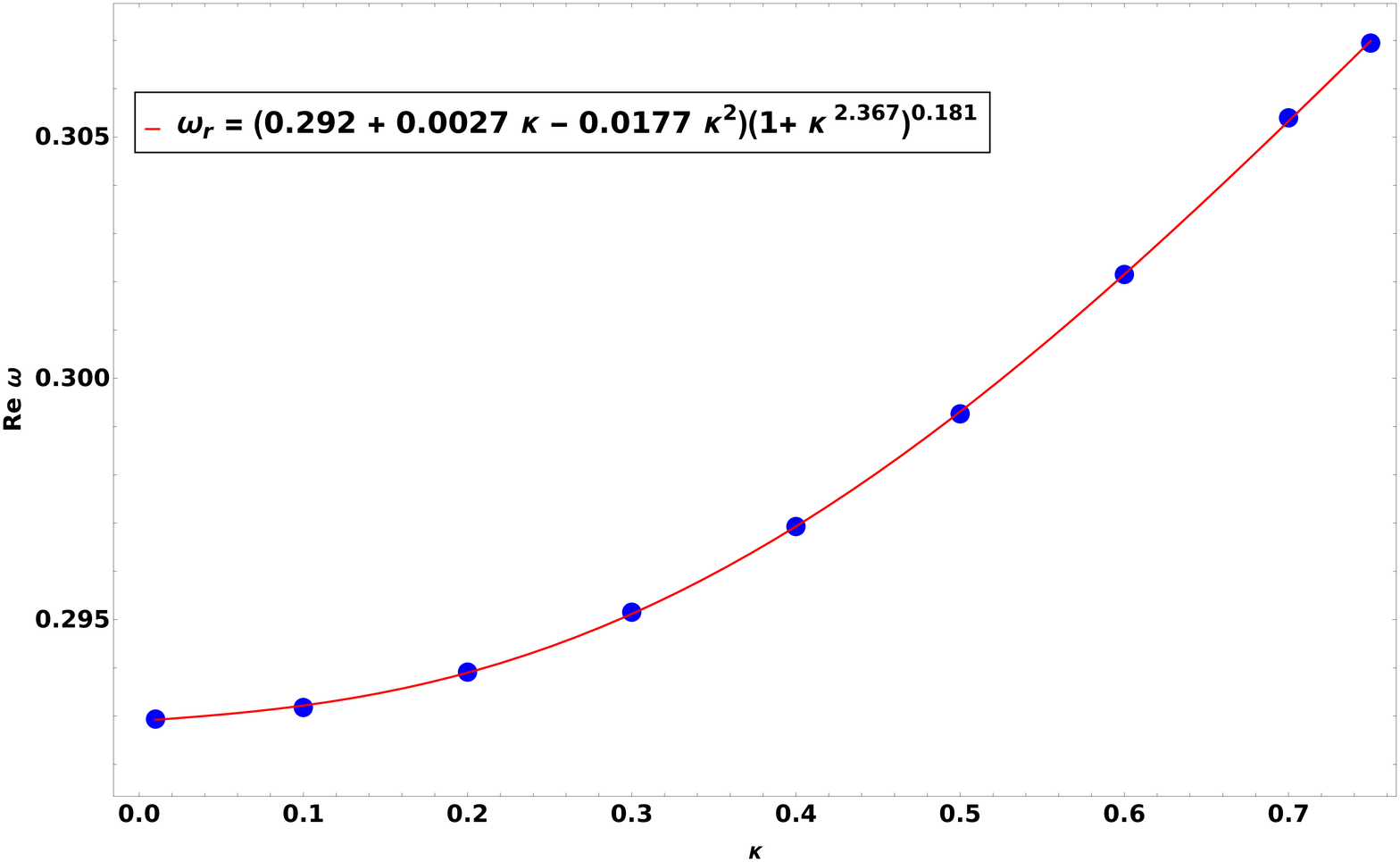}
 	\caption{Fitting $Re(\omega)$}
 \end{subfigure}
 \begin{subfigure}{0.46\textwidth}
 	\includegraphics[width=0.9\linewidth]{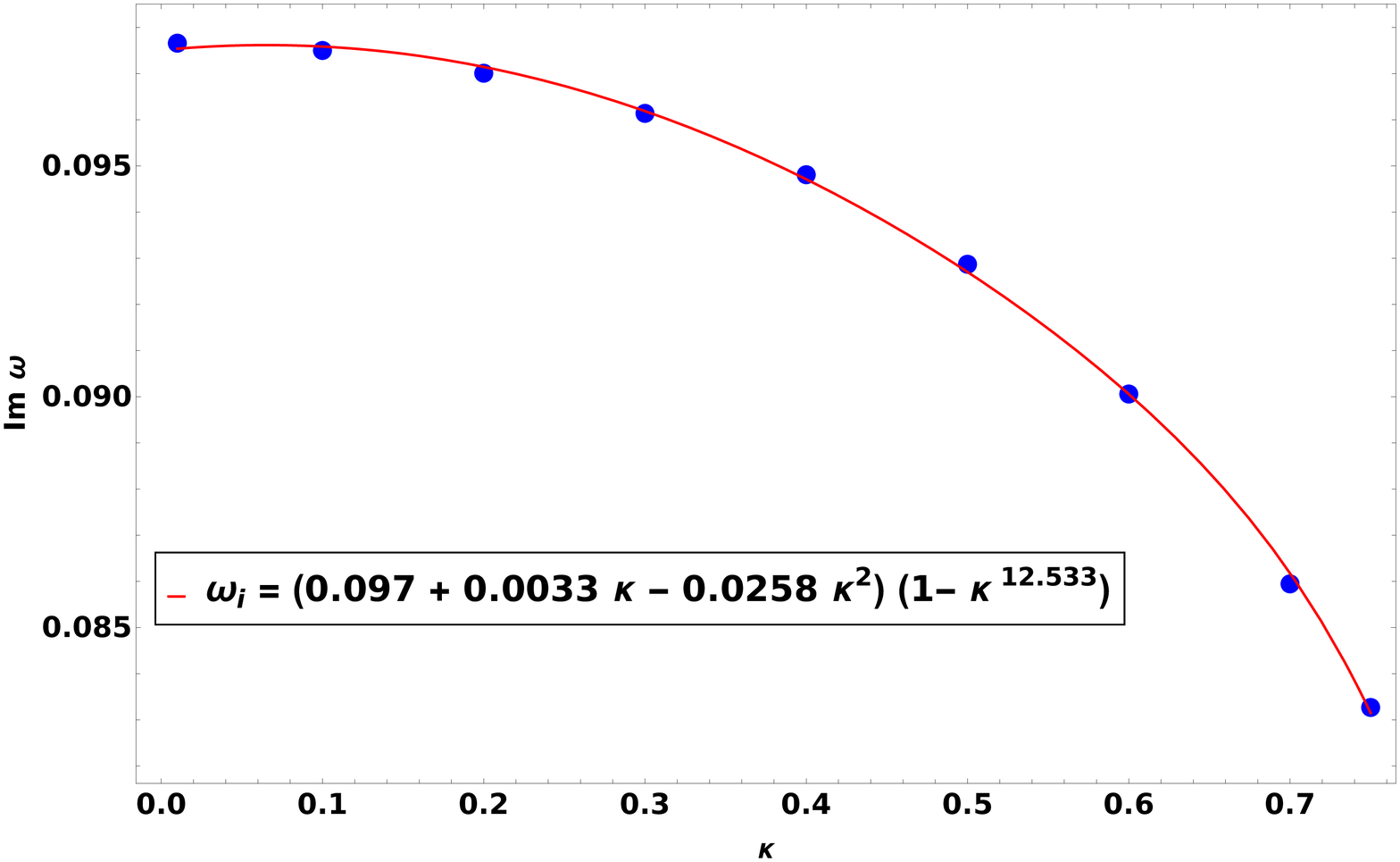}
 	\caption{ Fitting $Im (\omega)$}
 	\end{subfigure}
 	\caption{Analytical fitting functions mentioned in the inset boxes in the plots indicate the dependence of the Re($\omega$) and Im($\omega$) (in $M^{-1}$) on $\kappa$, for Hayward regular black hole and is shown by the red curve with $m=1$. }
 	\label{fig:CaseV_fit}
 \end{figure}

\noindent This case corresponds to the Hayward black hole which is regular at $r=0$. Once again we aim to study the dependence of the QNM frequencies on the parameter $\kappa$ as $\sigma$ is fixed to one. First, we consider the real part of $\omega$ and proceed by considering the fitting function plotted in Fig.(\ref{fig:CaseV_fit}) determined by {\em NonLinearModelFit} in {\em Mathematica}. Estimating the mass and horizon radius for say $\kappa=0.01$, we find $M=948 M_\odot$ and radius $2.79 \times 10^3$ km corresponding to 10 Hz frequency, while for higher 10 kHz we get mass and radius as $0.948 M_\odot$ and 2.79 km respectively. In essence, the black hole with $\kappa=0.01$ is much lighter when compared to a Case IV wormhole with same $\kappa$, for a particular frequency.\\
Similarly, for the imaginary part we obtain a fitting function that matches well with the calculated QNM values shown in Fig.(\ref{fig:CaseV_fit}). From the chi-square distribution we have also ensured the goodness of the above fitting functions --they turn out to be around $95\%$ and $93\%$ respectively.

\subsection{\bf{Case VI}: \texorpdfstring{$0<\sigma<1$, $0< \kappa \leq \frac{4}{3 \sqrt{3}}$}{$0<σ<1,0< κ ≤ 4/3√3$ } (Hayward-Damour-Solodukhin wormhole)} 

\begin{figure}
    \centering
    \includegraphics[width=0.55\linewidth]{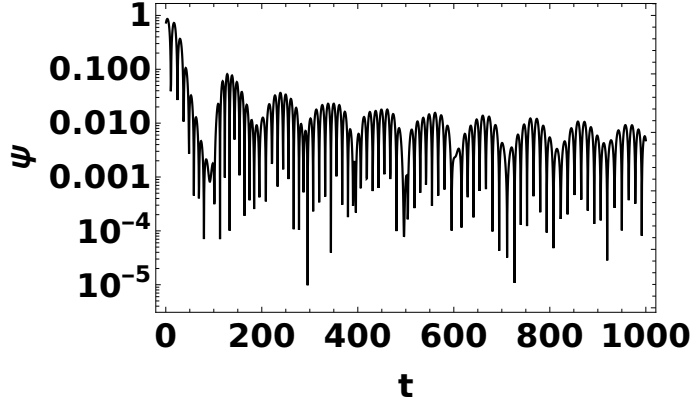}
    \caption{Time domain profile showing distinct echoes as observed at $r_*=30$ for $\sigma=0.9999$, $\kappa=0.5$ Case VI wormhole. The grid spacing $h=0.5$ and a Gaussian initial profile is used. }
    \label{fig:TD_echoes}
\end{figure}

\noindent We finally arrive at the most general parametrized wormhole metric which supports effective potentials with both single and double peaks depending on the value $\sigma$. Hence, the numerical routines mentioned earlier are better suited for calculating the QNMs. Nonetheless, for single barriers, the WKB method acts as a verification mechanism for the numerical results. Distinct echo patterns are also visible for wormholes that have double potential barriers as shown in Fig.(\ref{fig:TD_echoes}) for a particular wormhole example belonging to this case. Similar to Case III wormholes, the initial prompt damping is dominated by photon ring modes with the fundamental wormhole QNMs coming into play once the echoes have died down at very late times.

\noindent The study of the QNM dependence on parameters is more intricate for these wormholes as the metric is controlled by both $\sigma$ and $\kappa$.  To observe the QNM behaviour, we will keep one of the parameters fixed at a time and study the variation of the QNMs with the other. First, we fix $\sigma$ and choose different fitting functions to explore the dependence of QNMs on $\kappa$ when calculated in units of $M^{-1}$.\\



\begin{figure}[h]
\begin{subfigure}{0.45\textwidth}
	\includegraphics[width=0.85\linewidth]{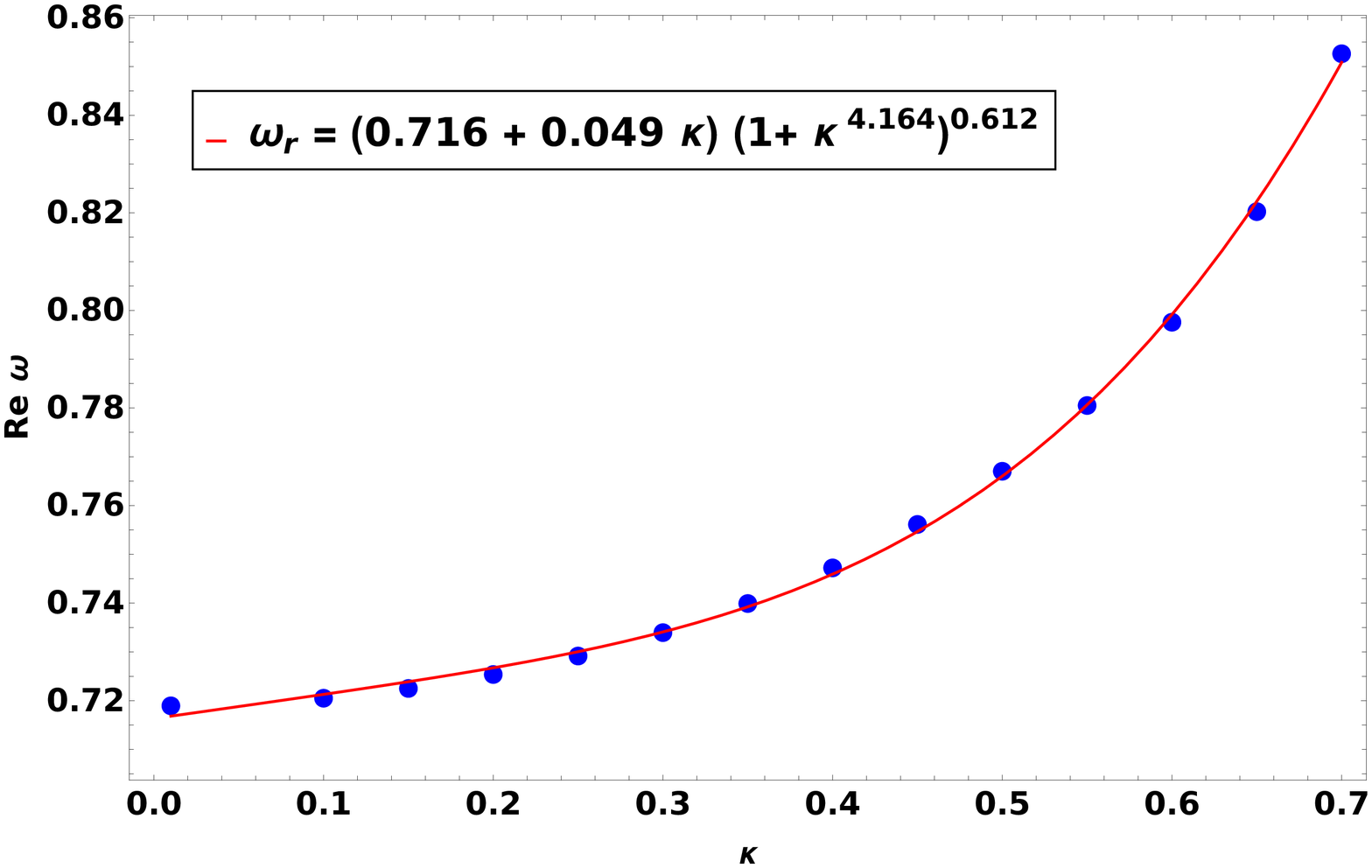}
 	\caption{Fitting $Re(\omega)$ for $\sigma=0.1$}
 \end{subfigure}
 \begin{subfigure}{0.45\textwidth}
 	\includegraphics[width=0.85\linewidth]{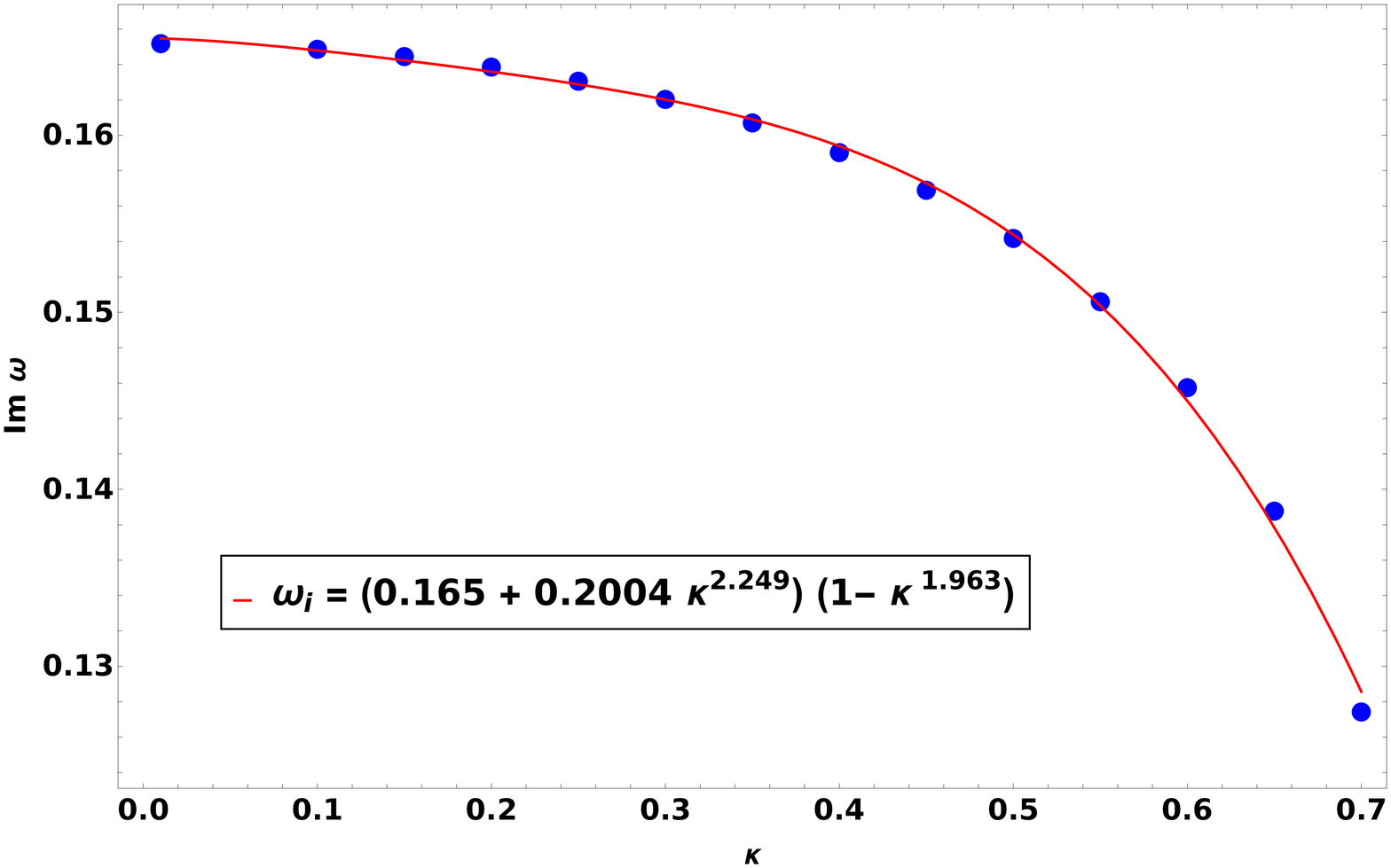}
 	\caption{ Fitting $Im (\omega)$ for $\sigma=0.1$}
 	\end{subfigure}
 \begin{subfigure}{0.45\textwidth}
	\includegraphics[width=0.85\linewidth]{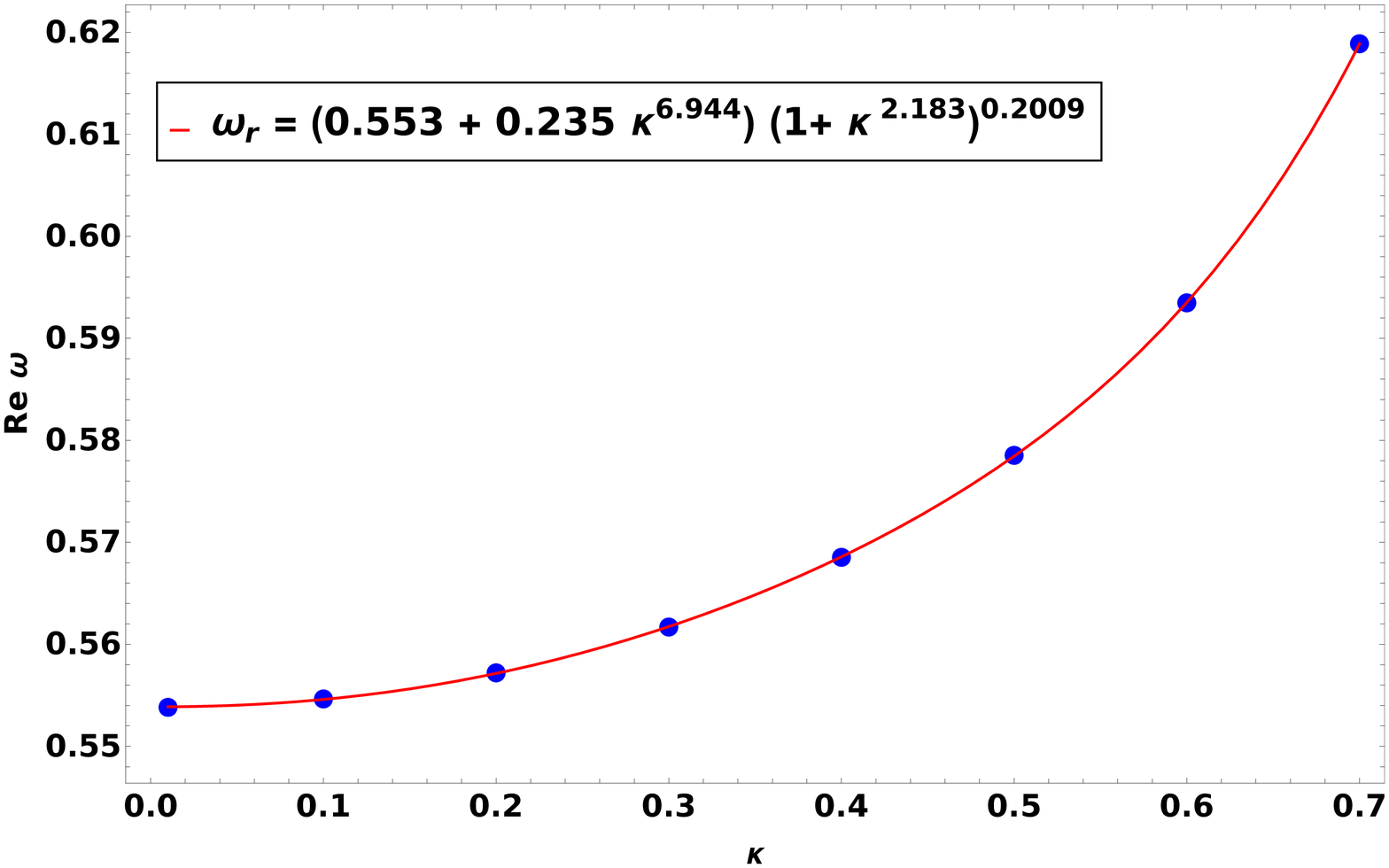}
 	\caption{Fitting $Re(\omega)$ for $\sigma=0.5$}
 \end{subfigure}
 \begin{subfigure}{0.45\textwidth}
 	\includegraphics[width=0.85\linewidth]{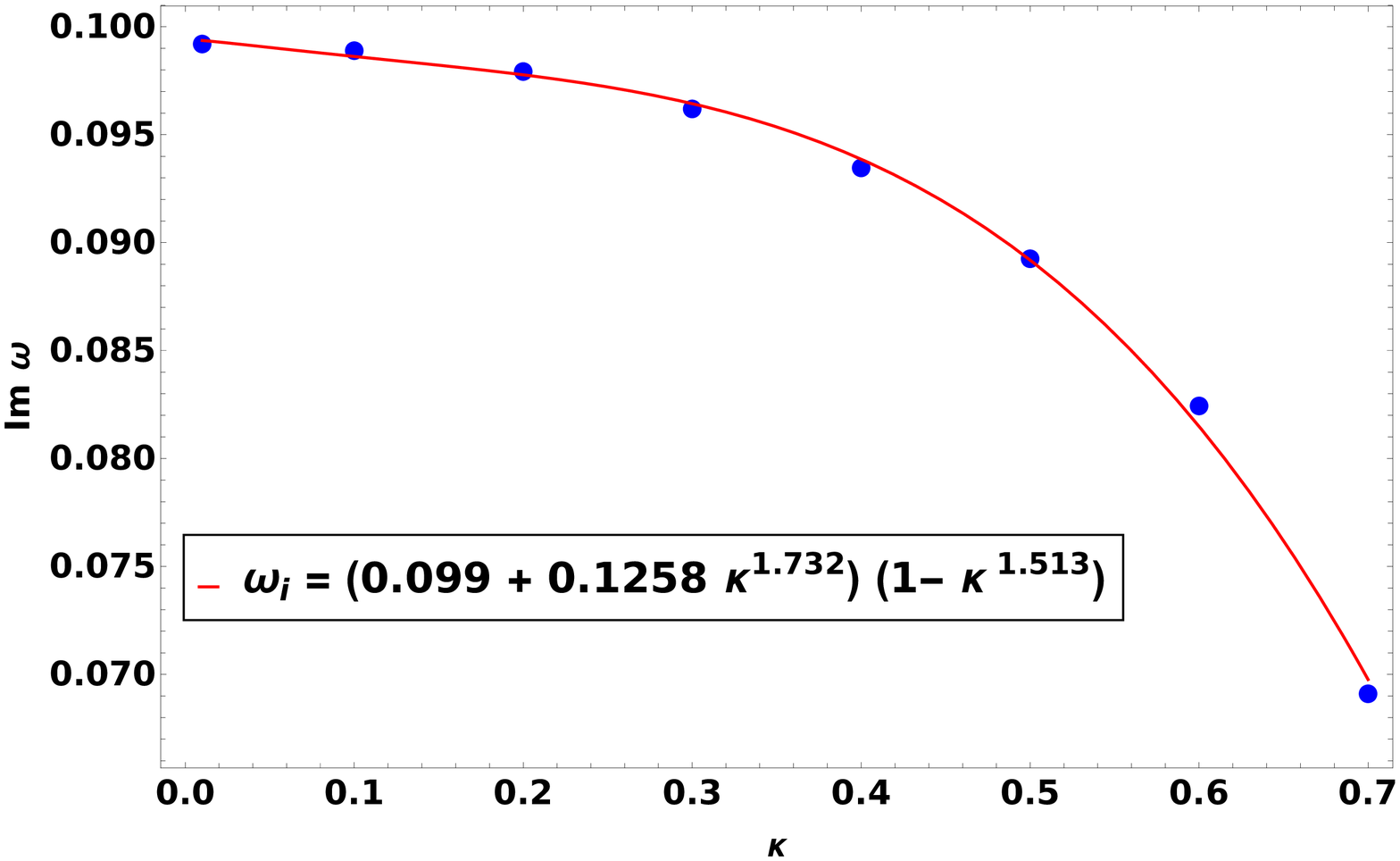}
 	\caption{ Fitting $Im (\omega)$ for $\sigma=0.5$}
 	\end{subfigure}	
 	\caption{Analytical fitting functions mentioned in the inset boxes in the figures indicate the dependence of the Re($\omega$) and Im($\omega$) (in $M^{-1}$) on $\kappa$ for Case VI wormholes and is shown by the red curve with $m=1$ and fixed $\sigma$.}
 	\label{fig:CaseVI_fit1}
 \end{figure}
 
\begin{figure}[h]
\begin{subfigure}{0.45\textwidth}
	\includegraphics[width=0.85\linewidth]{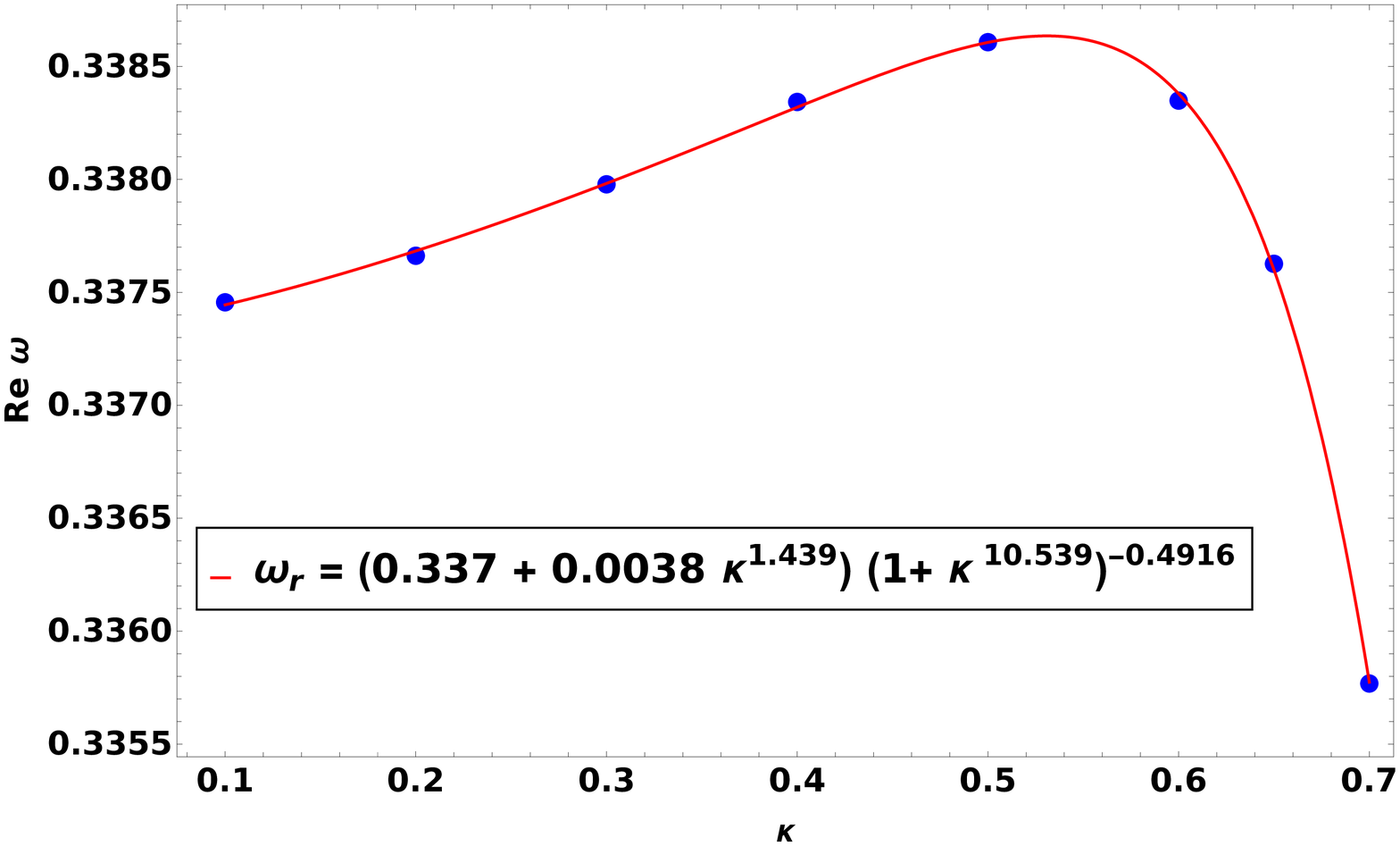}
 	\caption{Fitting $Re(\omega)$ for $\sigma=0.88$}
 \end{subfigure}
 \begin{subfigure}{0.45\textwidth}
 	\includegraphics[width=0.85\linewidth]{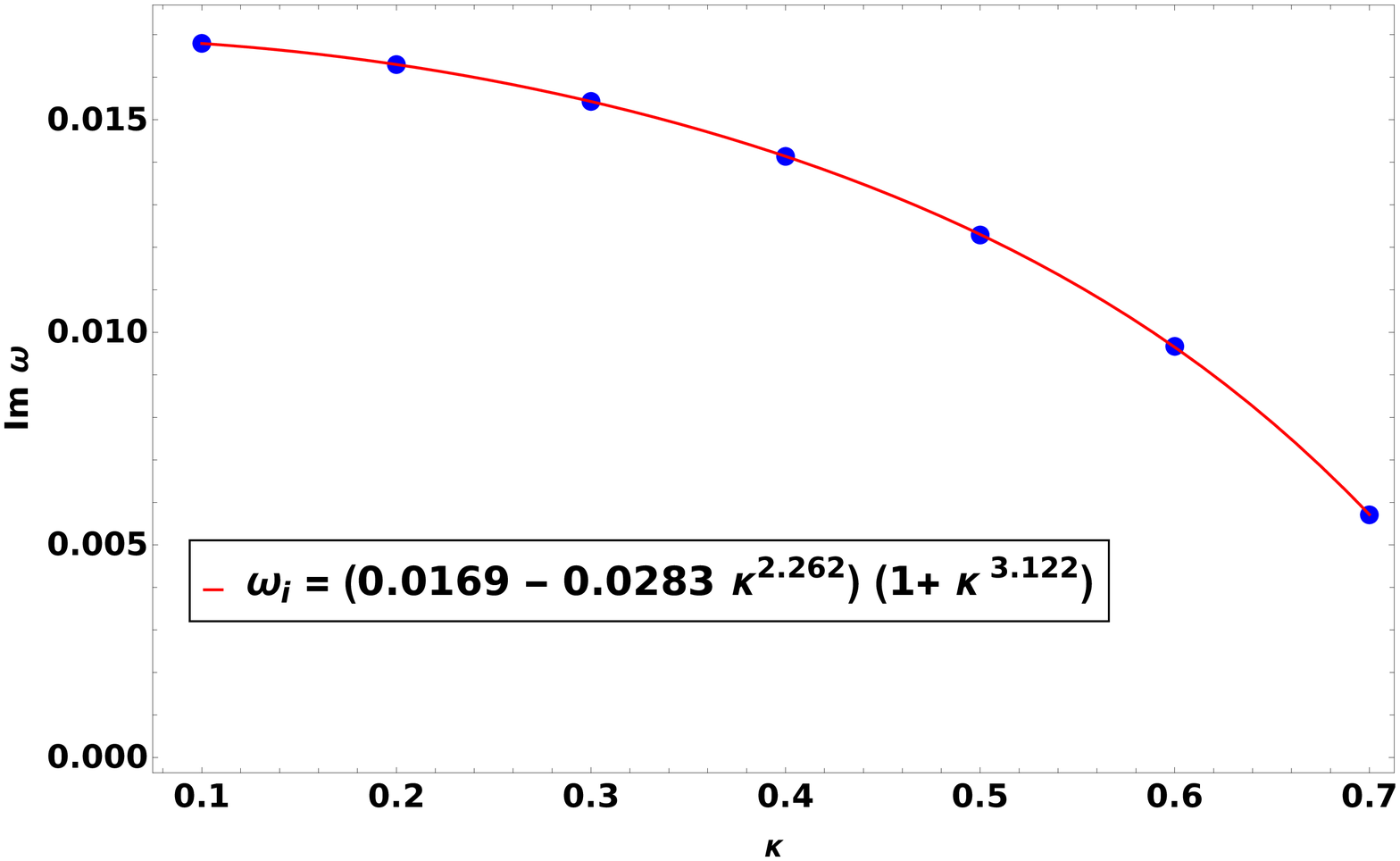}
 	\caption{ Fitting $Im (\omega)$ for $\sigma=0.88$}
 	\end{subfigure}
 \begin{subfigure}{0.45\textwidth}
	\includegraphics[width=0.85\linewidth]{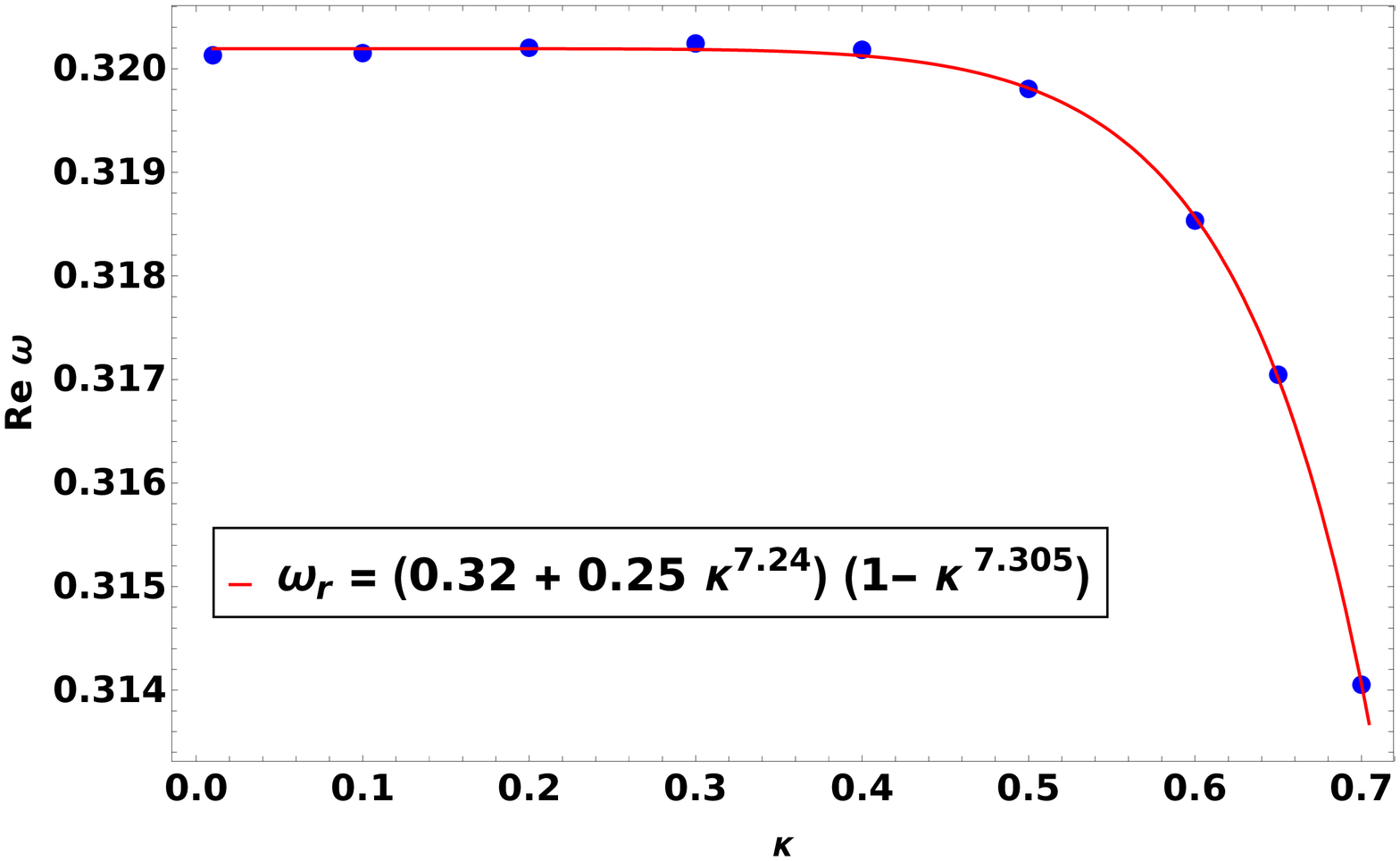}
 	\caption{Fitting $Re(\omega)$ for $\sigma=0.9$}
 \end{subfigure}
 \begin{subfigure}{0.45\textwidth}
 	\includegraphics[width=0.85\linewidth]{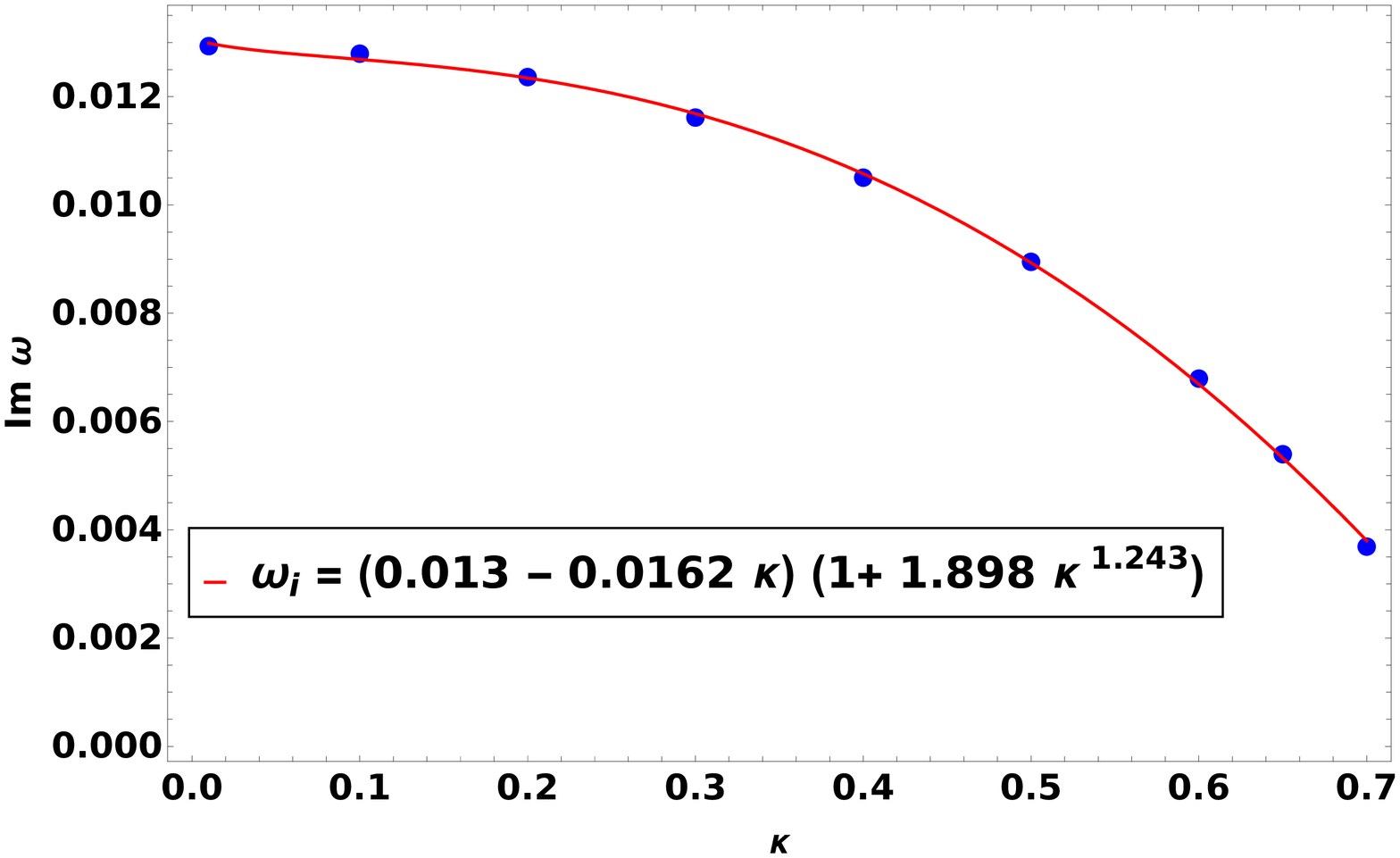}
 	\caption{ Fitting $Im (\omega)$ for $\sigma=0.9$}
 	\end{subfigure}	
 	\caption{The analytical fitting functions mentioned in the figures indicate the dependence of the Re($\omega$) and Im($\omega$) (in $M^{-1}$) on $\kappa$ for Case VI wormholes and is shown by the red curve with $m=1$ and fixed $\sigma$.}
 	\label{fig:CaseVI_fit2}
 \end{figure} 

\noindent The Fig.(\ref{fig:CaseVI_fit1}) and Fig.(\ref{fig:CaseVI_fit2}) show the fitting function as well as the variation of real and imaginary parts of QNM with parameter $\kappa$ for fixed $\sigma$ values. It is observed that while $\omega_i$ always decreases with increasing $\kappa$, there is no such general trend for $\omega_r$. For small values of $\sigma$, $\omega_r$ increases with increasing $\kappa$. But as we approach values of $\sigma$ close to unity,  $\omega_r$ varies slowly with the change occurring in third decimal place of two consecutive frequencies. Initially $\omega_r$ increases followed by a decrease in its value with $\kappa$. These behaviour are reflected in the fitting functions as well. Using the fitting function we estimate the mass and throat radius of the wormhole for different $\sigma$ values as shown in Tab.(\ref{tab:CaseVI}).

 \begin{table}[h]
  \centering
  \begin{tabular}{|c|c|c|c|}
      \hline
     $\sigma$ &  $\kappa$ & M $(M_\odot)$ & Throat Radius (km)\\ 
    \hline 
  0.1 & 0.01 & 2.32 & 6.84 \\
   & 0.7 & 2.75 & 6.63 \\
   \hline
  0.5 & 0.01 & 1.79 & 5.28 \\
   & 0.7 & 2.004 & 4.82\\
   \hline
   0.88 & 0.01 & 1.092 & 3.22\\
   & 0.7 & 1.087 & 2.61\\
   \hline
   0.9 & 0.01 & 1.037 & 3.05\\
   & 0.7 & 1.017 & 2.44\\
  \hline
\end{tabular}
\caption{\label{tab:CaseVI}Numerical estimates on mass and throat radius of Case VI wormholes for different sets of ($\sigma,\kappa$) values corresponding to $10$ kHz frequency obtained using the fitting functions mentioned earlier.}
\end{table}

\begin{figure}[h]
\begin{subfigure}{0.43\textwidth}
	\includegraphics[width=0.78\linewidth]{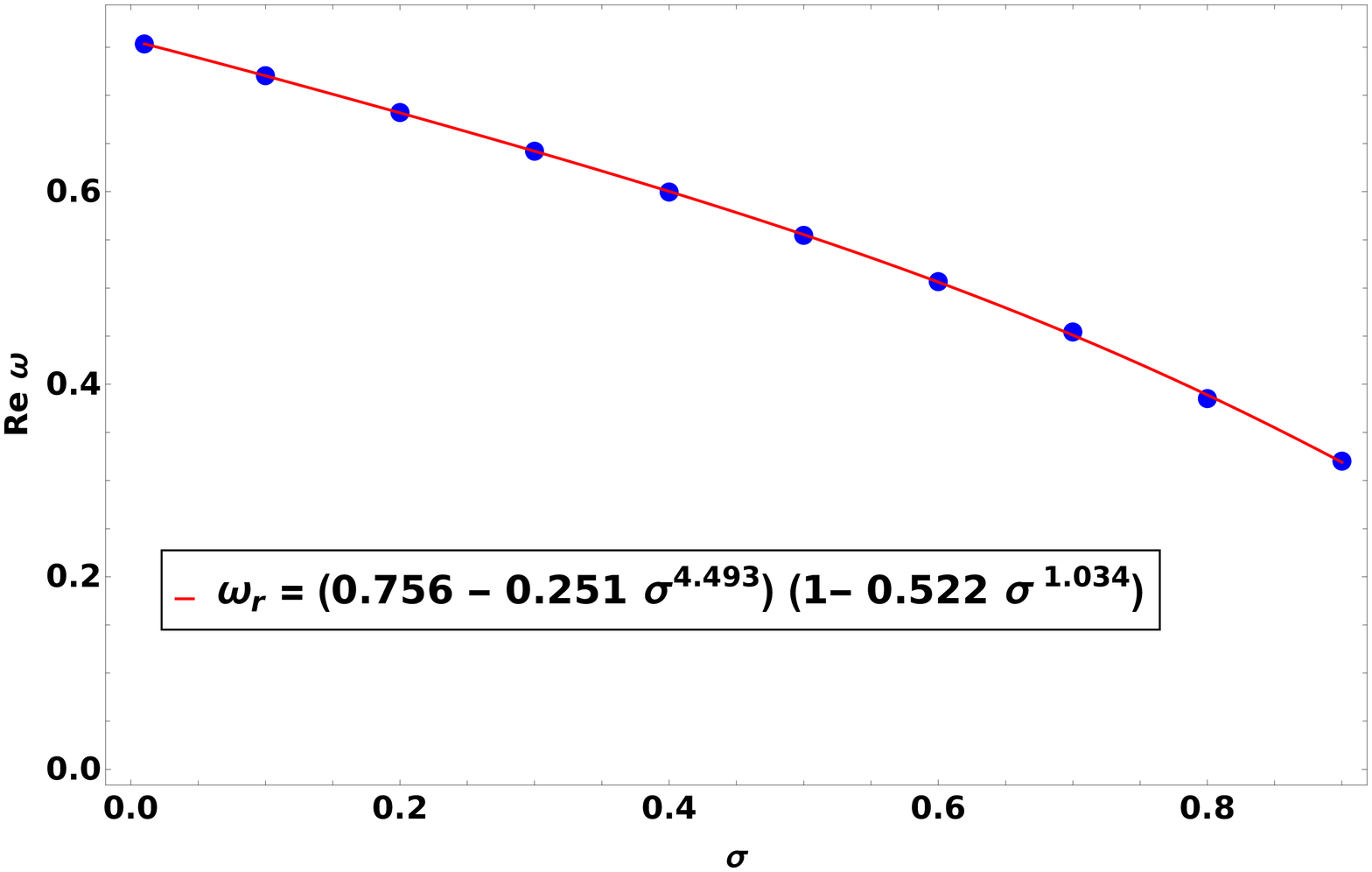}
 	\caption{Fitting $Re(\omega)$ for $\kappa=0.1$}
 \end{subfigure}
 \begin{subfigure}{0.43\textwidth}
 	\includegraphics[width=0.78\linewidth]{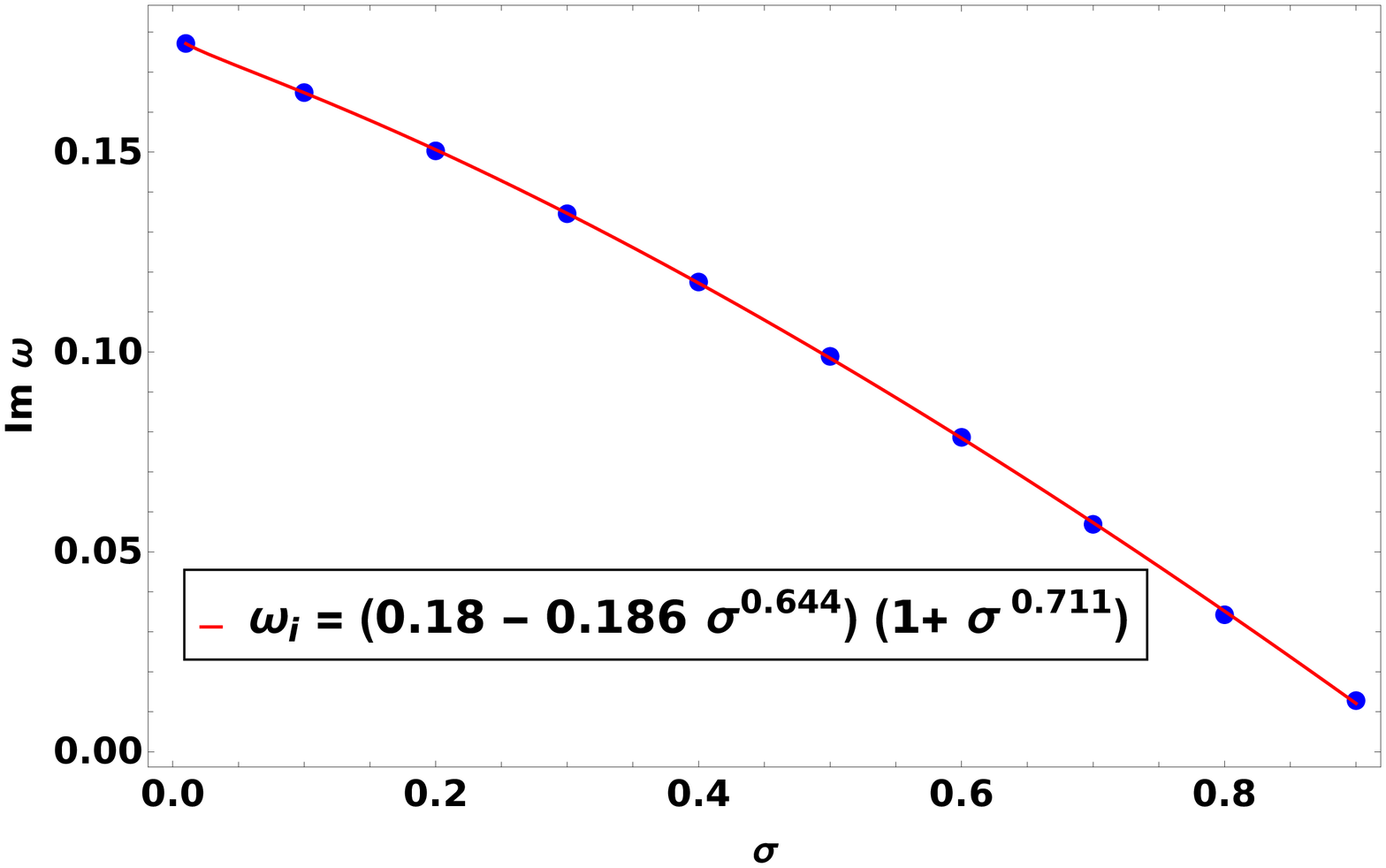}
 	\caption{ Fitting $Im (\omega)$ for $\kappa=0.1$}
 	\end{subfigure}
 \begin{subfigure}{0.43\textwidth}
	\includegraphics[width=0.78\linewidth]{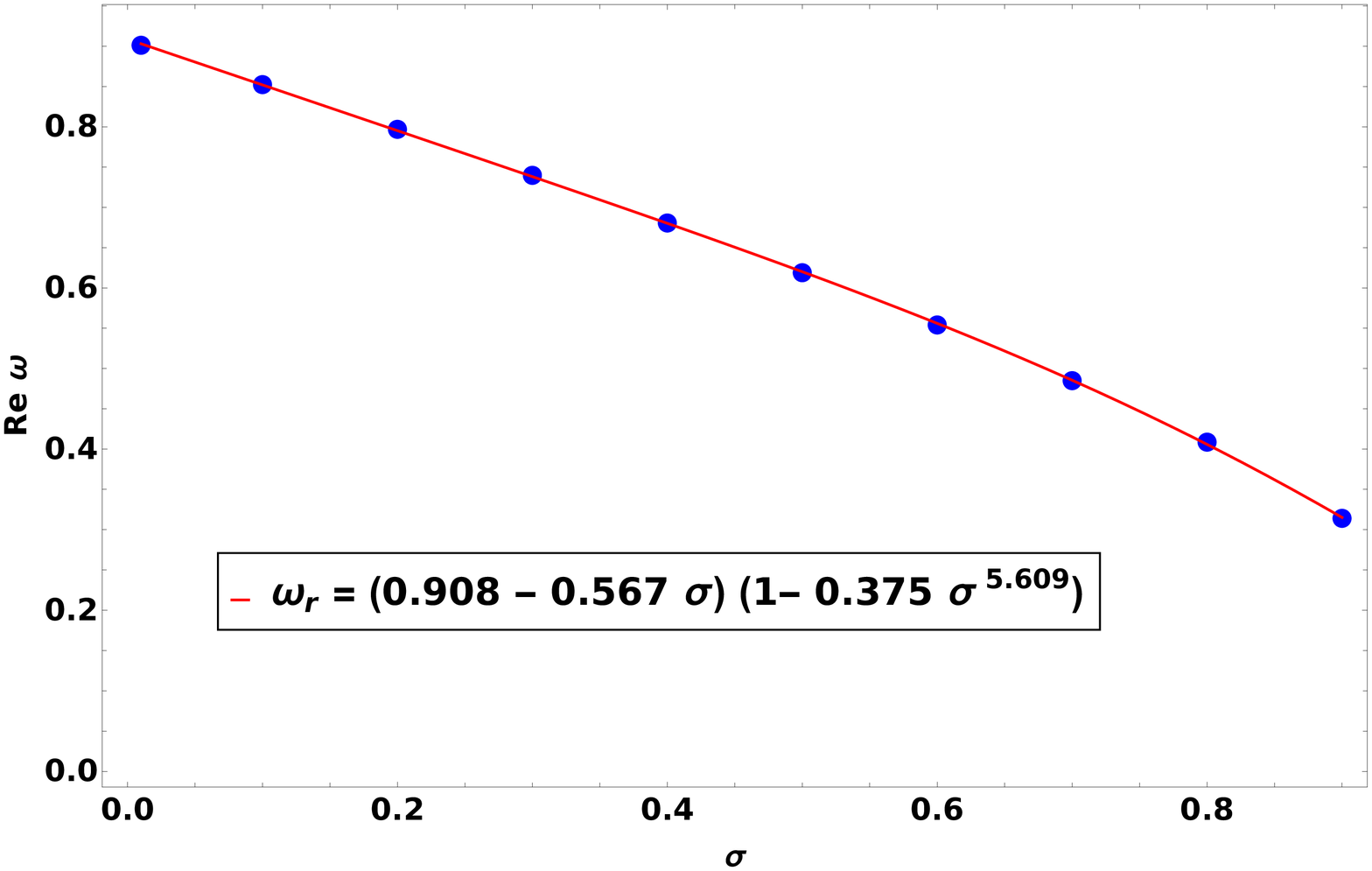}
 	\caption{Fitting $Re(\omega)$ for $\kappa=0.7$}
 \end{subfigure}
 \begin{subfigure}{0.43\textwidth}
 	\includegraphics[width=0.78\linewidth]{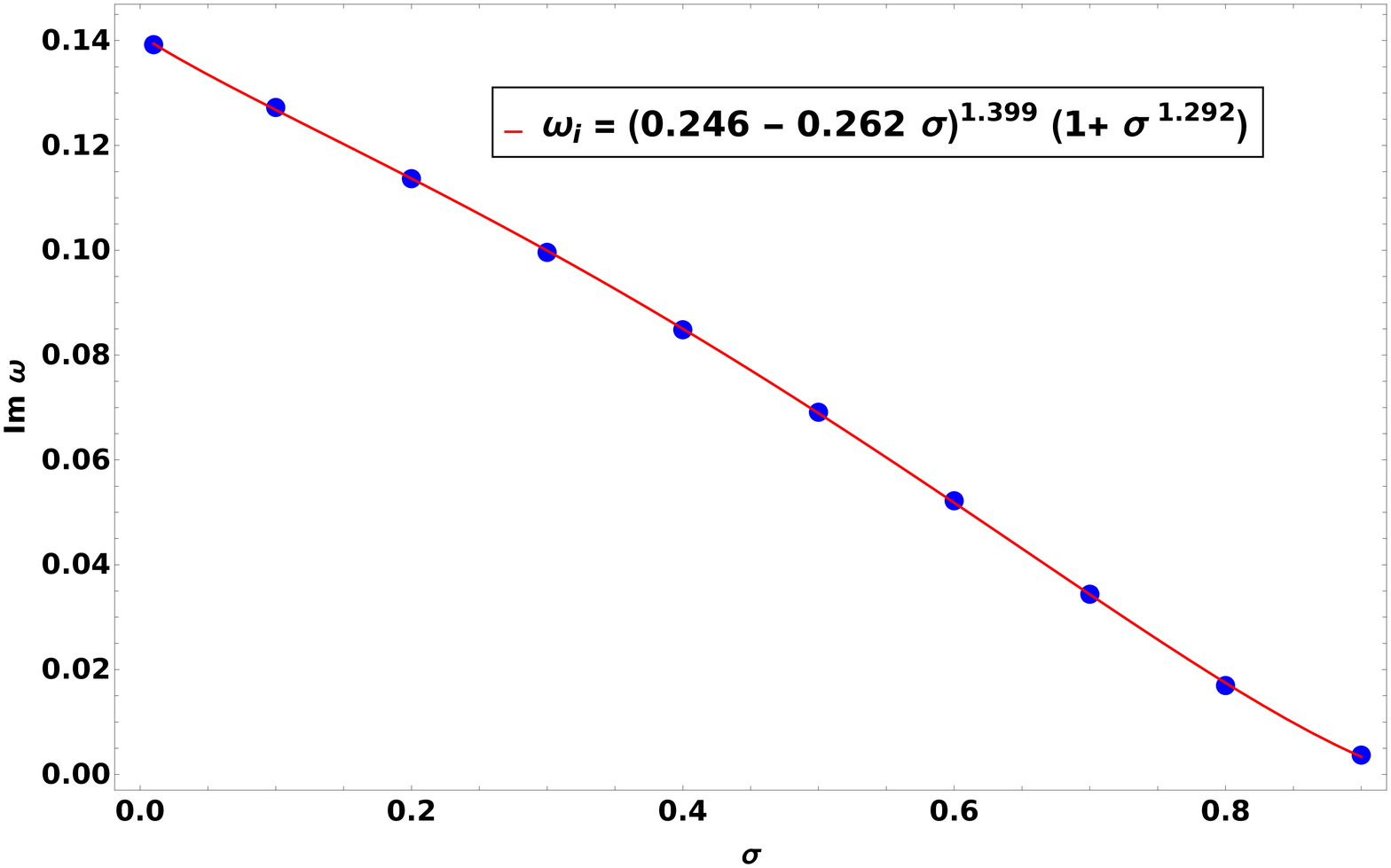}
 	\caption{ Fitting $Im (\omega)$ for $\kappa=0.7$}
 	\end{subfigure}	
 	\caption{Analytical fitting functions mentioned in the inset boxes in the figures indicate the dependence of the Re($\omega$) and Im($\omega$) (in $M^{-1}$) on $\sigma$ for Case VI wormholes and is shown by the red curve with $m=1$ and fixed $\kappa$.}
 	\label{fig:CaseVI_fit3}
 \end{figure}

\noindent A similar analysis can be performed by fixing $\kappa$ and observing the variation of QNM with $\sigma$. We consider $\kappa=0.1$ and $0.7$ and notice the general behavior of QNMs to be similar in both cases with variation of $\sigma$. As we see in Fig.(\ref{fig:CaseVI_fit3}), with increasing $\sigma$ both the real and imaginary parts of QNM decreases. 


\noindent The fitting function mentioned above for $\omega_r$ corresponding to $\kappa=0.7$ and $\sigma=0.5$ gives the mass of the wormhole as $2.008 M_\odot$ and throat radius $4.832$ km. Note that these estimates match with the values given in Tab.\ref{tab:CaseVI} for the same wormhole. This verifies the effectiveness of the fitting functions obtained in two different scenarios for the same wormhole geometries.
 
\section{Distinguishing between different types of spacetimes}

\begin{figure}
    \centering
    \includegraphics[width=0.78\linewidth]{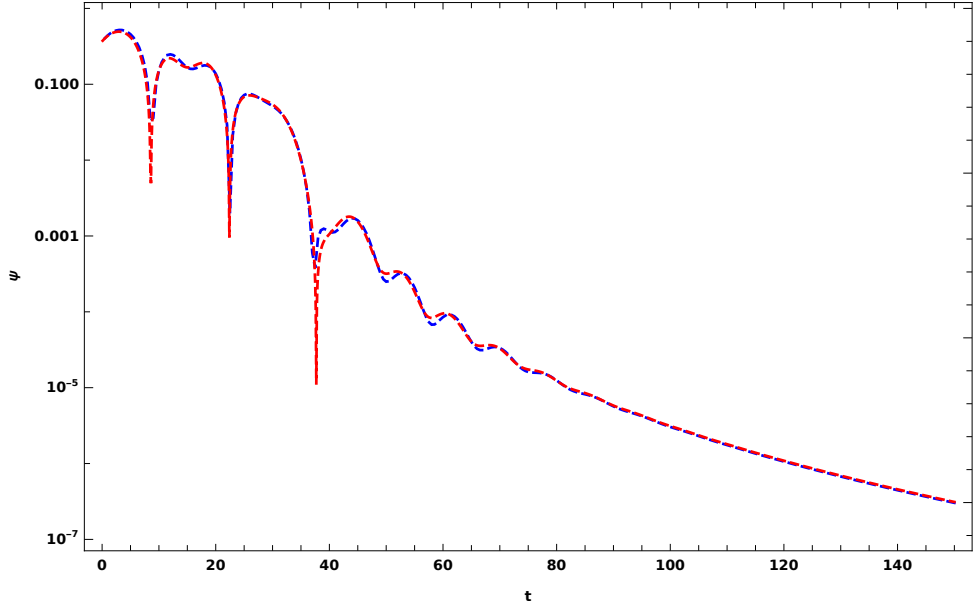}
    \caption{The time domain profiles for Case II wormhole (red) and Case III Damour-Solodukhin wormhole (blue) with $\sigma= 0.01$. The signal is observed at $r_*=5$, grid size $h=0.5$ and initial Gaussian signal of the form $e^{\frac{-(u-10)^2}{100}}$. With $\sigma$ values being close to 0, the two wormhole geometries are difficult to distinguish as evident from their TD profiles and QNM values.}
    \label{fig:Compare_II_III}
\end{figure}

\noindent In the previous section we have thoroughly explored the stability and corresponding quasi-normal modes of each spacetime arising in our parametrized family. We have also observed the dependence of the frequency and damping time on the metric parameters for each geometry and calculated the corresponding fitting functions. 
In some cases, values of the parameters and their change
correspond to switch-over from one geometry to another. An interesting feature of our results is that 
we are able to study those spacetimes where the parameters lie very close to
(above or below) those switch-over values. For example, taking values of $(\sigma,\kappa)$ as (0,0),(0,0.01),(0.01,0) and (0.01,0.01) will result in different wormhole geometries but with very identical ringdown behavior and QNMs. In essence, they can be thought as mimickers of each-other. Once a gravitational wave signal is detected, the signal is then probed to obtain information about the source. Quasi-normal modes, through parameter estimation, help in narrowing down the properties of the source. If two distinct astrophysical objects emit similar gravitational wave signal with QNM values almost identical, then the process of identifying the source becomes tricky. We explore this scenario assuming a scalar 
wave perturbing the wormhole.  We compare and contrast the 
scalar ringdown signal and the QNMs for parameter values that are around the 
above-mentioned switch-over values . Table \ref{tab:compare} shows the fundamental QNMs calculated using direct integration for four different wormhole spacetimes that result in near identical modes. In order to distinguish them, very high precision is required as the change occurs after second decimal place. Similarly, in Fig.(\ref{fig:Compare_II_III}) we compare the scalar ringdown behaviour of wormholes belonging to Case II and Case III with $\sigma=0.01$ which appear to be very identical. \\
\noindent In contrast, when we compare the scalar QNMs for wormholes which are close to black hole parameter values, we observe a distinct difference (see Table \ref{tab:compare1}). While the real component of the QNM is very similar for both geometries, the imaginary part of the QNM for the wormhole is very small. Also, if we compare wormholes belonging to Case VI with the Hayward regular black hole scenario, then we have $\sigma$ close to unity. This leads to distinct double potential barriers giving rise to echoes that dominate the ringdown signal of the wormhole. Hence, the wormhole and black hole geometries can be easily distinguished from the ringdown profile and also through their respective QNM values. \\
The above discussion once again highlights the significance of our metric. Apart from shifting between different wormhole and black hole spacetimes by choosing 
specific parameter values to study their ringdown behaviour, the form of the metric makes it suitable to compare between different spacetimes. We explored the possibility of using
scalar QNMs as a tool to distinguish various geometries. We found different wormholes to be very identical in their behaviour under scalar perturbations and thus will need very high precision to distinguish them only through such scalar quasi-normal modes. 

 \begin{table}[h]
  \centering
  \begin{tabular}{|c|c|c|c|c|}
      \hline
     m &  Case II & Case III $(\sigma=0.01)$ & Case IV ($\kappa =0.01$) & Case VI ($\sigma=\kappa=0.01)$\\ 
    \hline 
  1 & 0.755245 -i 0.1787874 & 0.7516804 -i 0.177469 & 0.7552616 -i 0.178785 & 0.751697 -i 0.17746636\\
  2 & 1.253101 -i 0.177527 & 1.246957 -i 0.1762021 & 1.2531305 -i 0.1775244 & 1.246987 -i 0.1761995\\
  3 & 1.752240 -i 0.1771437 & 1.7435064 -i 0.175854 & 1.752232 -i 0.1771796 & 1.743549 -i 0.175852\\
  \hline
\end{tabular}
\caption{\label{tab:compare} $\omega_{QNM}$ values in units of $M^{-1}$ for different modes calculated using DI.}
\end{table}

 \begin{table}[h]
  \centering
  \begin{tabular}{|c|c|c|}
      \hline
     m &  Case V ($\kappa=0.5$) & Case VI $(\sigma=0.9, \kappa=0.5)$\\ 
    \hline 
  1 & 0.2992639 -i 0.0928689 & 0.319806 -i 0.0089489 \\
  2 & 0.4936223 -i 0.09224176 & 0.4895156 -i 0.0020109 \\
  3 & 0.689167 -i 0.09208116 & 0.6578297 -i 0.00034295 \\
  \hline
\end{tabular}
\caption{\label{tab:compare1} Comparing $\omega_{QNM}$ ($M^{-1}$) values for Hayward regular black hole and Case VI wormhole calculated using DI.}
\end{table}

\section{Conclusions}

\noindent  Black hole solutions in GR have a singularity hiding behind the event horizon. One possible route towards 
avoiding the singularity in black hole spacetimes is to search for
regular spacetimes with or without event horizon(s). Such spacetimes, devoid of any singularity, were first introduced by Bardeen and
have been named as `regular' black holes. Later work by Hayward, Neves and Saa provided newer examples of such regular black holes. In our work
here, we begin with the Hayward spacetime and generalize the metric by introducing two {\em different} mass parameters $M_1$ and $M$ 
which appear in the $g_{tt}$ and $g_{rr}$ components of the metric,  respectively. The parametrized metric thus obtained not only gives regular black holes but also represents various wormhole geometries as
well as the singular Schwarzschild black hole, for specific choices of parameters. We have analysed in detail each  of the six distinct spacetimes (included in our general metric) for different values of the dimensionless parameters $(\sigma,\kappa)$. The allowed parameter space extends from $\sigma \in [0, 1]$ and $\kappa \in [0, \frac{4}{3\sqrt{3}}]$. 

\noindent As mentioned in the Introduction, we acknowledge the problems associated with non-singular spacetimes namely; energy-condition violating matter threading wormholes and the mass-inflation instability associated with the inner horizon of regular black holes. 
However, instead of being deterred by such issues, through our work, we have
tried to shed light on the stability question of these geometries under scalar perturbations, by calculating the scalar QNMs. The analysis of this
ringdown gives information about the
viability of these models as astrophysical entities. Our 
proposed generalized Hayward metric is especially suitable for this purpose as the behaviour of multiple types of regular spacetimes can be visualised 
simultaneously in a single framework.

\noindent In Section III, we discussed the nature of the spacetimes as determined by the behaviour of the corresponding curvature scalars, energy conditions and also through the embedding diagrams for the geometries.  The black hole solutions (both regular and singular) occur only when $\sigma=1$ while rest of the parameter space gives rise to distinct wormhole spacetimes. For $\sigma=\kappa=0$ we get the Schwarzschild wormhole, $0<\sigma<1, \kappa=0$ gives the well-known Damour-Solodukhin wormhole and the remaining parameter ranges lead to various unique wormhole solutions that have not been explored yet. The later half of our work deals with the stability analysis of these spacetimes under massless scalar wave propagation. The stability is deduced by obtaining the time domain profile and calculating the corresponding scalar quasi-normal modes of the spacetimes. We have found all the geometries to harbour modes with negative imaginary component indicating damping of the signal over time. Some wormholes also show double barrier potential that result in repetitive `echo' patterns in the signal. The dependence of the quasinormal modes on the metric parameters $(\sigma,\kappa)$ has also been explored by obtaining fitting functions for each spacetime and is summarized in Tab.\ref{tab:parameter_QNM}. The dependence on parameters is rather intriguing for the Hayward wormhole belonging to Case IV. The imaginary component of $\omega$ is found to initially decrease with $m$ for small $\kappa$ but increasing with $m$ for larger $\kappa$ values. The shift in behaviour occurs around $\kappa \approx 0.4$. Again for the Case VI wormholes, the dependence on the parameters is worth noting. Now the real part of $\omega$ is seen to increase with $\kappa$ for small $\sigma$. But as $\sigma$ 
approaches unity, $Re (\omega)$ starts to decrease with increasing $\kappa$. Such interesting dependences of the QNMs on the metric parameters are also reflected through the fitting functions mentioned within our work.

 \begin{table}[h]
  \centering
  \begin{tabular}{|c|c|c|c|c|}
      \hline
    \textbf{Spacetime} & \textbf{Re(\boldsymbol{$\omega$})} & \textbf{Im(\boldsymbol{$\omega$})} & \boldsymbol{$\sigma$} &  \boldsymbol{$\kappa$} \\ 
    \hline 
    \hline
  Schwarzschild BH & - & - & 1 & 0 \\
  \hline
  Schwarzschild WH & - & - & 0 & 0 \\
   \hline
  DS WH & $\downarrow$ & $\downarrow$ & Increasing & 0 \\
  \hline
  Hayward WH & $\uparrow$ & $\downarrow$ & 0 & Increasing\\
   \hline
   Hayward BH & $\uparrow$ & $\downarrow$ & 1 & Increasing\\
   \hline
   Case VI WH & $\uparrow$ & $\downarrow$ & Fixed ($\sigma < < 1$) & Increasing\\
   & $\downarrow$ & $\downarrow$ & Fixed ($\sigma \rightarrow 1$) & Increasing\\
   & $\downarrow$ & $\downarrow$ & Increasing & Fixed\\
  \hline
\end{tabular}
\caption{Table shows the dependence of the QNMs on the parameters of the metric for different spacetimes. $\uparrow$ and $\downarrow$ indicates respectively increasing and decreasing behaviour of the corresponding quantity with the change in parameter. For Case VI wormholes,  since both ($\sigma,\kappa$) change we have kept one fixed and have studied the dependence on the other. The 
inferences are for fixed $m$ value.}
\label{tab:parameter_QNM}
\end{table}

\noindent Some spacetimes lying close to each other in the parameter space are found to have near identical ringdown behaviour and quasinormal modes
which makes their possible detection even more involved. Through specific examples we have shown the how a similar ringdown profile 
arise for wormholes even if they belong to different parameter ranges.  Thus, in any real detection scenario, segregating such wormholes would demand high precision observations as the modes differ only after second decimal place. On the contrary, the wormholes close to black holes in the parameter space behave quite distinctly. Such wormholes possess well-separated double potential peaks that lead to echo patterns in the ringdown signal. The wormholes also have a very long damping time indicating  differences, in comparison to a black hole. \\
In conclusion, our parametrized metric provides an elegant way of probing multiple spacetimes with diverse characterisitics, within a 
single framework. The metric also facilitates comparison among the different spacetimes, generated from our parametrized metric, about their behaviour under scalar wave propagation. Our work can be extended in many possible directions. The wormhole spacetimes obtained can be studied for their geodesics and their stability can be explored under axial and polar gravitational perturbations. Future studies along different directions (perturbations or other)
will hopefully throw more light
on how such wormholes or black holes, as introduced here, may serve as possible candidates of
black hole mimickers.

\section*{Acknowledgements}

\noindent PDR thanks Indian Institute
of Technology, Kharagpur, India for support and for allowing her to use available facilities there.

\appendix

\section{Computation of tortoise coordinate for black holes and wormholes}

\noindent The tortoise coordinate defined in eq.(\ref{eq:tortoise}) is integrated numerically for any general values of parameter $(\sigma,\kappa)$ using {\em Mathematica}.  
Depending on the nature of the spacetime, the initial conditions are specified for solving the differential equation of the tortoise coordinate i.e. eq.(\ref{eq:tortoise}). For wormholes we set $r_*= 0 \hspace{0.1in} (r=r_0 + \epsilon)$ where $\epsilon << 1$. This ensures smooth numerical integration as the location of the throat is a root of the function $f(r)$ making eq.(\ref{eq:tortoise}) diverge exactly at the throat. The accuracy and precision of the integration is chosen such that changing them slightly does not affect our results providing robustness to the calculated $r_*$.
The tortoise coordinate for black holes rapidly goes to $-\infty$ as $r \rightarrow$ horizon. So the condition for integrating the tortoise coordinate is specified at some positive value of $r$ where the behavior of $r_*$ is not drastic. The authenticity of the computation has been verified by comparing results with Schwarzschild black hole where the tortoise coordinate is exactly known. The following figures (see Figs.(\ref{fig:A1}) and (\ref{fig:A2})) show the time domain profiles obtained for scalar potential with $m=1$ of the Schwarzschild black hole. We find the two integration schemes giving identical profiles validating the numerical integration approach for tortoise coordinate for general spacetimes.

\begin{figure}[!h]
\begin{subfigure}{0.43\textwidth}
	\includegraphics[width=0.78\linewidth]{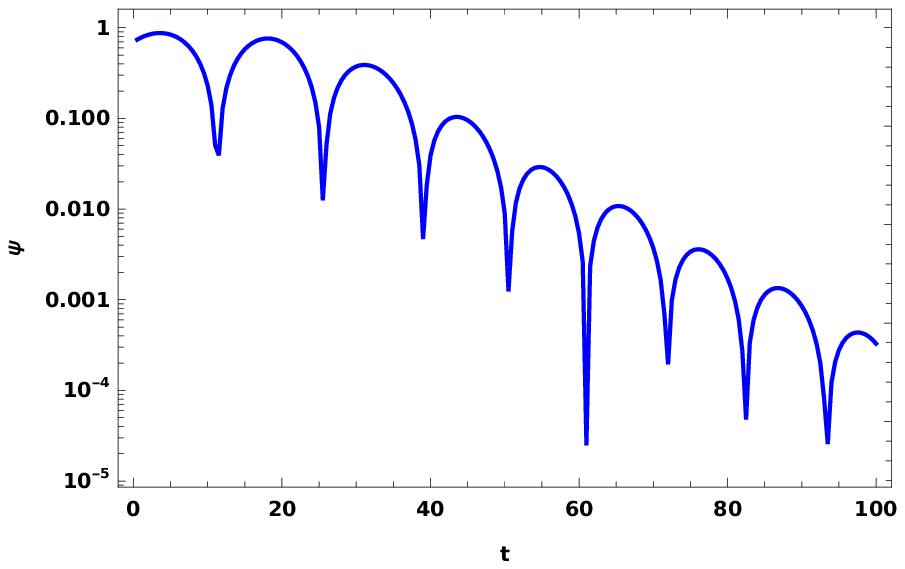}
 	\caption{TD using exact form of tortoise coordinate}
  \label{fig:A1}
 \end{subfigure}
 \begin{subfigure}{0.43\textwidth}
 	\includegraphics[width=0.78\linewidth]{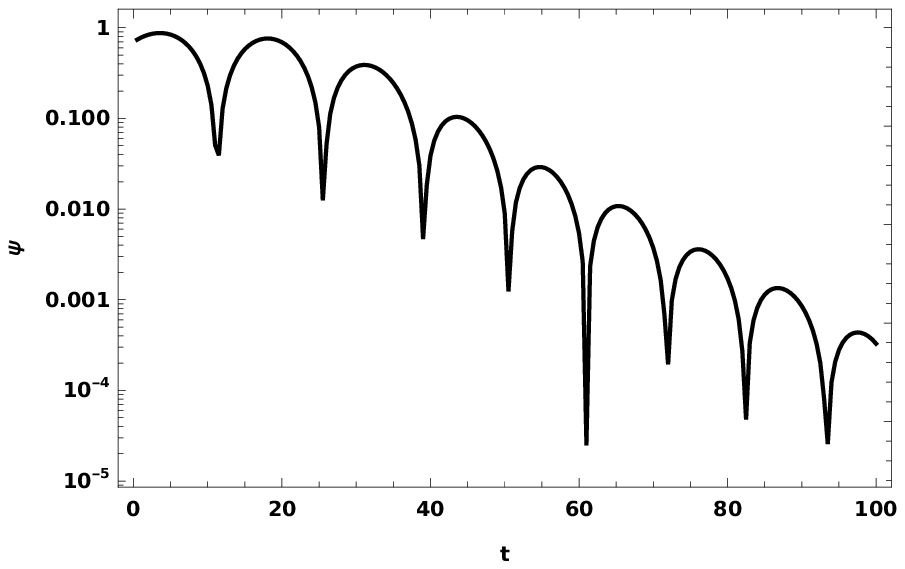}
 	\caption{TD using tortoise coordinate obtained by numerical interpolation}
 	 \label{fig:A2}
 	\end{subfigure}
\end{figure}
\bibliography{ref}

\end{document}